\documentclass[showpacs,preprintnumbers]{revtex4}
\usepackage{amsfonts}
\usepackage{amssymb}
\usepackage{amsmath}
\usepackage{epsfig}
\usepackage{tabularx}
\usepackage{array}
\allowdisplaybreaks
\begin{document}
\title{\bf Quasi-static evolution of axially and reflection symmetric large-scale configuration}

\author {Z. Yousaf$^{1}$ \thanks{zeeshan.math@pu.edu.pk}, Kazuharu Bamba$^2$
\thanks{bamba@sss.fukushima-u.ac.jp} M. Z. Bhatti$^{1}$ \thanks{mzaeem.math@pu.edu.pk}and
U. Farwa$^{1,3}$ \thanks{ume.farwa514@gmail.com\\ume.farwa@math.uol.edu.pk}\\
$^{1}$Department of Mathematics, University of the Punjab,\\
Quaid-i-Azam Campus, Lahore-54590, Pakistan\\
$^2$ Faculty of Symbiotic Systems Science,\\ Fukushima University, Fukushima 960-1296, Japan\\
$^3$Department of Mathematics and Statistics, University of Lahore,\\
1-KM Defence Road, Lahore, 54000, Pakistan}

\keywords{Axial spacetime; Modified gravity theory; Structure
scalars.}
\pacs{04.40.Nr, 04.50.Kd, 04.40.Dg}
\begin{abstract}
We review a recently offered notions of quasi-static evolution of the
axial self-gravitating structures at large-scales
and the criterium to characterize the corresponding evolutionary aspects under the influence of strong curvature regimes.
In doing so, we examine the axial source's dynamic and quasi-static behavior within the parameters of various modified gravity theories. We address the formalism of these notions and their possible implications in studying the dissipative and anisotropic configuration. We initiate by considering higher-order curvature gravity.
The Palatini formalism of $f(R)$ gravity is also taken into consideration to analyze the behavior of the kinematical as well as the dynamical variables of the proposed problem.
The set of invariant-velocities is defined to comprehend the concept
of quasi static-approximation that enhances the stability of the system in contrast to the dynamic mode. It is identified that
vorticity and distinct versions of the structure scalars $Y_{I}$, $Y_{II}$ and $Y_{KL}$ play an important role in revealing the significant effects of a fluid's anisotropy. As another example of evolution, we check the influence of Palatini based factors on the shearing motion of the object. A comparison-based study of the physical nature of distinct curvature factors on the propagation of the axial source is exhibited.
This provides the intriguing platform to grasp the notion of quasi-static evolution together with the distinct curvature factors at current time scenario. The importance of slowly evolving axially symmetric regimes will be addressed through the distinct modified
gravitational context. Finally, we share a list of queries that, we believe, deserve to be addressed in the near future.
\end{abstract}
\maketitle

\tableofcontents \clearpage

\section{Introduction}
The term evolution typically refers to the development of living beings; however, the mechanism through which the sun, stars, galaxies, and galaxy clusters originate and evolve over time is also a kind of evolution. There is a change over time in each of these circumstances. Objects that mainly interact via gravitational forces are said to be self-gravitating objects. Complex phenomena like tidal interactions, dynamical friction, and gravitational collapse are involved in physical mechanisms that control their evolution. Anisotropic matter, because of their influence on local pressure, are of special relevance to the researchers investigating the evolution of such self-gravitating objects. Astrophysics must have an in-depth grasp of how such entities form in order to better comprehend the dynamics and evolution of stellar bodies.
Stellar evolution describes the process by which an astronomical object changes over time. A star's mass, after passing through evolutionary phases, influences its fate, which can range from low to high.
The analysis of large-scale structures like stars and galaxies as well as their clusters, gives insights into the dynamics of the cosmos.
A considerable number of stars end up spending a lot of their active lifetimes in this state of equilibrium, fusing hydrogen into helium, yet it is the steady transition of elements through the fusion mechanism that allows their setup to change in any significant way.
Dynamical stability is the characteristic of an object or system to regain its stable position whenever it is disturbed due to inconsistencies and fluctuations. It has utmost relevance to the structure formation and evolution of self-gravitating bodies. The instability/stability of celestial bodies has great importance in Einstein's gravity as well as in modified gravity.
To investigate self-gravitating celestial bodies, we may take into consideration three feasible evolutionary regimes, namely: static evolution, quasi-static evolution, and dynamic evolution. In a static configuration, a coordinate structure can always be selected in such a way that all geometric as well as physical quantities are free from timelike coordinates. In this case, a time-like hyper-surface (also orthogonal) killing vector is revealed by spacetime. Afterwards, the system undergoes a complete dynamic phase,
where it is regarded as being out of equilibrium (either dynamic or thermal). In between the two phases mentioned above, there is quasi-static (QS) evolution. The system evolves slowly at every moment, so it may be regarded as in a state of equilibrium in such an evolution.
This indicates that the system evolves slowly on a time-scale larger than the typical one because the fluid responds to a small perturbative configuration of hydro-static equilibrium. The hydrostatic time scale is the term applied to this type of time scale \cite{1}. Consequently, one can say that the stellar objects are in hydro-static equilibrium in this phase. The quasi-static approach is very effective for several stages of the life of the celestial bodies \cite{2},
because the hydro-static time frame is of the order of $10^{-4}$ seconds for a neutron-star and $4.5$ seconds for a white-dwarf, and more important $27$ minutes for the sun. Hydro-static equilibrium is the state in which a gaseous objects internal pressure exactly balances its gravitational pressure, such as celestial objects.\\
\indent
Researchers have discovered a surprising fact: that the cosmos is
currently undergoing an accelerated-expansion phase, based on recent
astrophysical findings \cite{3,
4,5}. After this discovery, scientists focused their efforts on revealing the secret. They discovered that our observable cosmos may be uniformly constituted by a fluid with negative pressure, i.e., dark energy, which counteracts the gravitational attraction of fluid configurations. Nevertheless, the concept of dark energy with ordinary matter is not possible in Einstein's gravitational theory, making it very difficult to develop a framework for understanding cosmic evolution at all scales while also describing the formation
of dark energy. It is widely assumed that mysterious inexplicable factors known as dark energy and dark matter account for over $95\%$ of the energy composition in the cosmos. Their presence and attributes are still undetermined, implying that they need to examine a fundamentally large area of relativistic astronomy. The nature of the dark matter component is commonly acknowledged based on current large-scale cosmological data.
Another suggested elucidation proposed that Einstein's gravitational theory (EGT) needs to be modified at large-scale ranges. Such modifications or adjustments to EGT are named as alternative gravity theories (AGTs). In these alternative theories, the usual Einstein-Hilbert action is updated so that the leading generalized field equations include the factors that are significant at low-curvature regimes. The prime motive for employing such gravitational theories is to describe recent cosmological and astrophysical data on dark energy and dark matter. The source of the recently found acceleration in universal expansion is a hotly debated issue in today's age of current cosmology \cite{6}. This chapter covers the notion of quasi-static evolution of an axially symmetric system introduced in \cite{7} at large-scales under distinct massive gravity and its consequential influence within the domains of various scenarios. The modification in the action function contributes to these AGTs and their significance outcomes could be examined (for details, please see \cite{8,9,10,11,12,13,
14,15,16,17,18,19,20,
21,22,23,24,25,
26,27,28,29,30,31,32,
33,34,35}).\\
\indent
Gravitational collapse is an important aspect of stellar evolution and can be defined as a highly-dissipative phenomenon for the structure formation of celestial objects. It is the most vital aspect of star evolution. The relevance of the dissipative process in the analysis of gravitational collapse cannot be overstated. The study of the dynamic features of such things is a significant problem in current research.
Thus, the analysis of collapsing mechanism has been
widely examined (for instance, please see
\cite{36,37,38,39}).
Herrera and collaborators \cite{40,41}
addressed at the gravitational explosion of the distinct stellar systems under certain boundary constraints. The study of the evolution and instability of astrophysical objects after their initiation has acquired a lot of attention in theoretical physics,
and it is supposed to be a prevalent theme in AGTs.
The $f(R, T)$ gravity theory has gained great fascination in describing
the late-time accelerated speed up of our universe. There is a key
role of the matter distribution in the formation and evolution of a stellar system. Houndjo \cite{42} selected $f_1(R) + f_2(T )$ as $f(R, T)$ model and explored the transition from a matter-dominated epoch to an accelerating expansion phase.
In last decades, various aspects of the self-gravitating objects have been studied in AGTs (for details, please see \cite{43,44,45,46,47,
48,49,50,51,52,53,54,55,56,57,
58,59}).\\
\indent
In light of the notions presented above, a distinct scalar invariant
incorporating nonlinear elements of the stress-energy tensor might
be regarded to grab non-minimal geometry and matter coupling.
For the very first time, such an endeavor was made \cite{60},
allowing for a correction term
$f(R, T_{\gamma\lambda}T^{\gamma\lambda})$
in the action function of the theory. So-called the energy-momentum squared gravity, which is the vacuum analog of Einstein's gravitational theory. Its effects are only visible within the energy-matter distributions. The results of this theory are extremely obvious in domains with high curvature.
As a result, predicting Einstein's gravity deviations inside the
compact objects is only natural. Several researchers have looked
into this theory further. Akarsu \emph{et al.} \cite{61} discovered the nonlinear fluid aspects in the field equations to enhance cosmic evolutions at low-density domains, as per the specific type of the function $f$ and the variables under discussion. However, these aspects are also studied at high-density domains in \cite{62,63}.
Akarsu \emph{et al.} \cite{64}
presented the very first comprehensive investigation of $f(R,T^2)$ gravity via neutron stars, additionally by using realistic equation of states. By using the equation-of-state, they developed the hydro-static equilibrium configurations. The authors also solved them numerically to find out the mass radius relations for neutron stars. Nari and Roshan \cite{65} also studied compact
stars in $f(R, T^2)$ formalism. The various cosmological solutions have also been studied in the energy-momentum squared gravity \cite{66,67}. Akarsu and his collaborators
\cite{68} coupled the cold dark-matter with gravity by
taking into account the spatially-flat Robertson-Walker line element in
the framework of energy-momentum squared gravity. The axial perturbative configuration of a charged black holes is also evaluated in this gravity theory \cite{69}. Nazari
\emph{et al.} \cite{70} studied this gravity in
Palatini formalism and explored the cosmic behavior of this gravity
with a focus on bouncing solutions.
The current research inspires that $f(R, T^2)$ gravity demands more
attention, and there are numerous unresolved issues that could be
scrutinized. However, some recent advances in cosmological aspects could be seen in \cite{71,72,73,74,75,76,77,78}.\\
\indent
As compared to alternative models, the stellar structures that are proven to be stable in the context of minor perturbations have important physical properties. The set of physical quantities is the set of parameters used to characterize the self-gravitating fluids.
The anisotropic orientation of matter's contents plays a crucial
role in the formation and evolution of such objects.
Next, let us review the idea of structure scalars that have notable significance to apprehend the physical aspects of self-gravitating stellar systems. This concept has gained the attention of astrophysicists. In 2009,
Herrera et al. \cite{79} presented a
detailed study on the relativistic set of equations for spherical
configuration controlled by scalar-variables called as structure
scalars. After this, they \cite{80} utilized
the same idea for $(1+3)$ cylindrical formalism, and figured out
four-set of scalar variables, they also related these scalars to
the basic physical aspects of anisotropic matter contents.
Within the bounds of various high-curvature regimes, the evaluation of structure scalars has recently been the focus of numerous attempts for distinct studies \cite{81,82,83,84,85,86,87,88,
89,90,91,92,93,94,
95,96}.
An object is said to be axially symmetric if its presentation remains invariant when rotated about that axis. In mathematics, symmetry commonly leads to a structure that remains unchanged under certain adjustments, such as scaling, translation, rotation, and reflection. However, in the terminology of physics, symmetry has come to represent invariance, that is, the negation of change under a certain transformation. For instance, any coordinate-transformation \cite{97}. This idea has always been one of the most potent tools of theoretical physics, since it has become conclusive that almost all laws of the universe take place in symmetrical structures. The consideration of spherical symmetry in the analysis is highly significant for the evaluation of self-gravitating bodies because it helps to explain the presence of neutron stars, white-dwarfs and black-holes. However, the different space-times are known to obey the field equations for various physically significant stress-energy tensors. The research on the dynamical study of an axial-symmetric anisotropic matter contents exists in large numbers in the direction of EGT. However, it is a little bit inspiring to deal with such spacetime in
AGTs. Herrera \emph{et al.} \cite{98} studied the
dynamics of axial-symmetric and anisotropic relativistic system by
evaluating scalar-variables in a static configuration. After this
attempt, they \cite{99} also generalized the same
work in order to demonstrate the evolving axial and reflection-symmetric anisotropic stellar objects and reveal nice outcomes corresponding to physical aspects by means of these scalars. Their contributions delivered
gravitational-radiation, heat-dissipation and flow of super energy
associated with magnetic parts of the Weyl-tensor, heat-flux
vector, and vorticity, respectively. The various aspects of an axially symmetric configuration have been studied in other gravity theories \cite{100,101}. More recent developments of this subject under distinct scenarios
may be found in \cite{102,103,104,
105,106}. The consequential influence of extra curvature factors of $f(R,T),~f(R,T^2), \text{ and Palatini} f(R)$ theories
\cite{107,108,109} has been analyzed over the evolutionary regimes of axially and reflection symmetric source.\\
\indent
The quasi-static approximation is effective and sensible for distinct stages of the life of the stellar structures.
The evolution of axially and reflection symmetric systems within the regime of $f(R,T)$ gravity has been addressed in \cite{110}, in addition to that, its quasi-static evolutionary behavior is considered in \cite{111}. In terms of the generalized evolutionary pattern, the consequential influence of the additional curvature terms comprising the minimal-coupling of the matter field, is analyzed to explore the quasi-static impact over the system. However, in \cite{112} the
stronger effects of matter-curvature due to the presence
of the term, i.e., $T_{\gamma\lambda}T^{\gamma\lambda}=T^2$ as a quadratic contribution of fluid composition of axially symmetric structure has been carried out while its corresponding quasi-static study has considered in \cite{113}. It was shown that the effects of energy-momentum squared gravity are more obvious in high-curvature domains. The above-stated results are studied in a metric based approach in which the kinematical quantities get the identical expression as presented in \cite{114} but this scenario was also changed in a Palatini-based approach. The dynamical as well as kinematical factors of the under discussion system get the stronger influence of the additional curvature aspects arising due to Palatini $f(R)$ gravity \cite{115}.
In this article, we review the recently proposed
concept of quasi-static evolution of axially and reflection
symmetric anisotropic and dissipative system at large-scales, i.e.,
by accommodating distinct AGTs and checking the consequential influence of various additional curvature factors over the evolutionary behavior of the axial self-gravitating configuration.
This work offers a fascinating opportunity to understand the concept of quasi-static evolution in conjunction with unique curvature elements in the present-day context in a more precise scenario. In short, we will explore the significance of slowly changing axially symmetric conditions within the context of a modified gravitational framework.
For self-gravitating objects, the importance of quasi-static evolution resides in its potential to describe and examine the gradual alterations and equilibrium scenarios of these entities over time, especially within the context of gravitational interplay. We would like to highlight
its various noteworthiness characteristics which can be studied and and set forth for debate in near future together with the inclusion of
higher-order curvature attributes:
\begin{itemize}
\item
Formation and Structure: A quasi-static method lets us describe how self-gravitating objects like stars or galaxies come into being as well as how their internal structures evolve and alter over time as a consequence of gravitational mechanisms with the larger hydrostatic time-scale.
\item
Analysis of Stability: Self-gravitating object's stability and equilibrium can be investigated using QS evolution.
We can ascertain whether these entities will maintain a stable configuration or suffer substantial alterations as a result of gravitational pull by observing how slowly they evolve.
\item
Behavior Prevision: It assists in the long-term trend analysis of self-gravitating systems, such as stars, planets, galaxies, and etc. One may figure out the future behaviors and properties of these systems depending upon distinct modifications of field and matter profile by taking into account how they fluctuate over time, although assuming a QS manner.
\item
Comprehension of gravitational influence:
It is well acknowledged that self-gravitating bodies are affected by their own field of gravitation. By examining their QS behavior, one may get insight into how these objects or bodies respond to their own gravitational associations with many other objects within their vicinity, which reveal the distinct aspects of current cosmic scenarios.
\item
Energy and Thermodynamical Attributes:
This approach may also be significant and essential for analyzing the thermodynamical attributes and energy aspects of self-gravitating sources. This approach may permit one to track variations in physical parameters such as entropy and temperature as the systems evolve gradually or slowly.
\end{itemize}
In conclusion, QS evolution offers a helpful paradigm for understanding the behavior, stability, and long-term dynamics of self-gravitating objects, illuminating their physical characteristics and interactions in the cosmos.
One can easily comprehend the the notion of quasi-static evolutionary aspects of the axial system through the matter-geometry coupling context as well as by using Palatini based approach.
In this regards, we will also provide the summary table so that
the researcher can easily understand our findings just by looking at the table in the conclusion section.

\section{Axially and Reflection Symmetric Spacetime, Anisotropic and Dissipative Source}
To perform the quasi-static approximation on the evolution of compact stars, we are taking axial and reflection-symmetric compositions of collapsing fluid, that we suppose to be locally-anisotropic and enduring dissipation by means
of heat-flow or/and free-streaming radiations, enclosed by an axially
symmetric surface $\Sigma$. The metric is taken in Weyl spherical-coordinates
as \cite{7}.
\begin{align}\label{1}
ds^2=-A^2(t, r, \theta)dt^{2}+2G(t, r, \theta)
d\theta dt+B^2(t, r, \theta)(dr^{2}
+r^2d\theta^{2})+C^2(t, r, \theta)d\phi^{2},
\end{align}
where the geometric variables such as $A, B$ are dimensionless, and
at the same time $C$ and $G$ have dimensions of $r$.
To discuss the dissipation and anisotropy of compact objects, we consider the general stress energy-tensor in its canonical form as
\begin{align}\label{2}
T^{(m)}_{\gamma\lambda}=(P+\mu)V_{\gamma}V_{\lambda}+P
g_{\gamma\lambda}+\Pi_{\gamma\lambda}+q_{\lambda}V_{\gamma}+q_{\gamma}V_{\lambda},
\end{align}
where $T^{(m)}_{\gamma\lambda}$ indicates specific energy constituents, $\mu$
is the energy-density, $P$ the isotropic pressure, $\Pi_{\gamma\lambda}$
the anisotropic-tensor, $q_{\gamma}$ the heat flux, and
$V_{\gamma}$ denotes the four-velocity ascribed by a specific observer.
The fluid components are at rest in the
Eckart frame. We have taken co-moving fluid components into accounts in our situation. So,
\begin{align}\label{3}
V^{\gamma}=\left(\frac{1}{A},0,0,0\right);~~\quad
V_{\gamma}=\left(-A,0,\frac{G}{A},0\right).
\end{align}
The unit space-like vectors are now presented in component form as follows
\begin{align}\label{4}
K_{\gamma}=\left(0,B,0,0\right),\quad
L_{\gamma}=\left(0,0,\frac{\sqrt{A^2B^2r^2+G^2}}{A},0\right),\quad
S_{\gamma}=\left(0,0,0,C\right),
\end{align}
keeping the relationship
\begin{align}\label{5}
&V^{\gamma}V_{\gamma}=-K_{\gamma}K^{\gamma}=-L_{\gamma}L^{\gamma}
=-S_{\gamma}S^{\gamma}=-1,\\\label{6}
&V^{\gamma}K_{\gamma}=V^{\gamma}L_{\gamma}=V^{\gamma}S_{\gamma}
=S_{\gamma}K^{\gamma}
=K^{\gamma}L_{\gamma}=S_{\gamma}L^{\gamma}=0.
\end{align}
The unitary vectors $V_{\gamma}, K_{\gamma}, L_{\gamma},
S_{\gamma}$ form orthonormal-tetrad ($e^{\alpha}_{\gamma}$) such as
${e_{\gamma}}^{(0)}=V_{\gamma};~
{e_{\gamma}}^{(1)}=K_{\gamma};~
{e_{\gamma}}^{(2)}=L_{\gamma};~
{e_{\gamma}}^{(3)}=S_{\gamma}$,
here $\alpha=0, 1, 2,3$ and the dual-vector tetrad is represented by
$\eta_{(a)(b)}=g_{\gamma\lambda}e_{(a)}^{\gamma}e_{(b)}^{\lambda}$,
where $\eta_{(a)(b)}$ depicts the Minkowskian space-time. The
scalar functions are used to define anisotropic tensor as presented in
\cite{7}.
\begin{align}\nonumber
\Pi_{\gamma\lambda}&=\frac{1}{3}(2\Pi_{I}+\Pi_{II})\left(K_{\gamma}K_{\lambda}
-\frac{h_{\gamma\lambda}}{3}\right)+\frac{1}{3}(\Pi_{I}
+2\Pi_{II})\left(L_{\gamma}L_{\lambda}-\frac{h_{\gamma\lambda}}{3}\right)
+\Pi_{KL}K_{(\gamma}L_{\lambda)}; \text{ where}\\\nonumber
\Pi_{I}&=(2K^{\gamma}K^{\lambda}-L^{\gamma}L^{\lambda}
-S^{\gamma}S^{\lambda})T_{\gamma\lambda},\quad
\Pi_{II}=(2L^{\gamma}L^{\lambda}-K^{\gamma}K^{\lambda}
-S^{\gamma}S^{\lambda})T_{\gamma\lambda},\quad
\Pi_{KL}=T_{\gamma\lambda}L^{\gamma}K^{\lambda}.
\end{align}
This specific set of scalars is useful for evaluating the required equations
in a more straightforward and concise manner. The heat-flux vector is now
introduced in term of two scalars $q_{I}$ and $q_{II}$ as
\begin{align}\label{7}
&q_{\gamma}=q_{I}K_{\gamma}+q_{II}L_{\gamma}.
\end{align}
It is observed that $q^{\gamma}V_{\gamma}=0$, therefore in
coordinate-components \cite{99} that
$q_{\gamma}=\left(0,Bq_{I},\frac{\sqrt{A^2r^2B^2+G^2}q_{II}}{A},0\right)$.
In general, all of the variables described above are influenced by $t,r$, and $\theta$.
Kinematical variables have an important role in understanding the
evolution and physical meanings of self-gravitating entities. In
addition to the four-acceleration $(a_{\gamma})$, shear-tensor
$(\sigma_{\gamma\lambda})$ and expansion-scalar $(\Theta)$, we also
describe the vorticity-tensor $(\Omega_{\gamma\lambda})$. In this perspective, the four-acceleration and its non-vanishing components are defined as
\begin{align}\label{8}
a_{\gamma}&=V^{\lambda}V_{\gamma;\lambda}=a_IK_{\gamma}+a_{II}L_{\gamma};
\text{ along with}\\\label{9}
a_I&=\frac{A'}{AB};\quad a_{II}=\frac{A}{\sqrt{B^2r^2A^2+G^2}}\left[
-\frac{G}{A^2}\left(\frac{\dot{A}}{A}-\frac{\dot{G}}{G}\right)
+\frac{A_{,\theta}}{A}\right],
\end{align}
Through out the study, over dot and prime notations describe the derivatives of all quantities with time and radial coordinate, respectively,
whereas, the subscript $\theta$ denotes the derivative with
$\theta\text{-coordinate }$.
The expansion-scalar $\Theta=V_{;\gamma}^{\gamma}$ for axial
and reflection symmetric configuration is
\begin{align}\label{10}
\Theta=\frac{r^2AB^2}{r^2A^2B^2+G^2}\left[
\frac{G^2}{r^2A^2B^2}\left(\frac{\dot{B}}{B}-\frac{\dot{A}}{A}+\frac{\dot{G}}{G}
+\frac{\dot{C}}{C}\right)+2\frac{\dot{B}}{B}+\frac{\dot{C}}{C}\right].
\end{align}
If $\Theta < 0, \Theta > 0$, then it represents contracting and expanding nature of matter contents, respectively. However, $\Theta=0$ indicates the presence of vacuum cavity inside the matter contents. Moreover, the shear tensor quantifies the distortion in shape such that its volume remains
constant. It is defined by
\begin{align}\label{11}
\sigma_{\gamma\lambda}=\sigma_{(a)(b)}
e_{\gamma}^{(a)}e_{\lambda}^{(b)}=V_(\gamma;\lambda) +a_{(\gamma}
V_{\lambda)}-\frac{h_{\gamma\lambda}}{3}\Theta; \text{ where }
h_{\lambda}^{\gamma}=\delta^{\gamma}_{\lambda}+V^{\gamma}V_{\lambda}.
\end{align}
The non-vanishing components of shear-tensor are
\begin{align}\label{12}
\sigma_{11}&=\frac{1}{3(r^2B^2A^2+G^2)}\left[Ar^2B^4
\left(\frac{\dot{B}}{B}
-\frac{\dot{C}}{C}\right)-\frac{G^2B^2}{A}\left(-\frac{\dot{A}}{A}
-2\frac{\dot{B}}{B}
+\frac{\dot{C}}{C}+\frac{\dot{G}}{G}\right)\right],\\\label{13}
\sigma_{22}&=\frac{-1}{3A}\left[B^2r^2\left(-\frac{\dot{B}}{B}
+\frac{\dot{C}}{C}\right)
-\frac{G^2}{A^2}\left(2\frac{\dot{A}}{A}-\frac{\dot{C}}{C}
+2\frac{\dot{G}}{G}+\frac{\dot{B}}{B}
\right)\right],\\\label{14}
\sigma_{33}&=\frac{1}{3(r^2A^2B^2+G^2)}\left[2AB^2r^2C^2\left(
-\frac{\dot{B}}{B}
+\frac{\dot{C}}{C}\right)-\frac{G^2C^2}{A}\left(
-\frac{\dot{A}}{A}+\frac{\dot{B}}{B}
-2\frac{\dot{C}}{C}+\frac{\dot{G}}{G}\right)\right].
\end{align}
The above factors of the shear tensor are not mutually exclusive. As a result, two of its scalar functions, $\sigma_{I}$ and $\sigma_{II}$ can be used to express it as
\begin{equation}\label{15}
\sigma_{\gamma\lambda}=\frac{1}{3}(2\sigma_{I}
+\sigma_{II})\bigg(K_{\gamma}K_{\lambda}-\frac{h_{\gamma\lambda}}{3}\bigg)
+\frac{1}{3}(\sigma_{I}+2\sigma_{II})\bigg(L_{\gamma}L_{\lambda}
-\frac{h_{\gamma\lambda}}{3}\bigg).
\end{equation}
By using the tetrad relation $\sigma_{\gamma\lambda}=\sigma_{(a)(b)}
e_{\gamma}^{(a)}e_{\lambda}^{(b)}$ and Eqs. \eqref{12}--\eqref{14},
the shear-scalars may be expressed in form of geometric quantities
and their related derivatives as below
\begin{align}\label{16}
&2\sigma_I+\sigma_{II}=\frac{-3}{A}\left\{\frac{\dot{C}}{C}
-\frac{\dot{B}}{B}\right\},\\\label{17}
&\sigma_I+2\sigma_{II}=\frac{3A}{r^2A^2B^2+G^2}
\left[B^2r^2\left\{\frac{\dot{B}}{B}-\frac{\dot{C}}{C}
\right\}-\frac{G^2}{A^2}\left\{\frac{\dot{A}}{A}
-\frac{\dot{G}}{G}+\frac{\dot{C}}{C}\right\}\right].
\end{align}
The careful selection of these scalars is backed up by the
fact that they occur in the generalized scalars equations for
our relativistic system, which will be covered in the upcoming sections.
Finally, $\omega_{\gamma}$ can be used to represent the vorticity-vector by
$\omega_\gamma=\frac{1}{2}\eta_{\gamma\lambda\alpha\beta}
V^{\lambda;\alpha}V^{\beta}
=\frac{1}{2}\eta_{\gamma\lambda\alpha\beta}\Omega^{\lambda\alpha}V^{\beta}$,
here $\eta_{\gamma\lambda\alpha\beta}$ is the Levi-Civita tensor
whereas the vorticity-tensor is
$\Omega_{\gamma\lambda}=V_{[\gamma;\lambda]}+a_{[\gamma}V_{\lambda]}.$
So for, it can be expressed in form of unit-vectors as below
\begin{align}\nonumber
\Omega_{\gamma\lambda}=\left(L_{\gamma}K_{\lambda}
-L_{\lambda}K_{\gamma}\right)\Omega;~~\quad
\omega_{\gamma}=-S_{\gamma}\Omega.
\end{align}
In metric based approaches, it has only one non-zero component that is
\begin{align}\label{18}
\Omega_{12}&=-\frac{G}{2A}\left(-\frac{2A'}{A}+\frac{G'}{G}\right); \text{ with scalar function }
\Omega=-\frac
{G}{2B\sqrt{r^2A^2B^2+G^2}}\left(\frac{2A'}{A}-\frac{G'}{G}\right).
\end{align}
In the case of regularity assumptions at the center, the above equation suggests that $G=0\Leftrightarrow\Omega=0$.
A kinematical problem involves showing the geometry
and figuring out the initial conditions of certain
known configurations of the system.
Kinematics is used in astrophysics to illustrate the motion of
celestial objects including stars, galaxies, and a cluster of such
objects to investigate the large-scale structure of the universe.

\subsection{Some Specific Velocities Associated with Axial Self-gravitating Source}

For a complete understanding of the quasi-static evolution of proposed structures at large-scales, we have to address the specific velocities in addition to the kinematical characteristics provided above. In astrophysical cosmology, kinematics is applied in analyzing the dynamics of the stellar formation. Thereby, we now focus on the kinematical aspects that exhibit the fluid's motion. Now, we will focus on the kinematical factors that distinguish the motion of the fluid, as revealed in the previous section.
The formulation including the space-like triad $(e^{\gamma}_{(i)}
; i=1,2,3)$ can be used to establish the collection of particular-velocities
as in \cite{7}.
\begin{align}\label{19}
(\frac {D_T(\delta l)}{\delta l})_{(i,j)}=e^{\gamma}_{(i)}
e^{\lambda}_{(j)}\left(\sigma_{\gamma\lambda}
+\frac{h_{\gamma\lambda}}{3}\Theta
+\Omega_{\gamma\lambda}\right).
\end{align}
From Eq. \eqref{19}, we get
\begin{align}\label{20}
V_{(1)}&=K^{\lambda}K^{\gamma}\left(\sigma_{\lambda\gamma}
+\Omega_{\lambda\gamma}+\frac{h_{\lambda\gamma}}{3}\Theta\right);\quad
V_{(2)}=L^{\lambda}L^{\gamma}\left(\sigma_{\lambda\gamma}+\Omega_{\lambda\gamma}
+\frac{h_{\lambda\gamma}}{3}\Theta\right);\\\label{21}
V_{(3)}&=S^{\lambda}S^{\gamma}\left(\sigma_{\lambda\gamma}
+\Omega_{\lambda\gamma}
+\frac{h_{\lambda\gamma}}{3}\Theta\right);\quad
V_{(1,2)}=K^{\lambda}L^{\gamma}\left(\sigma_{\lambda\gamma}
+\Omega_{\lambda\gamma}
+\frac{h_{\lambda\gamma}}{3}\Theta\right);\\\label{22}
V_{(1,3)}&=K^{\lambda}S^{\gamma}\left(\sigma_{\lambda\gamma}
+\Omega_{\lambda\gamma}
+\frac{h_{\lambda\gamma}}{3}\Theta\right).
\end{align}
Using Eqs. \eqref{10}--\eqref{11} and \eqref{18}, we obtain
\begin{align}\label{23}
&V_{(1)}=\frac{1}{3}(\sigma_I+\Theta);\quad
V_{(2)}=\frac{1}{3}(\sigma_{II}+\Theta);\quad V_{(1,3)}=0;\quad
V_{(1,2)}=-\Omega; \quad
V_{(3)}=-\frac{1}{3}\left(\sigma_{I}+\sigma_{II}-\Theta\right),
\end{align}
satisfying the relation
\begin{align}\label{24}
\Theta=V_{(1)}+V_{(2)}+V_{(3)}&.
\end{align}
These specific parameters are observed to define the proper time-variation
of $\delta l$ (infinitesimal distance). Kinematical quantities and
unit space-like vectors, influence the geometrical and physical
illustration of these specific quantities, as shown in
Eqs.\eqref{20}--\eqref{22}.\\
\subsection{The Weyl Tensor and Structure Scalars: Notations and Basic Definitions}

The consideration of pressure anisotropy is premised on the idea
that local pressure-anisotropy can be induced by a wide range of
physical characteristics, that we anticipate observing in compact
objects \cite{106,116,
117}. Among all available scenarios of anisotropy,
there are main two that are especially relevant to our core concern.
One is a strong magnetic field that can be seen in compact stellar
objects like neutron-stars, white-dwarfs, or magnetized strange
quark-stars (for instance, please see \cite{118,
119}). It is a well-known fact
that Pressure-anisotropy is produced by a magnetic field operating
on a Fermi gas \cite{120}. In certain aspects, the
magnetic field may be regarded as anisotropy of fluid. The viscosity
is another cause of anisotropy thought to exist in neutron-stars or
generally in dense objects \cite{121,122}.
To study the electric and magnetic aspects of the Weyl-tensor, we
would like to introduce the electric $(E_{\gamma\lambda})$ and
magnetic $(H_{\gamma\lambda})$ parts of the Weyl-tensor
$(C_{\gamma\lambda\beta\alpha})$. These are usually defined as
\cite{7}.
\begin{align}\label{25}
E_{\gamma\lambda}&=C_{\gamma\beta\lambda\alpha}V^{\alpha}V^{\beta};~~~
\quad H_{\gamma\lambda}=\frac{1}{2}\eta_{\gamma\beta\epsilon\rho}
C^{~~~\epsilon\rho}_{\lambda\alpha}V^{\alpha}V^{\beta}.
\end{align}
Thus these might be expressed as
\begin{align}\label{26}
E_{\gamma\lambda}&=\frac{1}{3}\left(2\varepsilon_I
+\varepsilon_{II}\right)\left(K_{\gamma}K_{\lambda}
-\frac{h_{\gamma\lambda}}{3}\right)+\frac{1}{3}\left(\varepsilon_I
+2\varepsilon_{II}\right)
\left(L_{\gamma}L_{\lambda}-\frac{h_{\gamma\lambda}}{3}\right)
+\varepsilon_{KL}\left(K_{\gamma}L_{\lambda}
+K_{\lambda}L_{\gamma}\right),\\\label{27}
H_{\gamma\lambda}&=H_1\left(K_{\gamma}S_{\lambda}
+K_{\lambda}S_{\gamma}\right)+H_2
\left(L_{\gamma}S_{\lambda}+L_{\lambda}S_{\gamma}\right);
\text{ where } H_1,H_2=\text{ magnetic parts; }
\varepsilon_I,\varepsilon_{II},\varepsilon_{KL}= \text{ electric parts }
\end{align}
In the field of relativistic astrophysics, evolution of massive
objects has arisen as a major concern.
Its component preserve relationships that are
$E_{\lambda}^{\lambda}=E_{\alpha\lambda}V^{\lambda}=0$ and
$H_{\lambda}^{\lambda}=H_{\alpha\lambda}V^{\lambda}=0$.
Structure scalars are
significant in assessing the physical characteristics of fluid
contents. Therefore, for the evaluation of scalar variables, we
consider three-tensors $X_{\gamma\lambda},~Y_{\gamma\lambda}$ and
$Z_{\gamma\lambda}$ with the use of the Riemann-tensor.
\begin{align}\label{28}
X_{\gamma\lambda}&=\frac{1}{2}\eta^{~~~\epsilon\rho}_{\gamma\beta}
R^{*}_{\epsilon\rho\lambda\alpha} V^{\alpha}V^{\beta};~~\quad
Y_{\gamma\lambda}=R_{\gamma\beta\lambda\alpha}V^{\beta}V^{\alpha};~~\quad
Z_{\gamma\lambda}
=\frac{1}{2}\epsilon_{\gamma\epsilon\rho}
R^{~~~\epsilon\rho}_{\alpha\lambda}V^{\alpha},
\end{align}
with $\epsilon_{\gamma\lambda\rho}=\eta_{\beta\gamma\lambda\rho}V^{\beta}$
and $R^{*}_{\gamma\lambda\beta\alpha}
=\frac{1}{2}\eta_{\epsilon\rho\beta\alpha}
R^{~~\epsilon\rho}_{\gamma\lambda}$.
In the case of axial and reflection symmetric source, above tensors can be divided into their trace and trace-free components as follows
\begin{align}\label{29}
&X_{\gamma\lambda}=\frac{h_{\gamma\lambda}}{3}X_T
+\frac{1}{3}(2X_I+X_{II})\bigg(K_{\gamma}K_{\lambda}
-\frac{h_{\gamma\lambda}}{3}\bigg)
+\frac{1}{3}(X_I+2X_{II})\bigg(L_{\gamma}L_{\lambda}
-\frac{h_{\gamma\lambda}}{3}\bigg)+X_{KL}
(K_{\gamma}L_{\lambda}+L_{\gamma}K_{\lambda}),\\\label{30}
&Y_{\gamma\lambda}=\frac{h_{\gamma\lambda}}{3}Y_T
+\frac{1}{3}(2Y_I+Y_{II})\bigg(K_{\gamma}K_{\lambda}
-\frac{h_{\gamma\lambda}}{3}\bigg)
+\frac{1}{3}(Y_I+2Y_{II})\bigg(L_{\gamma}L_{\lambda}
-\frac{h_{\gamma\lambda}}{3}\bigg)+Y_{KL}
(K_{\gamma}L_{\lambda}+L_{\gamma}K_{\lambda}),\\\label{31}
&Z_{\gamma\lambda}=Z_IK_{\lambda}S_{\gamma}+Z_{II}K_{\gamma}S_{\lambda}
+Z_{III}L_{\gamma}S_{\lambda}+Z_{IV}S_{\gamma}L_{\lambda}.
\end{align}
These scalars are of extraordinary importance for
analyzing the impact of the electric Weyl-tensor components and matter anisotropy. In light of the ramifications of matter descriptions within the constraints of intense gravitational field interactions, we will express the impact of these scalars and the evolution of the conformal tensor in later sections.

\subsection{Generalized Heat Transport Equation and Its Non-vanishing Components}

Diffusion approximation is the most significant and widely used
approximation to understand the dynamics of cosmic rays. Dissipation
because of the emission of massless objects i.e., photons or
neutrinos is the distinctive process in the time-evolution of
compact stars. Indeed, neutrino emission seems the very well-suited
procedure to transfer the volume of the binding energy of that
collapsing star, leading to a black hole or neutron
star. For the study of the thermodynamical
behavior of our problem, the dissipative process is carried out via heat
transportation equation at large scales. This is obtained by using the usual dissipation theory
\cite{123,124}. The heat-flux
yields the following generalized transport equation
\begin{align}\label{32}
{(q^{\lambda})}^{\textrm{eff}}+\mathbb{D}\tau h^{\lambda}_{\mu}q_{;\alpha}^{\mu}V^{\alpha}
=-k\bigg[h^{\lambda\mu}\bigg(\mathbb{T}a_{\mu}+\mathbb{T}_{,\mu}\bigg)
+\frac{\mathbb{T}^2}{2}\bigg(\frac{\tau V^{\beta}}
{k \mathbb{T}^2}\bigg)_{;\beta}{(q^{\lambda})}^{\textrm{eff}}\bigg],
\end{align}
where $\mathbb{T}$ and $k$ demonstrate temperature and thermal conductivity,
respectively, while relaxation-time is represented by $\tau$, which
is the essential parameter corresponding to the theory of dissipation.
It can never be disregarded even the concerned issue have a smaller
time-scale in evolution states. On contraction of Eq. (\ref{32}) with
$K_{\lambda}$, also by making use of Eq. (\ref{18}), the resulting one is
\begin{align}\label{33}
\frac{\tau\mathbb{D}}{A}\bigg\{\dot{q_{I}}-A\Omega
q_{II}\bigg\}+{q_{I}}^{\textrm{eff}}
=-k\bigg[\frac{1}{B}\bigg(B\mathbb{T}a_I+\mathbb{T}'\bigg)
+\frac{\mathbb{T}^2}{2}\bigg(\frac{\tau
V^{\beta}} {k
\mathbb{T}^2}\bigg)_{;\beta}{q_{I}}^{\textrm{eff}}\bigg],
\end{align}
however, by contracting Eq. \eqref{32} with $L_{\lambda}$, we receive
\begin{align}\label{34}
&\frac{\tau\mathbb{D}}{A}\bigg\{\dot{q_{II}}+A\Omega
q_{I}\bigg\}+{q_{II}}^{\textrm{eff}}
=-k\bigg[\mathbb{T}a_{II}+\frac{1}{A}\bigg\{\frac{G\dot{\mathbb{T}}
+A^2\mathbb{T}_{,\theta}}{\sqrt{r^2A^2B^2+G^2}}\bigg\}
+\frac{\mathbb{T}^2}{2}\bigg(\frac{\tau V^{\beta}} {k
\mathbb{T}^2}\bigg)_{;\beta}{q_{II}}^{\textrm{eff}}\bigg].
\end{align}
The effective components of heat-flux could be evaluated within the context of considered AGTs at large scales. It is to
be noted that both components of Eq. \eqref{32} are carried out the
influence of extra curvature aspects and are also associated with vorticity
which requires some fascinating outcomes of thermodynamical behavior of the gravitating axial system. It is worthwhile to record the factor $\mathbb{D}$
for distinct massive gravity theories that are going to be addressed in this chapter. Thereby
\begin{itemize}
\item
For $f(R,T)$ gravity; $\mathbb{D}=\frac{1}{f_R}(1+f_T); \text{ with $R$= Ricci Scalar and T=trace of stress-energy tensor }.$
\item
For $f(R,T^2)$ gravity; $\mathbb{D}=\frac{1}{f_R}(1-2\mu f_{T^2})
\text{ where } \mu \text{ is the energy density of the source }.$
\item
For Palatini $f(R)$ gravity; $\mathbb{D}=\frac{1}{f_R}$
\text{ where $R$ } is the Ricci scalar, accommodating the additional curvature factors due to the consequential influence of Levi-Civita connection as well.
\end{itemize}

\subsection{Defining Quasi-static Approximation for Axial Self-gravitating
Source}

In astrophysics, understanding the evolutionary behavior of the self-gravitating systems is a highly important subject.
The QS evolution of such objects is considered
a field of definite interest
because it facilitates a more straightforward
understanding of how these systems behave over
extended periods of hydro-static equilibrium.
Consequently, it can be said that the system faces
variations slowly on a time-scale; this is a very long
compared to the typical time in which the system
responds to small perturbed configurations of hydrostatic.
Thus, we can conclude that in this phase, the system
is convenient to the hydrostatic state of equilibrium
and may be evaluated in the sequence of equilibrium-models.
The QS approximation is the very suitable
approach to deal such an evolution of stellar structures.
Let us now transform this approximation into conditions
for the kinematical quantities, additional curvature
factors that are arising due to the distinct AGTs
under review, and various concepts of specific
velocities provided in the prior section.
These conditions are entailed due to the fact that
the hydro-static time of the system under consideration
must be much larger than any characteristic time-scale of
that system. On account of this
\begin{itemize}
\item All quantities having order O$(\epsilon^2)$ and higher will be
neglected, where $\epsilon<<1.$
\item The specific-velocities such as $V_{(1),(2),(3)}$ and
$V_{(1,2)}$ defined in Eq. \eqref{20}--\eqref{22} are smaller
quantities, therefore have order O$(\epsilon).$
\item
It follows from Eqs.\eqref{20}--\eqref{22} that the scalars
$\Omega,\Theta,\sigma_{I,II}$ have order O$(\epsilon),$ also
Eqs. \eqref{10}, \eqref{16}-\eqref{17} indicate that $G, \dot{A}, \dot{B}$ and $\dot{C}$ are of  O$(\epsilon).$
\item
It is also observed from Eqs. \eqref{10}, \eqref{16}--\eqref{17} that $\tilde{\sigma}\equiv\sigma_{I}=\sigma_{II}$, having the same order i.e., O$(\epsilon)$ and
\begin{align}\label{35}
\Theta-2\tilde{\sigma}=\frac{3}{A}\bigg(\frac{\dot{C}}{C}\bigg);~~\quad
\tilde{\sigma}+\Theta=\frac{3}{A}\bigg(\frac{\dot{B}}{B}\bigg)
\end{align}
\item
The extra curvature factors accommodating from the gravitational interactions of the AGTs under consideration, for instance, $f_R\equiv \tilde{f}_R, \tilde{f}_{T^2} \equiv
f_{T^2},  f_T\equiv
\tilde{f}_T, q^{\textrm{eff}}_I\equiv \tilde{q_{I}}^{\textrm{eff}},
q^{\textrm{eff}}_{II}\equiv \tilde{q_{II}}^{\textrm{eff}},
\mu^{\textrm{eff}}\equiv \tilde{\mu}^{\textrm{eff}}$ and other
effective matter profiles in the QS approximation, accordingly.
\end{itemize}
Furthermore, the relaxation time in the evolution of generalized
transport equations is assumed to be ignored. In short, relaxing
time is the time required by the system to return to its stable
position intuitively after being suddenly taken away from it.
As nature of the QS approximation suggests that all mechanisms
innovate over a longer time-frame than that of transient-phenomena, implying
that we should expect generalized heat-fluxes to characterize a
steady heat flow with the influence of $f(R, T^2)$ constituents.
As a result, in both elements of the generalized transport
equation for relativistic fluid, the relaxation time is disregarded i.e., $\tau=0$. One can be easily calculated the QS approximated values of these
quantities by imposing the above defined constraints.
Henceforth, the scalar components of Eq. \eqref{9} are resulted to be of order
O$(\epsilon)$ after imposing these constraints, accordingly
\begin{align}\label{36}
a_{I}=\frac{A'}{A}\bigg(\frac{1}{B}\bigg);~~\quad a_{II}
=\bigg(\frac{1}{rB}\bigg)\frac{A_{,\theta}}{A}.
\end{align}

\section{Consequences of $f(R,T)$ Corrections Over the Quasi-static evolutionary Behavior of Axial Stellar Structures}

In order to check the QS evolutionary aspects of an axial self-gravitating system at large-scales, first we take into account $f(R, T)$ gravity which is proposed based on
non-minimal coupling between the system's geometry and its matter profile.
The Ricci scalar in Einstein's gravity action function is substituted
with its generic function of the Ricci scalar and trace of stress-energy tensor i.e., $f(R,T)$. It is highlighted that such modification in the Lagrangian may be noticed as the additional degrees of freedom. The equations of motion, that have developed from this type of Lagrangian, are different from EGT and accommodated the consequential influence of extra curvature factors in a better way. These are extremely important for studying the strong gravitational-field interaction concerns (for review, please see
\cite{6,125,126,
127,128}). The generalized
action for $f(R, T)$ gravity is expressed as \cite{119}
\begin{align}\label{1p1}
I_{f(R,T)}=\frac{1}{2\kappa} \int \sqrt{-g} \bigg[f(R,T)+
\textit{L}_m\bigg]d^4x,
\end{align}
where $\textit{L}_m$ is the relative Lagrangian-density of matter
contents. The stress energy-tensor is given as
$T_{\lambda\omega}=-\frac{2}{\sqrt{-g}}\frac{\delta\left(\sqrt{-g}
\textit{L}_m\right)}
{\delta{g^{\lambda\omega}}}$ and
applying variation on Eq. \eqref{1p1} with respect to metric tensor
$g_{\gamma\lambda}$, we receive the following set of equations.
\begin{equation}\label{1p2}
R_{\gamma\lambda}f_{R}-\frac{1}{2}g_{\gamma\lambda}f
+\bigg(g_{\gamma\lambda}\Box -\nabla_{\gamma}
\nabla_{\lambda}\bigg)f_{R}=\kappa
T_{\gamma\lambda}-f_{T}\bigg\{\hat{\Theta}_{\gamma\lambda}
+T_{\gamma\lambda}\bigg\},
\end{equation}
where $g$ is the determinant of metric tensor and $\nabla_{\lambda}$ is
the operator for covariant-derivative, while
$\Box=g^{\gamma\lambda}\nabla_\gamma\nabla_\lambda$ identifies d'Alembert's
operator. Also,
\begin{align}\nonumber
\hat{\Theta}_{\gamma\lambda}=g^{\alpha\beta}\frac{\delta{T_{\alpha\beta}}}
{\delta{g^{\gamma\lambda}}}=-2T_{\gamma\lambda}+g_{\gamma\lambda}\textit{L}_m
-2g^{\alpha\beta}\frac{\partial^2\textit{L}_m}{\partial
g^{\gamma\lambda}\partial g^{\alpha\beta}}.
\end{align}
By choosing  relativistic units $c=G=1$, so for $\kappa=8\pi$ and energy density $(L_m=\mu)$,
then
$\hat{\Theta}_{\gamma\lambda}=-2T_{\gamma\lambda}+\mu g_{\gamma\lambda}$.
From Eq. \eqref{1p2}, field equations in $f(R,T)$ gravity are
\begin{align}\label{1p3}
G_{\gamma\lambda}&={{T}_{\gamma\lambda}}^{\textrm{eff}}
=\frac{1}{f_R}\bigg[(1+f_{T})T_{\gamma\lambda}^{(m)}+\mu
g_{\gamma\lambda}f_{T}+\bigg(\frac{f} {2}-\frac{R}{2}f_R\bigg)
g_{\gamma\lambda}+\nabla_{\gamma}\nabla_{\lambda}{f_R}
-g_{\gamma\lambda}\Box{f_R}\bigg],
\end{align}
where $f\equiv f(R, T)$, $f_{R}= \frac{\partial f}{\partial R}$,
$f_{T}= \frac{\partial f}{\partial T}$, and $G_{\lambda\omega}$
represents Einstein-tensor.
In literature, there have been fruitful discussions on
cosmological reconstruction, viscous solution, stability of
homogeneous universe and dynamical instability in $f(R, T)$ gravity
\cite{129,130,131}. The
$f(R, T)$ gravity theory has gained great fascination in describing
the late-time accelerated to speed up our universe.

\subsection{$f(R,T)$ Structure Scalars and Their Physical Meanings}

The physical nature of self-gravitating matter contents can be
smoothly explained by structure scalars. For the formation and
evolution of celestial objects, many astrophysicists have been using
this idea. To calculate these scalars for our problem under discussion, let us take into account Eq. \eqref{28} through Riemann-tensor for the evaluation of these scalars. So,
the explicit form of these tensors for our study is calculated as
\begin{align}\label{1p4}
X_{\gamma\lambda}=-E_{\gamma\lambda}-\frac{\kappa}{f_R}\bigg\{(1
+f_T)\frac{\Pi_{\gamma\lambda}}{2}
-\mu\frac{h_{\gamma\lambda}}{3}-\bigg(f-Rf_R\bigg)\frac{h_{\gamma\lambda}}
{6}\bigg\}+\psi_1,
\end{align}
comparing this with Eq. \eqref{29}, we get the corresponding four structure scalars as follows
\begin{align}\label{1p5}
X^*_T&=\frac{\kappa}{f_R}\bigg\{\mu+\frac{1}{2}\bigg(f-Rf_R\bigg)\bigg\}
+\psi^{*}_1;\quad\quad
X^*_{I}=-\varepsilon_I-\frac{\kappa}{2f_R}\bigg(1+f_T\bigg)\Pi_I,\\\label{1p6}
X^*_{II}&=-\varepsilon_{II}-\frac{\kappa}{2f_R}\bigg(1+f_T\bigg)\Pi_{II};\quad
\quad X^*_{KL}
=-\varepsilon_{KL}-\frac{\kappa}{2f_R}\bigg(1+f_T\bigg)\Pi_{KL}.
\end{align}
The scalar $X_T$ describes the trace part of $X_{\gamma\lambda}$ that is showing the effects of energy density and $f(R,T)$ corrections.
Whereas, remaining ones are relating to unit space-like
vectors and includes the consequential influence of tidal force and pressure anisotropy within the regime of $f(R,T)$ gravity. Similarly
\begin{align}\label{1p7}
Y_{\gamma\lambda}&=E_{\gamma\lambda}-\frac{\kappa}{2f_R}\bigg[(1
+f_T)\Pi_{\gamma\lambda}+\psi_{2}\bigg]
+\frac{\kappa}{3f_R}\bigg[\mu+3P+3(\mu+P)f_T+2(f-Rf_R)\bigg]; \text{ with }\\\label{1p8}
Y^*_T&=\frac{\kappa}{2f_R}\bigg[(\mu+3P)(1+f_T)+4\bigg\{2{\mu
f_T}+(f-Rf_R)\bigg\}+\psi_3\bigg];\quad
Y^*_{I}=\varepsilon_I-\frac{\kappa}{2f_R}\bigg(1+f_T\bigg)\Pi_I,\\\label{1p9}
Y^*_{II}&=\varepsilon_{II}-\frac{\kappa}{2f_R}\bigg(1+f_T\bigg)\Pi_{II},\quad
Y^*_{KL}=\varepsilon_{KL}-\frac{\kappa}{2f_R}\bigg(1+f_T\bigg)\Pi_{KL},
\end{align}
where the scalar $Y_T$ demonstrates the mutual consequence of pressure, energy density, and the strong gravitational-field interaction factors that are emerged due to $f(R,T)$ gravity. However, the others depict the impact of
tidal forces due to the Weyl tensor and fluid's anisotropy along with
$f(R,T)$  corrections.
Finally
\begin{align}\label{1p10}
Z_{\gamma\lambda}&=H_{\gamma\lambda}+\frac{\kappa}{2f_R}
\bigg(1+f_T\bigg)q^{\omega}\epsilon_{\gamma\lambda
\omega}+\frac{\kappa}{2}\psi_4,
\end{align}
comparing this with Eq. \eqref{31}, we get
\begin{align}\label{1p11}
Z^*_{I}&=H_1-\frac{\kappa}{2f_R}\bigg(1+f_T\bigg)q_{II};~\quad Z^*_{II}
=H_1+\frac{\kappa}{2f_R}\bigg(1+f_T\bigg)q_{II},\\\label{1p12}
Z^*_{III}&=H_2-\frac{\kappa}{2f_R}\bigg(1+f_T\bigg)q_{I};~\quad
Z^*_{IV}=H_2+\frac{\kappa}{2f_R}\bigg(1+f_T\bigg)q_{I}.
\end{align}
The expressions of $\psi_i's$ are given in Appendix A.
The motivation to demonstrate such an analysis and to insight further
these structure scalars in the evolution of self-gravitating compact objects
arise from their various physical aspects.
Dissipation effects on the interior
region of stellar objects is defined by
generalized structure scalars as presented in Eqs. \eqref{1p11} and \eqref{1p12},
obtained from $Z_{\gamma\lambda}$. Consequently, we can say that the
incorporation of $Z^*_{I, II, III, IV}$ has a direct correlation with
the magnetic effects of the Weyl-tensor and heat dissipation. Whereas
the evolution of expansion and shearing rate for axial and reflection symmetric anisotropic and dissipative fluid is controlled by $Y^*_{T}$ and $Y^*_{I, II}$, respectively (as expressed in Eqs. (A8)--(A10) in
Appendix A).
We argue that, besides Einstein's gravity structure scalars, the generalized form of such scalars are also significant in describing compact galactic configuration. It is significant to note that super-massive and enormous compact galactic structures in the universe exclusively exist in $f(R, T)$ gravity. The specific choice of these scalars, is to evaluate the quasi-static behavior of basic modified scalar-equations, which are presented in Appendix A. In short, it could be said that these scalars are the decisive parameters for the evolution of the distinct relativistic consequences of the
axial system.
To discuss the astrophysical application of our work, we would
like to discuss the evolution of expansion and shear in various
configurations. Firstly, we would like to discuss the dynamical instability
of those collapsing objects which evolve with zero expansion i.e., $\Theta=0$
Therefore we consider the Raychaudhuri
equation (A8) and when the system is expansion-free
then from Eq. (A8), we obtain
\begin{align}\nonumber
2\left(\sigma^2-\Omega^2 \right)+Y^*_T=a_{;\mu}^{\mu},
\end{align}
we can rewrite it as
\begin{align}\nonumber
Y^*_{T}=a_{;\mu}^{\mu}+2\left(\Omega^2-\sigma^2\right),
\end{align}
if $\Theta=0$ and $a_{\mu}=0$, illustrating the motion of no
expansion system along a geodesic. In geodesic case
$Y^*_T$ governs the vorticity and shearing motion of the system.
Under these constraints, the system would experience two very
fascinating dynamical processes. The above equation describe the role of
structure scalar $Y^*_{T}$ in terms of $\Omega$ and $\sigma$
for the following types of stellar systems.
\begin{itemize}
\item
The condition $\Theta=0$ gives the formation of two
different boundaries within the stellar systems. First boundary lies at
the outside that distinguishes relativistic matter content
from the exterior vacuum metric, and other boundary stays at inside that
distinguishes the central Minkowskian-core from the matter
gravitational source. The matter content is involved without being
squeezed under the zero expansion scalar. For example, changes in
the volume of a spherical stellar gradient cause a similar expansion
in the outward hyper-surface, counterbalancing similar interior
surface expansion. As a result of the zero expansion scalar, a
specific type of system evolution occurs in which the innermost
shell drags away from the central point, resulting in the vacuum
core. Based on this concept, zero expansion matter populations could
be effective in explaining voids. In an astrophysical sense, the
zero expansion evolving system leads to the formation of a
vacuum-cavity inside the consideration system, illustrating voids as
the vacuum-cavities in between two clusters. Therefore such a study could be helpful at cosmological scales.
\item
The collapsing expansion-free fluid, upon advancing toward the center point, underwent a shear scalar blowup. The significant shear effects generate obstructions in the emergence of an apparent horizon. In cosmology, it depicts the isotropic and relative motion of decamping galaxies; however, in the presence of strong effects of gravity, it becomes the source of the emergence of the naked singularity \cite{132}. Thus, in nature, the zero expansion condition and the naked singularity are weaved together. To gain a thorough understanding of how naked singularity appears, Virbhadra \emph{et al.} \cite{133} constructed generic formalism. Further, Virbhadra and Ellis \cite{134} also linked this remarkable naked singularity event to gravitational lensing and reported some preliminary findings. Due to zero expansion, matter sources could be useful in explaining voids. Hence, our results for zero expansion conditions hold in explaining cosmic voids. Voids are so-called under dense zones that hold a lot of information about the cosmic environment \cite{135}. Voids are reputable resources for discussing the appearance of cosmic structure at enormous dimensions. Their extremely basic shape and structure make them valuable tools for obtaining the value of several cosmic factors, including the influence of dark energy, presumably. The clean environment of voids not only provides a suitable testing ground for evaluating the significance of the environment in galaxies' creation and evolution, but the scarcity of dwarf galaxies may also provide a major challenge to the mainstream view of cosmic formation mechanisms.
About $95$ percent of the entire volume in the galaxy distribution is made up of voids \cite{136}. The cosmological significance of voids is that they hold a lot of data about the underlying cosmic scenario as well as global cosmological characteristics. Also, the outflow velocities and corresponding red-shift distortions show notable cosmic effects \cite{137}. The pristine low-density setting of voids shows a suitable and pure context for studying the formation of the galaxy and the impact of the cosmic background on galaxy formation. As compared to EGT, voids are more rich in modified gravity
\cite{138}.
In case the of constant curvature constraints, the cosmic model
$$f(R, T)=R+\beta RT\Rightarrow
f(\tilde{R}, \tilde{T})=\tilde{R}+\beta \tilde{R}\tilde{T};~~~  \beta>0 $$
where tilde represents the evolution of quantities under constant curvature constraints. This equation indicates the Ricci scalar algebraic equation for a certain specific $f(R, T)$ model.
Thus, we have $R_-^+|=0$ also $\dot{f}=f'=f_{\tilde{R}}=f_{\tilde{T}}=0$
then by use of these conditions, the $f(R, T)$ generalized
Euler's equation (A7) for axially and reflection symmetric source under constant curvature constraints gets the form
\begin{align}\nonumber
&a_{\mu}(\tilde{\mu}+P)+h^{\gamma}_{\mu}\left(P_{;\gamma}
+\Pi^{\nu}_{\gamma;\nu}+q_{\gamma;\nu}U^{\nu}\right)+
q^{\gamma}\left(\sigma_{\mu\gamma}+\Omega_{\mu\gamma}
+\frac{4}{3}h_{\mu\gamma}\Theta\right)
-q^{\gamma}\left(\sigma_{\mu\gamma}+\Omega_{\mu\gamma}\right)
=-\frac{1}{2}f,
\end{align}
we may also rewrite it as follows
\begin{align}\nonumber
\Theta=\frac{3}{4}q^{\mu}\left[-\frac{1}{2}f-a_{\mu}(\tilde{\mu}+P)
+h^{\gamma}_{\mu}\left(P_{;\gamma}
+\Pi^{\nu}_{\gamma;\nu}+q_{\gamma;\nu}U^{\nu}\right)
-q^{\gamma}\left(\sigma_{\mu\gamma}+\Omega_{\mu\gamma}\right)\right],
\end{align}
showing that $\Theta$ can be expressed in terms of shear-tensor,
vorticity, and anisotropic fluid contents. However, for non-dissipative source, this results
implies that $\Theta=0$. It means that the existence of vacuum
core necessarily implies the existent of adiabatic stellar systems.
\end{itemize}
Now we would like to explain few applications of our equations.
Indeed, as Eq. (A11) shows that the zero-vorticity condition
entails $Y^*_{KL}=\frac{\kappa}{2f_R}f_T \Pi_{KL}$ in the case of
geodesic, which demonstrates the influence of $f(R, T)$ corrections
on the evolution of axial and reflection symmetric systems. However,
in Einstein's gravity theory, this condition implies $Y^*_{KL}=0$.
Since our line element is non-diagonal and so for $L^0$ is non-zero.
On the other hand, it is deduced from Eq. (A12), the zero-vorticity
condition is not stable in the presence of usual dissipative fluxes,
yet the extra constituents of modified heat fluxes commence the
evolving system into a more stable configuration. Consequently, in
Einstein's gravity, the shear-free and geodesic conditions are too
restrictive \cite{139} as compared to $f(R, T)$
gravity. Also the system obeys
\begin{align}\nonumber
&2\sigma_I+\sigma_{II}=\frac{-3}{A}\left\{-\frac{\dot{B}}{B}
+\frac{\dot{C}}{C}\right\},\\\nonumber
&\sigma_I+2\sigma_{II}=\frac{3A}{A^2r^2B^2+G^2}
\left[r^2B^2\left\{\frac{\dot{B}}{B}-\frac{\dot{C}}{C}
\right\}-\frac{G^2}{A^2}\left\{\frac{\dot{A}}{A}
-\frac{\dot{G}}{G}+\frac{\dot{C}}{C}\right\}\right],
\end{align}
from above two equations, it is to be followed that the vanishing of
share condition infers
\begin{align}\nonumber
G=CAf(r, \theta),
\end{align}
where $f(r, \theta)$ is an arbitrary function of independent
variables $r, \theta$. Since in evolutionary regimes the geometric
variables $A$ and $C$ can never vanish, therefore from this
condition, one can deduce that the shear free configuration that is
initially vorticity free, and will be remained unaltered during the
evolution in geodesic case.
A similar situation holds for
vorticity free-configuration (as seemed from Eq. (A11)) in that case
as well. Hence, the above-mentioned factors as well as the extra
constituents of $f(R, T)$ gravity are responsible for the stable
configuration of zero-vorticity condition. This result is in turn,
in complete agreement with previous work
\cite{99} by choosing $\beta=0\Rightarrow f(R,
T)=R$.
Also, as it implies from Eq. (A27), the inverse holds for
non-dissipative matter contents, and it is the generalization of the
result by Glass \cite{130}, demonstrating that necessary
and sufficient requirement for shear free perfect matter contents to
be irrotational is that there is no contribution of magnetic
components of the Weyl-tensor i.e., the Weyl-tensor is entirely
electric. So, this is the generalization of the Glass result to
anisotropic matter contents in $f(R, T)$ gravity. Hence, it is
observed that in the scenario of dissipative fluids the disappearing
of magnetic-part of the Weyl-tensor does not necessarily mean the
disappearing of the vorticity.

\subsection{Consequence of Quasi-static Conditions on $f(R, T)$ Relativistic
Equations}

Since it is deduced from the nature of the QS approximation that all the processes evolve on a larger time-scale than the time taken for
transient-phenomena, inferring that we are firstly expecting the modified
heat-fluxes to characterize a constant heat flow along with the
effect of $f(R, T)$ corrections. Therefore, the relaxation-time $\tau$ is
neglected in both components of Eq. \eqref{32} that are
presented in Eqs. \eqref{33} and \eqref{34} for
our relativistic-system then the results are obtained as
\begin{align}\label{1p13}
&{\tilde{q_{I}}}^{\textrm{eff}}=-\frac{\kappa}{B}\left(B\mathbb{T}a_I
+\mathbb{T}'\right),\\\label{1p14}
&{\tilde{q_{II}}}^{\textrm{eff}}=\frac{\kappa}{A}\left(A\mathbb{T}a_{II}
+\frac{G\dot{\mathbb{T}}+A^2\mathbb{T}_{,\theta}}{rAB}\right).
\end{align}
Since $\dot{\mathbb{T}}$ has order $O(\epsilon)$, and using
thermal-equilibrium conditions \cite{131}.
From above equations, we receive the following expressions in the quasi
static-regime
\begin{align}\label{1p15}
\bigg(\mathbb{T}A\bigg)'=\frac{1}{rB\tilde{f_R}}\chi^{approx}_1, \quad
\bigg(\mathbb{T}A\bigg)_{,\theta}=\frac{1}{rAB\tilde{f_R}}\chi^{approx}_2,
\end{align}
where ``approx" is used to illustrate the QS approximation on corresponding
quantities. One can be easily evaluated the QS approximated values
of these quantities, presented in Appendix A, by using the above defined conditions.

\subsection{$f(R,T)$ Field, Hydro-dynamical Equations and QS approximation}

In this section, we would like to evaluate the QS approximation over the
modified field (A1)--(A5) as well as the
conservation equations (A6)--(A7) that are
presented in Appendix A. These are obtained by putting the proposed constraints as specified earlier. Therefore, we have
\begin{align}\label{1p16}
G_{00}&=\frac{\kappa
A^2}{\tilde{f_R}}\bigg[\mu-\frac{1}{2}\bigg(\tilde{f}-\tilde{R}\tilde{f_R}
\bigg)\bigg]
+\frac{\kappa}{\tilde{f_R}}\bigg[\frac{1}{r^2B^2}\tilde{f}_{R,\theta\theta}
+\tilde{f}'_R\bigg\{\frac{A^2}{rB^2}+\frac{A^2}{B^2}\frac{C'}{C}\bigg\}
+\tilde{f}_{R,\theta}\frac{A^2}{r^2B^2}\bigg(\frac{C_{\theta}}{C}\bigg)
\bigg],\\\label{1p17}
G_{01}&=\frac{\kappa}{\tilde{f_R}}\bigg[-AB\bigg(1
+\tilde{f_T}\bigg)q_{I}\bigg],\\\label{1p18}
G_{02}&=\frac{-ABr\kappa}{\tilde{f}_R}\bigg[\frac{\mu
G}{ABr}+\bigg(1+\tilde{f}_T\bigg)q_{II}\bigg] +\frac{\kappa
G}{2\tilde{f_R}}(\tilde{f}-\tilde{R}\tilde{f_R})-\frac{\kappa
G}{2r^2B^2\tilde{f}_R} \bigg[\tilde{f}_{R,\theta\theta}
+\frac{C_{,\theta}}{C}\tilde{f}_{R,\theta\theta}\bigg],\\\label{1p19}
G_{12}&=\frac{\kappa}{\tilde{f}_R}\bigg[(1+\tilde{f}_T)\bigg(B^2r\Pi_{KL}\bigg)
+\frac{BG}{A}q_I+\tilde{f'}_{R,\theta}-\frac{B_{\theta}}{B}
\tilde{f'}_{R}-\frac{(Br)'}{Br}\bigg],\\\label{1p20}
G_{11}&=\frac{\kappa
B^2}{f_R}\bigg[(1+\tilde{f}_T)\bigg(P+\frac{\Pi_{I}}{3}\bigg)
+\mu\tilde{f}_T+\frac{1}{2}(\tilde{f}-\tilde{R}\tilde{f_R})
+\frac{1}{r^2B^2}\bigg\{\tilde{f}_{R,\theta\theta}
+\bigg(\frac{A_{\theta}}{A}-\frac{B_{\theta}}{B}
+\frac{C_{\theta}}{C}\bigg)\tilde{f}_{R,\theta}
\bigg\}\bigg],\\\label{1p21}
G_{22}&=\frac{\kappa
r^2B^2}{\tilde{f}_R}\bigg[(1+\tilde{f}_T)\bigg(P
+\frac{\Pi_{II}}{3}\bigg)+\mu\tilde{f}_{T}
+\frac{1}{2}(\tilde{f}-\tilde{R}\tilde{f_R})
-\frac{1}{r^2B^2}\bigg\{\tilde{f}_{R,\theta\theta}
+\bigg(\frac{A_{\theta}}{A}-\frac{B_{\theta}}{B}
+\frac{C_{\theta}}{C}\bigg)\tilde{f}_{R,\theta}
\bigg\}\bigg],\\\label{1p22}
G_{33}&=\frac{\kappa
C^2}{\tilde{f}_R}\bigg[(1+\tilde{f}_T)\bigg\{P
-\frac{1}{3}\bigg(\Pi_I+\Pi_{II}\bigg)\bigg\}
+\mu\tilde{f}_{T}+\frac{1}{2}(\tilde{f}-\tilde{R}\tilde{f_R})
-\frac{1}{r^2B^2}\bigg\{\tilde{f}_{R,\theta\theta}
+\bigg(\frac{A_{\theta}}{A}\bigg)\tilde{f}_{R,\theta}\bigg\}\bigg].
\end{align}
The dynamical equations describe the change in the parameters of a physical
system for time. These equations are related to the study
of motion of celestial objects which is supported by stress-energy tensor. As a result of gravitational collapse,
static celestial objects become unstable. To deal with such a situation, the gravitational field equations are helpful to provide the dynamical equations.
Therefore, we want to execute the evolution of dynamical equations in the quasi-static constraints. Thereby, the QS context of equation
(A6), is given as
\begin{align}\nonumber&
\frac{1}{\tilde{f_R}}\bigg[(1+\tilde{f_T})\bigg\{\frac{\dot{\mu}}{A}
+\Theta(\mu+P)
+\frac{1}{9}\bigg(\Pi_I(2\sigma_I
+\sigma_{II})+\Pi_{II}(\sigma_I+2\sigma_{II})\bigg
)+\frac{q'_I}{B}
+\frac{1}{Br}\bigg(q_{,\theta}
+\frac{G}{A^2}\dot{q_{II}}\bigg)+
2\bigg(q_Ia_I\\\label{1p23}&+q_{II}a_{II}\bigg)
+\frac{q_I}{B}\bigg(\frac{C'}{C}+\frac{(Br)'}{Br}\bigg)
+\frac{q_{II}}{Br}\bigg(\frac{B_{,\theta}}{B}
+\frac{C_{,\theta}}{C}\bigg)\bigg\}\bigg]=\frac{1}{\tilde{f_R}}\chi^{approx}_7.
\end{align}
From the modified Euler-lagrange equation (A7), following two
equations are attained in the QS scenario
\begin{align}\nonumber
&\frac{1}{\tilde{f_R}}\bigg[(1+\tilde{f_T})\bigg\{\frac{1}{B}\bigg(P
+\frac{\Pi_I}{3}\bigg)'+
\frac{1}{Br}\bigg(\Pi_{KL,\theta}+\frac{G}{A^2}\dot{\Pi}_{KL}\bigg)
+\bigg(\mu+P+\frac{\Pi_I}{3}\bigg)a_I
+a_{II}\Pi_{KL}+\frac{\Pi_I}{3B}\bigg(\frac{2C'}{C}+\frac{(Br)'}{Br}\bigg)
\\\label{1p24}&+\frac{\Pi_{II}}{3B}\bigg(\frac{C'}{C}-\frac{(Br)'}{Br}\bigg)
+\frac{\Pi_{KL}}{Br}\bigg(\frac{2B_{,\theta}}{B}+\frac{C_{,\theta}}{C}\bigg)
+\frac{\dot{q}_I}{A}\bigg\}\bigg]
=\bigg(-\frac{1}{2\tilde{f_R}}(\tilde{f}
-\tilde{R}\tilde{f_R})\bigg)'+
\frac{1}{B\tilde{f_R}}\chi^{approx}_8; \\\nonumber
&\frac{1}{\tilde{f_R}}\bigg[(1+\tilde{f_T})\bigg\{\frac{1}{Br}\bigg((P
+\frac{\Pi_I}{3})_{,\theta}+\frac{G}{A^2}(\dot{P}+\frac{\dot{\Pi}_{II}}{3})\bigg)
+\frac{\Pi'_{KL}}{B}+\bigg(\mu+P+\frac{\Pi_{II}}{3}\bigg)a_{II}
+a_{I}\Pi_{KL}
+\frac{\Pi_I}{3Br}\bigg(-\frac{B_{,\theta}}{B}
+\frac{C_{,\theta}}{C}\bigg)
\\\label{1p25}&+\frac{\Pi_{II}}{3Br}\bigg(\frac{B_{,\theta}}{B}
+\frac{2C_{,\theta}}{C}\bigg)
+\frac{\Pi_{KL}}{B}\bigg(\frac{C'}{C}+\frac{(Br)'}{Br}\bigg)
+\frac{\dot{q_{II}}}{A}\bigg\}\bigg]
=\frac{1}{\tilde{f_R}}\bigg[\frac{1}{Br}(\tilde{R}\tilde{f_R}
-\tilde{f})-\chi^{approx}_9\bigg].
\end{align}
The emergence of extra terms due to $f(R,T)$ corrections are named $\chi_7-\chi_9$, which are given in Appendix A. These equations represent the simplest mode of the energy content associated with the axial structures within high curvature domain.

\subsection{$f(R,T)$ Scalar Equations and QS Approximation}

In the QS regime, it follows Eq. (\ref{18}) that the time
derivative of the vorticity-scalar i.e., $\dot{\Omega}$ has order
$O(\epsilon^2)$. The evolution of Eqs. (A15) and (A16) in QS constraints yields, respectively.
\begin{align}\label{1p26}
&\frac{2}{3B}\Theta'-\frac{\Omega}{Br}\bigg(
\frac{2A_{,\theta}}{A}+\frac{C_{,\theta}}{C}\bigg)
-\frac{\Omega_{,\theta}}{Br}-\frac{\tilde{\sigma}'}{3B}
-\frac{\tilde{\sigma}C'}{BC}=\kappa
\tilde{q_{I}}^{\textrm{eff}},\\\label{1p27}
&\frac{2}{3Br}\Theta_{,\theta}+\frac{\Omega'}{B}-\frac{\Omega}{B}\bigg(
\frac{2A'}{A}+\frac{C'}{C}\bigg)-\frac{\tilde{\sigma}_{,\theta}}{3Br}
-\frac{\tilde{\sigma}C_{,\theta}}{rBC}=\kappa
\tilde{q_{II}}^{\textrm{eff}}.
\end{align}
It depicts that dissipative fluxes have also order $O(\epsilon)$. So
for, we summarize all the outcomes deduced from the QS approximation as
\begin{itemize}
\item
The order of $\dot{\Omega}, \dot{G}$ is $O(\epsilon^2)$.
Whereas, components of the matter profile and $ \dot{a_I}, \dot{a_{II}}$ all are of order $O(\epsilon)$.
\end{itemize}
Since the hydro-static equilibrium state can be detained at any
time, therefore the corresponding equations containing $\sigma_{22}, \sigma_{33}$
components of $\sigma_{\lambda\omega}$ hold \cite{99}.
It is obtain from above mentioned equations, respectively
\begin{align}\label{1p28}
&\dot{\Pi}_{KL}\approx O(\epsilon);\quad \dot{q}_{I}\approx
O(\epsilon^2); \quad \ddot{C}\approx O(\epsilon^2); \quad
\ddot{B}\approx O(\epsilon^2);\quad \dot{\Pi}_{II}\approx
O(\epsilon); \quad \dot{P}\approx O(\epsilon); \quad
\dot{q}_{II}\approx O(\epsilon^2).
\end{align}
It has been imposed the fact that $P, \Pi_{I}, \Pi_{II}$ include
terms with $\ddot{C}$ and $\ddot{B}$ other than the terms involving
some spatial-coordinate derivatives of corresponding line-element.
Now, it is followed immediately from Eq. \eqref{21}
\begin{align}\label{1p29}
&\dot{\Theta}\approx O(\epsilon^2); \quad \dot{\tilde{\sigma}}\approx O(\epsilon^2).
\end{align}
By using Eqs. \eqref{21} and \eqref{1p26} turned out to be
\begin{align}\label{1p30}
&2V'=\frac{\Omega}{r}\bigg[\ln(\Omega CA^2)\bigg]_{,\theta}+\tilde{\sigma}\bigg[\ln(\tilde{\sigma}C)\bigg]'
+\kappa B{\tilde{q_{I}}}^{\textrm{eff}},
\end{align}
where $V\equiv V_{1}\equiv V_{2}$, after integration we attain
\begin{align}\label{1p31}
&V=V_{\Sigma}-\frac{1}{2}\int^{r_\Sigma}_{r}\bigg\{\frac{\Omega}
{r}\bigg[\ln(\Omega CA^2)\bigg]_{,\theta}
+\tilde{\sigma}\bigg[\ln(\tilde{\sigma}C)\bigg]'+\kappa B{\tilde{q_{I}}}^{\textrm{eff}}\bigg\}dr,
\end{align}
it can also be expressed as
\begin{align}\label{1p32}
&V_{3}=V_{(3)\Sigma}-\frac{1}{2}\int^{r_\Sigma}_{r}
\bigg\{\frac{\Omega}{r}\bigg[\ln(\Omega CA^2)\bigg]_{,\theta}
+\tilde{\sigma}\bigg[\ln\bigg(\frac{C}{\tilde{\sigma}}\bigg)\bigg]'
+\kappa B{\tilde{q_{I}}}^{\textrm{eff}}\bigg\}dr,
\end{align}
here, the equation $r=r_\Sigma$ described the surface boundary of
the source and we have also used the fact that
$V{3}=V-\tilde{\sigma}$. In similar fashion, Eq. \eqref{1p27} may be
written as follows
\begin{align}\label{1p33}
&2V_{,\theta}=-\Omega r\bigg[\ln\bigg(\frac{CA^2}{\Omega}\bigg)\bigg]'
+\tilde{\sigma}\bigg[\ln(\tilde{\sigma}C)\bigg]_{,\theta}
+\kappa Br{\tilde{q_{II}}}^{\textrm{eff}},
\end{align}
generating
\begin{align}\label{1p34}
V=V_\Sigma-\frac{1}{2}\int^{\theta_\Sigma}_{\theta}\bigg\{-\Omega
r\bigg[\ln\bigg(\frac{CA^2}{\Omega}\bigg)\bigg]'
+\tilde{\sigma}\bigg[\ln(\tilde{\sigma}C)\bigg]_{,\theta} +\kappa
Br{\tilde{q_{II}}}^{\textrm{eff}} \bigg\}d\theta,
\end{align}
or
\begin{align}\label{1p35}
V_{(3)}=V_{(3)\Sigma}-\frac{1}{2}\int^{\theta_\Sigma}_{\theta}
\bigg\{-\Omega r\bigg[\ln\bigg(\frac{CA^2}{\Omega}\bigg)\bigg]'
+\tilde{\sigma}\bigg[\ln\bigg(\frac{C}{\tilde{\sigma}}\bigg)\bigg]_{,\theta}
+\kappa Br{\tilde{q_{II}}}^{\textrm{eff}}
\bigg\}d\theta,
\end{align}
in this case, the boundary surface is given by the equation
$\theta=\theta_\Sigma$.
Next, it could be feasible to evaluate the QS approximation
of Eqs. (A21)--(A29) to check the slowly evolving regimes of the
self-gravitating axial source in the following manner:
\begin{align}\nonumber
&\frac{1}{3A}\bigg[\varepsilon_I+\frac{\kappa}{2\tilde{f_R}}(1
+\tilde{f_T})(\Pi_I+\mu)\bigg]^{\dot{}}
+\frac{1}{3}(\Theta\varepsilon_I+\tilde{\sigma}\varepsilon_{II})
-\Omega\bigg(\varepsilon_{KL}+\frac{\kappa}{2\tilde{f_R}}(1
+\tilde{f_T})\Pi_{KL}\bigg)
-\frac{1}{Br}
\bigg(H_{1,\theta}+H_1\frac{C_{,\theta}}{C}\bigg)
\\\label{1p36}&-\frac{H_2}{B}\bigg(\frac{C'}{C}-\frac{(Br)'}{Br}\bigg)
=2a_{II}H_1-\frac{\kappa}{6}(\Theta+\tilde{\sigma})
\bigg(\tilde{\mu}^{\textrm{eff}}
+(\tilde{P}+\frac{\tilde{\Pi_I}}{3})^{\textrm{eff}}\bigg)
-\kappa a_{I}\tilde{q_{I}}^{\textrm{eff}}-\frac{\kappa}{2B}
\bigg[\tilde{q_{I}}^{\textrm{eff}}
+\frac{B_{,\theta}}{B}\tilde{q_{II}}^{\textrm{eff}}\bigg],\\\nonumber
&\frac{1}{A}\bigg[\varepsilon_{KL}+\frac{\kappa}{2\tilde{f_R}}(1
+\tilde{f_T})\Pi_{KL}\bigg]^{\dot{}}
+\frac{\Omega}{6}\bigg[\varepsilon_{I}-\varepsilon_{II}
+\frac{\kappa}{2\tilde{f_R}}(1+\tilde{f_T})(\Pi_{I}-\Pi_{II})
\bigg]-(a_{II}H_2
-a_{I}H_1)-\bigg\{\varepsilon_{KL}
+\frac{\kappa}{2\tilde{f_R}}\\\nonumber&\bigg(1+\tilde{f_T}\bigg)\Pi_{KL}\bigg\}
(\tilde{\sigma}-\Theta)-\frac{1}{2B}\bigg[H_1\bigg(\frac{(Br)'}{Br}
-\frac{2C'}{C}\bigg)-H'_1\bigg]
-\frac{1}{2Br}\bigg[H_{2,\theta}
-H_2\bigg(\frac{B_{,\theta}}{B}
-\frac{2C_{,\theta}}{C}\bigg)\bigg]
=-\frac{2\kappa}{6}\bigg(2\tilde{\sigma}\\\label{1p37}&
-\Theta\bigg)\tilde{\Pi_{KL}}^{\textrm{eff}}
-\frac{\kappa}{2}
\bigg(a_{II}\tilde{q_{I}}^{\textrm{eff}}
+a_{I}\tilde{q_{II}}^{\textrm{eff}}\bigg)
-\frac{\kappa}{4B\tilde{f_R}}(1
+\tilde{f_T})\bigg(q'_{II}-q_{II}\frac{(Br)'}{Br}\bigg)
-\frac{\kappa}{4rB\tilde{f_R}}(1+\tilde{f_T})\bigg\{q_{I,\theta}
-q_{I,\theta}\frac{B_{,\theta}}{B}\bigg\},\\\nonumber
&\frac{1}{3A}\bigg[\varepsilon_{II}+\frac{\kappa}{2\tilde{f_R}}(1
+\tilde{f_T})(\Pi_{II}+\mu)\bigg]^{\dot{}}
+\frac{1}{3}(\Theta\varepsilon_{II}+\tilde{\sigma}\varepsilon_{I})
+\Omega\bigg(\varepsilon_{KL}+\frac{\kappa}{2\tilde{f_R}}(1
+\tilde{f_T})\Pi_{KL}\bigg)
+2H_2a_{I}+\frac{1}{B}\bigg(\frac{H_2C'}{C}
+H'_2\bigg)\\\label{1p38}&+\frac{H_1}{rB}
\bigg(\frac{C_{,\theta}}{C}-\frac{B_{,\theta}}{B}\bigg)=
-\frac{\kappa}{6}(\Theta+\tilde{\sigma})\bigg(\tilde{\mu}^{\textrm{eff}}
+(\tilde{P}+\frac{\tilde{\Pi_{II}}}{3})^{\textrm{eff}}\bigg)
-\kappa a_{I}\tilde{q_{I}}^{\textrm{eff}}
-\frac{\kappa}{2\tilde{f_R}}(1+\tilde{f_T})\bigg[\frac{q_{II}}{Br}
+\frac{q_{I}}{B}
\frac{(Br)'}{Br}\bigg],\\\nonumber
&\frac{1}{3A}\bigg[-\frac{\kappa}{2\tilde{f_R}}(1
+\tilde{f_T})(-\mu+\Pi_{I}
+\Pi_{II})-(\varepsilon_{II}+\varepsilon_{I})\bigg]^{\dot{}}
+\frac{\kappa}{18\tilde{f_R}}(1+\tilde{f_T})(\Pi_{I}
+\Pi_{II})(2\tilde{\sigma}-\Theta)
-2(a_IH_2-a_{II}H_1)\\\nonumber&-\frac{1}{3}(\Theta
+\tilde{\sigma})(\varepsilon_{I}+\varepsilon_{II})
+\frac{1}{B}\bigg(H_2\frac{(Br)'}{Br}+H'_2\bigg)+\frac{1}{rB}
\bigg(H_{1,\theta}+\frac{B_{,\theta}}{B}H_1\bigg)
=-\frac{\kappa}{6}(\tilde{\mu}^{\textrm{eff}}
+\tilde{P}^{\textrm{eff}})(\Theta-2\tilde{\sigma})
-\frac{\kappa\tilde{q_{I}}^{\textrm{eff}}}{2B}\frac{C'}{C}
\\\label{1p39}&-\frac{\kappa\tilde{q_{II}}^{\textrm{eff}}}{2Br}
\frac{C_{,\theta}}{C},\\\nonumber
&\frac{1}{3B}\bigg[\varepsilon_I+\frac{\kappa}{2\tilde{f_R}}(1
+\tilde{f_T})\Pi_I\bigg]'
+\frac{1}{Br}\bigg(\frac{\kappa}{2\tilde{f_R}}(1
+\tilde{f_T})\Pi_{KL}+\varepsilon_{KL}\bigg)_{,\theta}
+\frac{1}{3B}\bigg(\varepsilon_I\frac{\kappa}{2\tilde{f_R}}(1+
+\tilde{f_T})\Pi_I\bigg)
\bigg(\frac{2C'}{C}+\frac{(Br)'}{Br}\bigg)
\\\nonumber&+\frac{1}{3B}\bigg(\varepsilon_{II}+\frac{\kappa}{2\tilde{f_R}}(1
+\tilde{f_T})\Pi_{II}\bigg)
\bigg(\frac{C'}{C}-\frac{(Br)'}{Br}\bigg)
+\frac{1}{Br}
\bigg(\varepsilon_{KL}+\frac{\kappa}{2\tilde{f_R}}(1
+\tilde{f_T})\Pi_{KL}\bigg)
\bigg(\frac{C_{,\theta}}{C}-\frac{B_{,\theta}}{B}\bigg)
=\frac{2\kappa}{6B\tilde{f_R}}\bigg(1\\\label{1p40}&+\tilde{f_T}\bigg)\mu',\\\nonumber
&\frac{1}{3Br}\bigg(\varepsilon_{II}+\frac{\kappa}{2\tilde{f_R}}(1
+\tilde{f_T})\Pi_{II}\bigg)_{,\theta}
+\frac{1}{B}\bigg(\frac{\kappa}{2\tilde{f_R}}(1+\tilde{f_T})\Pi_{KL}
+\varepsilon_{KL}\bigg)'
+\frac{1}{3Br}\bigg(\varepsilon_I+\frac{\kappa}{2\tilde{f_R}}
(1+\tilde{f_T})\Pi_I\bigg)
\bigg(\frac{C_{,\theta}}{C}-\frac{B_{,\theta}}{B}\bigg)
\\\nonumber&+\frac{1}{3Br}\bigg(\varepsilon_{II}+\frac{\kappa}{2\tilde{f_R}}(1
+\tilde{f_T})\Pi_{II}\bigg)
\bigg(\frac{2C_{,\theta}}{C}+\frac{B_{,\theta}}{B}\bigg)
+\frac{1}{B}
\bigg(\frac{\kappa}{2\tilde{f_R}}(1
+\tilde{f_T})\Pi_{KL}+\varepsilon_{KL}\bigg)
\bigg(\frac{2C'}{C}+\frac{(Br)'}{Br}\bigg)
=\frac{2\kappa}{6B\tilde{f_R}}\bigg(1\\\label{1p41}&
+\tilde{f_T}\bigg)\mu_{,\theta},\\\nonumber
&-\frac{1}{B}\bigg[H_1\bigg(\frac{(Br)'}{Br}
+\frac{2C'}{C}\bigg)+H'_1\bigg]
-\frac{1}{rB}\bigg[H_{2,\theta}+H_2\bigg(\frac{B_{,\theta}}{B}
+\frac{2C_{,\theta}}{C}\bigg)\bigg]
=\frac{\kappa}{2B\tilde{f_R}}(1+\tilde{f_T})
\bigg(q_{II}\frac{(Br)'}{Br}
+q'_{II}\bigg)+\Omega\bigg[\kappa(\tilde{\mu}^{\textrm{eff}}
\\\label{1p42}&+\tilde{P}^{\textrm{eff}})-(\varepsilon_{I}
+\varepsilon_{II})+\frac{\kappa}{6\tilde{f_R}}
(1+\tilde{f_T})(\Pi_{I}+\Pi_{II})\bigg]
-\frac{\kappa}{2rB\tilde{f_R}}
(1+\tilde{f_T})\bigg(q_{I,\theta}
+q_{I}\frac{B_{,\theta}}{B}\bigg),\\\nonumber
&-\frac{1}{B}\bigg[\frac{\kappa}{6\tilde{f_R}}(1
+\tilde{f_T})\Pi_{KL}\bigg]'
+\frac{1}{rB}\bigg(\frac{\kappa}{3\tilde{f_R}}(1
+\tilde{f_T})\Pi_{KL}\bigg)_{,\theta}
-\frac{\varepsilon_I}{3rB}\bigg(\frac{C_{,\theta}}{C}
+\frac{2A_{,\theta}}{A}\bigg)
-\frac{\varepsilon_{II}}{3rB}
\bigg(\frac{2C_{,\theta}}{C}
+\frac{A_{,\theta}}{A}\bigg)
-\frac{\kappa}{B\tilde{f_R}}\bigg(1
\\\label{1p43}&+\tilde{f_T}\bigg)\Pi_{KL}\frac{(Br)'}{Br}
-\frac{\varepsilon_{KL}}{B}
\bigg(\frac{C'}{C}-\frac{A'}{A}\bigg)+\frac{\kappa}{6rB\tilde{f_R}}(1
+\tilde{f_T})\bigg(\Pi_{I}
-\Pi_{II}\bigg)\frac{B_{,\theta}}{B}+\frac{1}{A}\dot{H_1}
=-\frac{2\kappa}{6B\tilde{f_R}}\bigg(1+\tilde{f_T}\bigg)\mu_{,\theta},\\\nonumber
&-\frac{1}{B}\bigg[\frac{\kappa}{6\tilde{f_R}}(1+\tilde{f_T})\Pi_{II}
-(\varepsilon_{I}-\varepsilon_{II})\bigg]'
+\frac{\kappa}{2rB\tilde{f_R}}(1+\tilde{f_T})
\bigg(\Pi_{KL}\frac{2B_{,\theta}}{B}+\Pi_{KL,\theta}\bigg)
+\frac{1}{A}\dot{H_2}
+\frac{\varepsilon_{I}}{3B}\bigg(\frac{2C'}{C}+\frac{A'}{A}\bigg)
\\\label{1p44}&+\frac{\varepsilon_{II}}{3B}\bigg(\frac{C'}{C}
+\frac{2A'}{A}\bigg)
+\frac{\kappa}{6B\tilde{f_R}}(1+\tilde{f_T})(\Pi_{I}
-\Pi_{II})\frac{(Br)'}{Br}+\frac{\varepsilon_{KL}}
{rB}\bigg(\frac{C_{,\theta}}{C}
-\frac{A_{,\theta}}{A}\bigg)
=\frac{\kappa}{6B\tilde{f_R}}\bigg(1+\tilde{f_T}\bigg)\mu',
\end{align}
where $\tilde{q_{II}}^{\textrm{eff}}$ shows the QS evolution of effective component of heat-flux $(q_{II}^{\textrm{eff}})$. So,
one can easily computed the quasi static-evolution of effective components
of relativistic-fluid by using the conditions of the QS approximation as
defined in above section.
Equations \eqref{1p40} and \eqref{1p41} are representing the QS nature of propagation Eqs. (A25) and (A26) listed in Appendix A, which are turned out due to the contraction of Eqs. (A18) with the vectors with $\textbf{K}$ and $\textbf{L}$. The Eqs. \eqref{1p40} and \eqref{1p41} are
significant in describing the homogeneity of the energy-density. In
fact, it implies at once from Eqs. \eqref{1p40} and \eqref{1p41} disappearing of $X_I, X_{II}, X_{KL}$ and $Z_I, Z_{II}, Z_{III}, Z_{IV}$
\cite{99}, for our system within the bounds of $f(R,T)$
gravity, they imply
\begin{align}\nonumber
X^*_{I}&=-\frac{\kappa}{2f_R}f_T\Pi_I;\quad
X^*_{II}=-\frac{\kappa}{2f_R}f_T\Pi_{II};\quad X^*_{KL}
=-\frac{\kappa}{2f_R}f_T\Pi_{KL};\quad
Z^*_{I}=-\frac{\kappa}{2f_R}f_Tq_{II};\quad\\\nonumber
Z^*_{II}
&=\frac{\kappa}{2f_R}f_Tq_{II};\quad
Z^*_{III}=-\frac{\kappa}{2f_R}f_Tq_{I};\quad
Z^*_{IV}=\frac{\kappa}{2f_R}f_Tq_{I},
\end{align}
describing the homogeneity of the energy-density as well as the
effects of $f(R, T)$ corrections on the anisotropy of matter contents.
However, the disappearance of $(X^{*}_{I, II, KL})$ and
$(Z^{*}_{I, II, III, IV})$ from Eqs. \eqref{1p40} and \eqref{1p41} may allow us to identify
homogeneity of the energy-density in $f(R, T)$ gravity. On the other
hand, the Eqs. (A21)-(A24), (A28) and (A29) are useful to determine the evolution of modified scalars, mentioned above, their QS behavior is reported in Eqs.\eqref{1p36}--\eqref{1p39}, \eqref{1p43} and \eqref{1p44}.

\subsection{Magnetic Part of the Weyl Tensor and QS approximation}

Two basic anisotropy scenarios are entirely pertinent to our
main issue among the many that are available. The first is a
strong magnetic field, which may be observed in compact celestial
bodies such as neutron stars, white dwarfs, and magnetized
strange quark stars \cite{111}.
The Weyl-tensor can be defined in terms of its electric and magnetic parts.
Merely, the electric part of the Weyl-tensor may be determined for spherically symmetric objects \cite{79}. However, the contribution of
its magnetic parts is also observed for axially and reflection symmetric
sources \cite{7}.
Let us investigate the order of the magnetic part of the
Weyl-tensor from Eqs. (A30)--(A31) in Appendix A.
It is highly important to report that the final outcomes in that case
are same as presented in \cite{7}.
\begin{align}\label{1p45}
&H_1=-a_{I}\Omega-\frac{1}{2Br}\bigg(\frac{\tilde{\sigma}
C_{,\theta}}{C}+\tilde{\sigma}_{,\theta}\bigg)
+\frac{1}{2B}\bigg(\frac{\Omega C'}{C}-\Omega'\bigg)\\\label{1p46}
&H_2=-a_{II}\Omega+\frac{1}{2Br}\bigg(\frac{\Omega
C_{,\theta}}{C}-\Omega_{,\theta}\bigg)
+\frac{1}{2B}\bigg(\frac{\tilde{\sigma}
C'}{C}+\tilde{\sigma}'\bigg),
\end{align}
inferring that the $H_1$ and $H_2$ are of order O$(\epsilon)$. It is
worth observing that $G=0=\Omega$ i.e, in vorticity free case, it
implies from Eqs. \eqref{1p45} and \eqref{1p46}
\begin{align}\label{1p47}
H_1=-\frac{(\tilde{\sigma C}_{,\theta})}{2rBC}; \quad \quad
\quad H_2=-\frac{(\tilde{\sigma C})'}{2BC}.
\end{align}
Then from Eqs. \eqref{35}, \eqref{1p26} and  \eqref{1p27} with the
assumption $\Omega=0$ yield, respectively.
\begin{align}\label{1p48}
&2\left(\frac{\dot{B}}{AB}\right)'-\frac{(\tilde{\sigma}C)'}{C}=\kappa
B\tilde{q_{I}}^{\textrm{eff}}; \quad \quad\quad
2\left(\frac{\dot{B}}{AB}\right)_{,\theta}
-\frac{(\tilde{\sigma}C)_{,\theta}}{C}=\kappa
Br\tilde{q_{II}}^{\textrm{eff}}.
\end{align}
Now, combining both above equations with Eq. \eqref{1p47}, we attain
the following relations, respectively.
\begin{align}\label{1p49}
&H_1+\frac{1}{rB}\bigg(\frac{\dot{B}}{AB}\bigg)_{,\theta}
=\frac{\kappa}{2}{\tilde{q_{II}}^{\textrm{eff}}};\quad \quad\quad
H_2-\frac{1}{B}\bigg(\frac{\dot{B}}{AB}\bigg)'
=-\frac{\kappa}{2}{\tilde{q_{I}}^{\textrm{eff}}}.
\end{align}
Thus, from Eq. \eqref{1p47} it is followed that the disappearance of shear
$(\Omega=0)$ is the necessary as well as sufficient condition for
the matter to be purely-electric in the QS approximation. The expressions of modified heat-fluxes and the other fluid contents are given in Appendix A. However, the modified heat-fluxes have also played an effective role in that case.
\section{Consequences of $f(R, T^2)$ Corrections Over the Quasi-static evolutionary Behavior of Axial Stellar Structures}

Now, we will examine the effects of $f(R,T^2)$ corrections over the quasi-static evolutionary aspects of an axial self-gravitating system
provided in Eq. \eqref{1}. In $2014$, a simple modification of Einstein's gravity theory was presented \cite{60}, allowing a correction term $T_{\gamma\lambda}T^{\gamma\lambda}$ in the action function of the
theory. This gravity theory is comparable to Einstein's gravity in the
vacuum, and its effects occur solely inside the energy-matter
distribution. More precisely, its impacts are more noticeable in
high curvature domains. Because of the matter-dominated era, it is
believed that some valuable results could be achieved to examine the
problems of dark energy and current cosmic accelerated expansion. In
the case of $f(R,T^2)=f(R)$, one can see the field-equations of
$f(R)$ gravity. However, the consequences of EGT would
be achieved for $f(R,T^2)=R$.
For $f(R, T^2)$ gravity, the generic action is given as follows \cite{60}.
\begin{align}\label{2p1}
I_{f(R,T^2)}=\frac{1}{2\kappa} \int \sqrt{-g}f(R,T^2)d^4x+ \int
\sqrt{-g}\textit{L}_m d^4x,
\end{align}
here the relative Lagrangian-density of matter distribution is
denoted by $\textit{L}_m$. For the sake of convenience, we treat
$\kappa=8\pi$. Equation \eqref{2p1} implies that there are more degrees
of freedom in this theory. The stress-energy tensor is
same as provided in the first case.
when the Lagrangian-density of matter exclusively determine by
metric elements and not by their derivatives, then we receive
\begin{align}\nonumber
&T_{\omega\lambda}=g_{\omega\lambda}\textit{L}_m-2\frac{\partial
\textit{L}_m}{\partial g^{\omega\lambda}}.
\end{align}
The following set of gravitational field equations are acquired by
varying the Eq. \eqref{2p1} with the inverse metric-tensor
\begin{align}\label{2p2}
&R_{\omega\lambda}f_{R}-\frac{1}{2}g_{\omega\lambda}f+g_{\omega\lambda}\Box
f_R-\nabla_{\omega} \nabla_{\lambda}f_{R} =\kappa
T_{\omega\lambda}-f_{T^2}\hat{\Theta}_{\omega\lambda},
\end{align}
with
\begin{align}\nonumber
&\hat{\Theta}_{\omega\lambda}=-2\textit{L}_m\bigg(T_{\omega\lambda}
-\frac{T}{2}g_{\omega\lambda}\bigg)
-T_{\omega\lambda}T+2T_{\omega}^{\beta}T_{\beta\lambda}
-4\frac{\partial^2 \textit{L}_m}{\partial g^{\omega\lambda}\partial
g^{\beta\gamma}}T^{\beta\gamma};~~\text{ where }
f=f(R, T^2),~~ f_R=\frac{\partial f}{\partial R},~~
f_{T^2}=\frac{\partial f}{\partial T^2}.
\end{align}
rearrangement of Eq. \eqref{2p2} yields
\begin{align}\label{2p3}
G_{\omega\lambda}=\frac{\kappa}{f_R}\left[T_{\omega\lambda}^{(m)}
+T_{\omega\lambda}^{(c)}\right]; \text{ where }T_{\omega\lambda}^{(m)}=
\text{usual stress-energy tensor. }
\end{align}
Whereas $T_{\omega\lambda}^{(c)}$ specifies the $f(R, T^2)$
corrections and is calculated as follows:
\begin{align}\label{2p4}
&T_{\omega\lambda}^{(c)}=\frac{1}{2}(f-Rf_R)g_{\omega\lambda}
+\nabla_{\omega}\nabla_{\lambda}{f_R}-g_{\omega\lambda}\Box{f_R}
-f_{T^2}\hat{\Theta}_{\omega\lambda}.
\end{align}

\subsection{$f(R,T^2)$ Structure Scalars and Their Physical Meanings}

In this section, we would like to describe the structure scalars and
their corresponding physical meanings within the bounds of
$f(R,T^2)$ gravity. In doing so, the three tensors presented in Eq.
\eqref{28} are evaluated with the help of the Riemann
tensor, whose description is offered below.
\begin{align}\label{2p5}
X_{\omega\lambda}=-E_{\omega\lambda}-\frac{8\pi}{f_R}\left[\bigg(1-2\mu
f_{T^2}\bigg)\frac{\Pi_{\omega\lambda}}{2} +7\bigg(-\mu
+3P\bigg)^2f_{T^2}\frac{h_{\omega\lambda}}{3}+\frac{1}{6}\bigg(f-Rf_R\bigg)
h_{\omega\lambda}\right]+\psi_1,
\end{align}
with the four scalar variables
\begin{align}\label{2p6}
X_T&=\frac{8\pi}{f_R}\bigg[7(-\mu+3P)^2+\frac{1}{2}(f-Rf_R)\bigg]
+\psi^{*}_1;\quad
\quad X_{I}=-\varepsilon_I-\frac{4\pi}{f_R}\bigg(1-2\mu
f_{T^2}\bigg)\Pi_I,\\\label{2p6}
X_{II}&=-\varepsilon_{II}-\frac{4\pi}{f_R}\bigg(1-2\mu
f_{T^2}\bigg)\Pi_{II};\quad \quad X_{KL}
=-\varepsilon_{KL}-\frac{4\pi}{f_R}\bigg(1-2\mu
f_{T^2}\bigg)\Pi_{KL}.
\end{align}
In this scenario, the scalar $X_T$ describes the mutual effects of
inhomogeneity of the energy-density and fluid pressure in a higher
degree manner due to the domains of $f(R,T^2)$ gravity as compared
to the previous case. Whereas, the other scalars include the
consequential influence of tidal forces and fluid's anisotropy
together with the higher order curvature terms, that would describe
the evolutionary aspects of the axial system in a better way. In the
similar fashion
\begin{align}\nonumber
&Y_{\gamma\lambda}=E_{\gamma\lambda}-\frac{8\pi}{f_R}\bigg[\psi_2
+\bigg(1-2\mu
f_{T^2}\bigg)\bigg\{\bigg(-\mu+3P\bigg)V_{\gamma}V_{\lambda}
+\bigg(\mu+P\bigg)g_{\gamma\lambda}+\Pi_{\gamma\lambda}\bigg\}
\bigg]+\frac{8\pi}{3f_R}h_{\gamma\lambda}\bigg[(1-2\mu
f_{T^2})\\\label{2p8}&\times\bigg(-\mu+3P\bigg)
-28\bigg(-\mu+3P\bigg)^2f_{T^2}-2\bigg(Rf_R-f\bigg)\bigg]; \text{ with } \\\label{2p9}
&Y_T=\frac{4\pi}{f_R}\bigg[(1-2\mu
f_{T^2})\bigg(\mu+3P\bigg)-28\bigg(-\mu+3P\bigg)^2f_{T^2}
-2(Rf_R-f)+\frac{1}{2}\psi_3\bigg];~~~
Y_{I}=\varepsilon_I-\frac{4\pi}{f_R}\bigg(1-2\mu
f_{T^2}\bigg)\Pi_I,\\\label{2p10}
&Y_{II}=\varepsilon_{II}-\frac{4\pi}{f_R}\bigg(1-2\mu
f_{T^2}\bigg)\Pi_{II}; \quad
Y_{KL}=\varepsilon_{KL}-\frac{4\pi}{f_R}\bigg(1-2\mu f_{T^2}\bigg)\Pi_{KL}.
\end{align}
In this case, the physical aspects of $Y_T$ and $Y_{T_i}$'s are also
a variant, as explained in the first case. More precisely, it can be
well to say that strong gravitational effects of the system
\eqref{1} could be revealed by these dynamical variables and
the related QS evolution of the anisotropic axial system could be paved
the better understanding due to the higher order curvature terms
with matter profile appearing in a quadratic manner. Finally,
\begin{align}\label{2p11}
Z_{\gamma\lambda}&=H_{\gamma\lambda}+\frac{4\pi}{f_R}(1
+f_T)q^{\varrho}\epsilon_{\gamma\lambda
\varrho}+4\pi\psi_4,
\end{align}
along with corresponding scalar variables
\begin{align}\label{2p12}
Z_{I}&=H_1-\frac{4\pi}{f_R}\bigg(1+2\mu f_{T^2}\bigg)q_{II};~~\quad Z_{II}
=H_1+\frac{4\pi}{f_R}\bigg(1+2\mu f_{T^2}\bigg)q_{II},\\\label{2p13}
Z_{III}&=H_2-\frac{4\pi}{f_R}\bigg(1+2\mu f_{T^2}\bigg)q_{I};~~\quad
Z_{IV}=H_2+\frac{4\pi}{f_R}\bigg(1+2\mu f_{T^2}\bigg)q_{I}.
\end{align}
We may conclude that the magnetic effects of the Weyl-tensor and
heat dissipation are directly related to the inclusion of $Z_{I, II,
III, IV}$ in this case as well. The values of $\psi_i's$ are given
in Appendix B. The particular choice of these structure scalars is
to evaluate the QS approximation of $f(R,T^2)$ scalar equations,
which are expressed in Appendix B. The above-stated groups of
scalars define various physical attributes for the evolution of
self-gravitating compact structures.

\subsection{Consequence of Quasi-static Conditions on $f(R, T^2)$ Relativistic
Equations}
In this section, we would like to check the nature of QS
evolution of the evolving system \eqref{1} within the regime of
$f(R, T^2)$ gravity. For this regime, the proposed constraints
\eqref{35} and \eqref{36} are also held. Furthermore, the consequences of heat fluxes \eqref{1p13}--\eqref{1p14} can be
carried out in the presence of heat fluxes that are describing the
effects of $f(R, T^2)$ curvature terms. The expressions of these
fluxes are enlisted in Appendix B, which provide
\begin{align}\label{2p14}
\bigg(\mathbb{T}A\bigg)'=\frac{1}{rB\tilde{f_R}}\chi^{approx}_1;~~~
\quad
\bigg(\mathbb{T}A\bigg)_{,\theta}=\frac{1}{rAB\tilde{f_R}}\chi^{approx}_2.
\end{align}
It is noticed that equations \eqref{1p30}--\eqref{1p35} are accommodated with $f(R, T^2)$ heat fluxes. However their corresponding physical meanings are
illustrated as
\begin{itemize}
\item
The sign of $V$ and $V_{(3)}$ is equal to the sign of
$V_\Sigma$ and $V_{(3)\Sigma}$, when the fluid contents are
considered in a dissipationless, irrotational, and shear-free
configurations, but the impact of effective constituents illustrates
a continuous heat-flow.
\item
Besides that in the appearance of any of the above
constituents like shear, vorticity, and generalized heat-fluxes, the
system may reach to a situation where sign of velocity flips in
fluid composition relative to its sign on boundary of surface with
the components of generalized heat-flux vector.
\item
The behavior of these velocity functions for the QS evolution of compact stars can be better understood using the generalized
constituents of fluid under discussion. As a result, it is possible
that the outer areas move in the opposite direction as the inside
ones, which propagate in a single direction.
\end{itemize}

\subsection{$f(R, T^2)$ Field, Hydro-dynamical Equations and QS approximation}

We want to assess $f(R, T^2)$ field and conservation equations in
the QS approximation in this section. So, the QS approximation over
Eqs. (B1)--(B4) in Appendix B, are turned
out to be as follows:
\begin{align}\nonumber
&G_{00}=\frac{8\pi
A^2}{\tilde{f_R}}\bigg[\mu-\frac{1}{2}(\tilde{f}-\tilde{R}\tilde{f_R})
-\bigg\{2\mu^2-7\bigg(-\mu+3p\bigg)^2\bigg\}\tilde{f}_{T^2}
+\frac{1}{A^2r^2B^2}\tilde{f}_{R,\theta\theta}+f'_{R}\bigg\{\frac{1}{B^2}
\bigg(\frac{1}{r}\bigg) \frac{1}{B^2}\bigg(\frac{C'}{C}\bigg)\bigg\}
\bigg]\\\label{2p15}&+f_{R,\theta}\bigg\{\frac{1}{r^2B^2}
\bigg(\frac{C,\theta}{C}\bigg)\bigg\};
\quad \quad
G_{01}=-\frac{8\pi}{\tilde{f_R}}\bigg[AB\bigg(1+2\mu\tilde{f}_{T^2}\bigg)
q_{I}\bigg],\\\nonumber
&G_{02}=\frac{8\pi}{\tilde{f}_R}\bigg[-\mu
G+q_{II}\bigg(ArB\bigg)+\frac{G}{2}\bigg(\tilde{f}-\tilde{R}\tilde{f_R}\bigg)+
\tilde{f}_{T^2}\bigg\{2\mu^2+2q_{II}\mu\bigg(ArB\bigg)+7G\bigg(-\mu+3p\bigg)^2
\bigg\}
+\frac{G}{r^2B^2}\bigg\{\tilde{f}_{R,\theta\theta}
\\\label{2p16}&+\tilde{f}_{R,\theta}\bigg(\frac{C_{,\theta}}{C}\bigg)\bigg\}
\bigg];~
G_{12}=\frac{8\pi}{\tilde{f}_R}\bigg[(1+2\mu\tilde{f}_{T^2})\bigg(
B^2r\Pi_{KL}+Bq_{I}\frac{G}{A}\bigg)+\tilde{f}'_{R,\theta}
-\frac{B_{,\theta}}{B}\tilde{f'}_{R}-\tilde{f}_{R,\theta}
\frac{(Br)'}{Br}\bigg],\\\label{2p17}
&G_{11}=\frac{8\pi
B^2}{\tilde{f}_R}\bigg[(1+2\mu\tilde{f}_{T^2})\bigg(P+\frac{\Pi_{I}}{3}\bigg)-
7\bigg(-\mu+3p\bigg)^2\tilde{f}_{T^2}+\frac{1}{r^2B^2}\bigg\{\tilde{f}_{R,\theta\theta}
-\tilde{f}_{R,\theta}\bigg(\frac{A_{,\theta}}{A}-\frac{B_{,\theta}}{B}
+\frac{C_{,\theta}}{C}\bigg)\bigg\}\bigg],\\\nonumber
&G_{22}=\frac{8\pi
}{\tilde{f}_R}\bigg[(1+2\mu\tilde{f}_{T^2})\bigg\{\frac{r^2B^2}{A^2}
(P+\frac{\Pi_{II}}{3})+2GrBq_{II}\bigg\}
-7r^2B^2\tilde{f}_{T^2}\bigg(-\mu+3p\bigg)^2+\frac{1}{2}(\tilde{f}
-\tilde{R}\tilde{f_R})
-\tilde{f}_{R,\theta}\bigg\{\frac{A_{,\theta}}{A}
\\\nonumber&-\frac{B_{,\theta}}{B}
+\frac{C_{,\theta}}{C}\bigg\}\bigg];\quad
G_{33}=\frac{8\pi
C^2}{\tilde{f}_R}\bigg[(1+2\mu\tilde{f}_{T^2})\bigg(P-\frac{1}{3}(\Pi_I
+\Pi_{II}\bigg)
+\frac{1}{2}(\tilde{f}-\tilde{R}\tilde{f_R})-7\tilde{f}_{T^2}\bigg(-\mu+3p\bigg)^2
-\frac{1}{r^2B^2}\\\label{2p18}&\bigg(\tilde{f}_{R,\theta\theta}
+\tilde{f}_{R,\theta}\frac{A_{,\theta}}{A}\bigg)\bigg].
\end{align}
Dynamics is the study of the motion of objects under the influence
of certain forces. The motions of galaxies are thought to be caused
by enormous and invisible stuff known as dark matter. The present
state of matter in the universe suggests that many star systems are
in a higher-dimensional era, making it harder to investigate their
dynamical structure. To solve this difficulty, a relativistic
mathematical formalism based on some invariant suppositions may be
required. However, these findings do not fully highlight the
invisible precise aspects of higher dimensional
gravitational-collapse. Such solutions, on the other hand, may lead
to some useful views. In this case, fluid dynamics could aid in the
creation of a methodology for studying the evolution of astronomical
objects. To examine the hydrodynamics of compact stars, we want to
evaluate the evolution of conservation equations in the QS
approximation within the bounds of $f(R,T^2)$ gravity. Therefore,
QS study of these equations is expressed below.
\begin{align}\nonumber
&\frac{1}{\tilde{f_R}}\bigg[(1+2\mu\tilde{f}_{T^2})\bigg\{\frac{\dot{\mu}}{A}
+\Theta(P+\mu)
+\frac{1}{9}\bigg\{\Pi_{II}(\sigma_I+2\sigma_{II})+\Pi_I(2\sigma_I
+\sigma_{II})\bigg\}+\frac{q'_I}{B}
+\frac{1}{rB}\bigg(q_{I,\theta}
G\frac{\dot{q_{II}}}{A^2}\bigg)
+2\bigg(q_{II}a_{II}\\\label{2p19}&+q_Ia_I\bigg)
+\frac{q_I}{B}\bigg(\frac{(Br)'}{Br}+\frac{C'}{C}\bigg)
+\frac{q_{II}}{rB}\bigg(\frac{C_{,\theta}}{C}+\frac{B_{,\theta}}{B}
\bigg)\bigg\}\bigg]=\frac{1}{\tilde{f_R}}\chi^{approx}_7
+7\frac{G}{A^2r^2B^2+G^2}\bigg(-\mu+3P\bigg)^2\tilde{f}_{T^2}.
\end{align}
From $f(R,T^2)$ Euler equation (B6), the
following two equations are obtained in the QS approximation.
\begin{align}\nonumber&
\frac{1}{\tilde{f_R}}\bigg[(1+2\mu\tilde{f}_{T^2})\bigg\{\frac{1}{B}\bigg(P
+\frac{\Pi_I}{3}\bigg)'+
\frac{1}{rB}\bigg(\Pi_{KL,\theta}+G\frac{\dot{\Pi}_{KL}}{A^2}\bigg)
+\bigg(\mu+P+\frac{\Pi_I}{3}\bigg)a_I
+\Pi_{KL}a_{II}
+\frac{\Pi_I}{3B}\bigg(\frac{2C'}{C}+\frac{(Br)'}{Br}\bigg)
\\\nonumber&-\frac{\Pi_{II}}{3B}\bigg(\frac{(Br)'}{Br}-\frac{C'}{C}\bigg)
+\frac{\Pi_{KL}}{rB}\bigg(2\frac{B_{,\theta}}{B}+\frac{C_{,\theta}}{C}\bigg)
+\frac{\dot{q_I}}{A}\bigg\}\bigg]= -\frac{A}{B}\mu
q_{I}\tilde{f}_{T^2}
\bigg(-\frac{1}{2\tilde{f_R}}(\tilde{f}
-\tilde{R}\tilde{f_R})\bigg)'-\frac{7}{rB}\bigg(-\mu+3p\bigg)^2\tilde{f}_{T^2}
\\\label{2p20}&+\frac{1}{B\tilde{f_R}}\chi^{approx}_8,\\\nonumber
&\frac{1}{\tilde{f}_R}\bigg[(1+2\mu\tilde {f}_{ T^{2}})\bigg\{\frac{1}{r B}\bigg(\frac{G}{A^2}\bigg(\dot{P}+\frac{\Pi_{I I}}{3}\bigg)+\bigg(P+\frac{\Pi_I}{3}\bigg)_{, \theta}\bigg)+\frac{\Pi_{K L}^{\prime}}{B}+\bigg(\mu+P+\frac{\Pi_{I I}}{3}\bigg) a_{I I}+\Pi_{K L} a_I+\frac{\Pi_I}{3Br}\bigg(-\frac{B_{, \theta}}{B}\\\label{2p21}&+\frac{C_{, \theta}}{C}\bigg)+\frac{\Pi_{I I}}{3 B r}\bigg(\frac{B_{, \theta}}{B}+\frac{2 C_{, \theta}}{C}\bigg)+\frac{\Pi_{K L}}{B}\bigg(\frac{C^{\prime}}{C}+\frac{(B r)^{\prime}}{B r}\bigg)+\frac{q_{I I}}{A}\bigg\}\bigg]=\frac{1}{\tilde{f}_R}\bigg[-\frac{1}{rB}(\tilde{f}
-\tilde{R}\tilde{f_R})-\chi^{approx}_9\bigg].
\end{align}
describing that the energy content of the anisotropic and
dissipative axial system are not conserved because of extra
curvature terms owing to the matter-field coupling in $f(R,T^2)$
context. The quantities $\chi_7$ to $\chi_9$ appearing in
Eqs. \eqref{2p19}--\eqref{2p20} could be found in Appendix B.

\subsection{$f(R,T^2)$ Scalar Equations and QS Approximation}

By enforcing the QS definitions of the kinematical quantities,
$f(R,T^2)$ Ricci-Identities have been evaluated in the same manner
as in the case of $f(R,T)$ gravity, except that the heat fluxes are
invoked the strong gravitational interactions of the matter-field
coupling due to the quadratic effects of the generalized matter
profile. These results also show that QS evolution of $f(R,T^2)$
heat fluxes get order $O(\epsilon)$. Thereby, conditions
\eqref{1p28}--\eqref{1p29} are also entertained in this case
as well. Furthermore, it is found that the QS evolution of the Weyl
tensor's magnetic constituents is exactly the same as in the first
case \cite{105} and Einstein's gravity case
\cite{7}. Subsequently, the key note aspects of the
QS evolution of $H_1$ and $H_2$ are that these components have
merely the inclusion of the geometrical constituents. The presence
of these components are the exclusive features of the axial geometry
as compared to the spherical one \cite{132}.
Therefore, equations \eqref{1p45}--\eqref{1p46} could get the same mathematical interpretation associated
with geometrical variables but physical meanings of the heat fluxes
would include the generic effects of energy-momentum squared gravity
i.e., $f(R,T^2)$ gravity. In order to discuss the evolution
equations for the Weyl-tensor constructed from the
Bianchi-Identities
$\nabla_{[\omega}R_{\lambda\alpha]\beta\delta}=0$, which are
represented as \cite{80}
\begin{align}\label{2p22}
 C^{~~~\gamma}_{{\rho\alpha\beta}~;\gamma}=R_{\beta[\rho;\alpha]}
 -\frac{1}{6}g_{\beta[\rho}R_{,\alpha]}
\end{align}
In $f(R,T^2)$ gravity bounds, above equation comes out to be as
follows
\begin{align}\label{2p23}
C^{~~~\gamma}_{{\rho\alpha\beta}~;\gamma}=\kappa
{T_{\beta[\rho;\alpha]}}^{\textrm{eff}}
-\frac{\kappa}{3}g_{\beta[\rho}{T_{,\alpha]}}^{\textrm{eff}}.
\end{align}
Contraction of the above said equation with the distinct
combinations of the four-velocity vector and projection tensor, read
the propagations (B7)--(B10).
 The effective terms of fluid contents appearing in
$f(R,T^2)$ field as well as scalar equations are exhibited in
Appendix B. Thereby, it could be said that QS evolution of the axial
stellar structures enhances its stability as compared to its fully
dynamic scenario, displayed by (B13)--(B16) in
Appendix B. Therefore, it follows immediately from
Eq. \eqref{1p28} (likewise, see
Eq. \eqref{1p48}) that
$\dot{H_2}\approx\dot{H_1}\approx O(\epsilon^2)$ (at least), and
hence are virtually unnoticeable in the QS approach. This means that
if the magnetic part of the Weyl-tensor vanishes at any point in
time, it will do so again at any time. Alternatively, no form of
radiation for a specific duration is foreseen in the QS approach,
the energy transfer is carried on by the related effective
dissipative flux in the evolution of the axial compact structures.
It is important to remark that energy transformation is conducted by
means of modified heat-fluxes ($q_{I}^{\textrm{eff}}$ and
$q_{II}^{\textrm{eff}}$), and these effective forms reveal the
significance of $f(R, T^2)$ corrections along with the usual matter
distribution. It is illustrated from \eqref{1p31}--\eqref{1p32} and \eqref{1p34}--\eqref{1p35}, that, when the
fluid contents are evaluated in non-dissipative context i.e.,
$q_{I}=0=q_{II}$, irrotational, and shear-free configurations, the
sign of $V$ and $V_{(3)}$ is identical to the sign of $V_\Sigma$ and
$V_{(3)\Sigma}$, but the contribution of $f(R,T^2)$ heat fluxes is
presented in that case as well, because $q^{\textrm{eff}}_I\neq0$
and $q^{\textrm{eff}}_{II}\neq0$, as shown below
\begin{align}\nonumber
q^{\textrm{eff}}_I&=\frac{1}{f_R}\bigg[q_I(1-2\mu
f_{T^2})-\frac{1}{AB}\chi_1\bigg],\\\nonumber
q^{\textrm{eff}}_{II}&=\frac{1}{f_R}\bigg[q_{II}(1-2\mu
f_{T^2})+\frac{1}{\sqrt{r^2A^2B^2+G^2}}\bigg(\frac{G}{A^2}\chi_0
+\chi_2\bigg)\bigg].
\end{align}
The extra correction terms emerging due to $f(R, T^2)$ gravity are
presented in $\chi_0$-$\chi_2$, whose values are displayed in
Appendix B.

\section{Consequences of Palatini $f(R)$ Corrections Over the
Evolutionary Behavior of Axial Stellar Structures}

In the preceding few years, the researchers offered a variety of
cosmological concepts that might communicate inflation and other
aspects of the universe that are still unidentified. To determine
the equations of motion, this method offers several possible
contexts: The first method is called metric interpretation (also
known as metric $f(R)$ theory), which induces the fourth-order field
equations by varying the generic action with the metric tensor. The
second method, known as the Palatini $f(R)$ approach, varies the
generic action by metric tensor and connection independently which
provides second-order field equations inducing more convenient
outcomes. The name for the third formulation is metric-affine $f(R)$
gravity. Since Einstein's gravity is conserved and the same notion
holds in these theories as well, permitting
$T_{\gamma\lambda}^{\textrm{eff}}$ to be divergence free. In this
section, we want to analyze the impact of Palatini $f(R)$
corrections over the QS approximation of the proposed axial system.
The geometric part of the action function can be generalized to obtain
$f(R)$ gravity as
\begin{align}\label{3p1}
I_{f(R)}=\frac{1}{2\kappa}\int\sqrt{-g}f(\hat{R})d^4x+
\mathcal{L}_m,
\end{align}
where $\kappa$ is the coupling constant, $\mathcal{L}_m$ represents
the matter action. It is important to note that in the above action,
the scalar curvature $\hat{R}$ is constructed by the contraction of
the relating metric-tensor with that of Ricci-tensor linked with
connection symbol, that is,
$\hat{R}:=g^{\alpha\gamma}R_{\alpha\gamma}$ which depicts $\hat{R}$
by geometric connections \cite{133}. According
to the Palatini approach, the connections and the metric tensor are
geometrical variables that are independent from each other in
variational-principle. This method contributes to the novel
modifications to the general relativity to elucidate large-scale
astronomical configurations and dark-energy aspects. By keeping the
relation $\Gamma_{\gamma\alpha}^{\lambda}\neq
\Gamma_{\alpha\gamma}^{\lambda}$ in mind, and varying Eq. \eqref{3p1}
by $g_{\alpha\gamma}$ and $\Gamma_{\alpha\gamma}^{\lambda}$,
respectively yield
\begin{align}\label{3p2}
&f_{R}(\hat{R})\hat{R}_{\alpha\gamma}-\frac{1}{2}[g_{\alpha\gamma}f_{R}(\hat{R})]
=\kappa T_{\alpha\gamma},\\\label{3p3}
&\hat{\nabla}_{\lambda}\bigg(g^{\alpha\gamma}f_{R}(\hat{R})\sqrt{-g}\bigg)=0,
\end{align}
By taking trace of Eq. \eqref{3p2}, we have
\begin{align}\label{3p4}
\hat{R}f_{R}(\hat{R})-2f(\hat{R})=\kappa T.
\end{align}
The subscript $R$ on the corresponding mathematical values has been
used to describe the application of $df/d\hat{R}$. We calculate the
connection from Eq. \eqref{3p2} to continue our study with the
quadratic order metric equations, and which after utilizing in
Eq. \eqref{3p3} gives
\begin{align}\label{3p5}
&\hat{R}_{\alpha\gamma}-\frac{1}{2}g_{\alpha\gamma}\hat{R}
=\frac{1}{f_R}\bigg(\hat{\nabla}_{\alpha}\hat{\nabla}_{\gamma}
-g_{\alpha\gamma}\hat{\Box}\bigg)f_R
+\frac{\kappa}{f_R}T_{\alpha\gamma}+\frac{1}{2f_R}g_{\alpha\gamma}(f-\hat{R}f_R)
-\frac{3}{2f^2_R}\bigg[\hat{\nabla}_{\alpha}f_R
\hat{\nabla}_{\gamma}f_R
-\frac{1}{2}g_{\alpha\gamma}(\hat{\nabla}f_R)^2
\bigg],
\end{align}
We can rewrite Eq. \eqref{3p5} in a more simple form as follows
\begin{align}\nonumber
&\hat{G}_{\alpha\gamma}=\frac{\kappa}{f_R}\bigg(T_{\alpha\gamma}+\mathcal{T}_{\alpha\gamma}
\bigg)\equiv T^{(eff)}_{\alpha\gamma}; \text{ along with }~
\mathcal{T}_{\alpha\gamma}=\frac{1}{\kappa}\bigg(\hat{\nabla}_{\alpha}\hat{\nabla}_{\gamma}
-g_{\alpha\gamma}\hat{\Box}\bigg)f_R+\frac{1}{2\kappa}g_{\alpha\gamma}(f-\hat{R}f_R)
\\\label{3p6}&-\frac{3}{2f_R}\bigg[\hat{\nabla}_{\alpha}f_R
\hat{\nabla}_{\gamma}f_R
-\frac{1}{2}g_{\alpha\gamma}(\hat{\nabla}f_R)^2 \bigg],
\end{align}
describing the gravitational effects due to Palatini $f(R)$
approach. In this case, the Einstein tensor $\hat{G}_{\alpha\gamma}
=\hat{R}_{\alpha\gamma}-\frac{1}{2}g_{\alpha\gamma}\hat{R}$, and
$\hat{\nabla}_{\gamma}$ describes the covariant derivative with
respect to the Levi-Civita connection of the corresponding
metric-tensor. It is important to mention that both $f$ and $f_R$
are the functions of $R(\Gamma)\equiv
g^{\alpha\gamma}R_{\alpha\gamma}(\Gamma)$. The main constituent of
$f(R)$ gravity is that more degrees-of-freedom can be assembled into
an effective type of stress energy-tensor, that can lead to dark
source aspects.

\subsection{Impact of Palatini $f(R)$ Correction Over Kinematical
Quantities and Structure Scalars}

In contrast to Einstein's gravity
and metric $f(R)$ counterparts, the Palatini corrections influenced
the kinematical quantities. Thereby, the consequences of Palatini
formalism are observed on the shear, expansion and vorticity
evolution as well as compared to the previous two scenarios that are
$f(R,T)$ gravity and $f(R,T^2)$ gravity. In the Palatini context,
equations \eqref{9}--\eqref{14} are turned out
in the way given below.
\begin{align}\label{3p7}
&a_{I}=\frac{1}{2B}\bigg(\frac{2A'}{A}+\frac{f'_R}{f_R}\bigg),\\\label{3p8}
&\sigma_{I}=-\frac{1}{3A}\bigg(\frac{\dot C}{C}-\frac{\dot
B}{B}\bigg)-\frac{G^2}{3A(r^2A^2B^2+G^2)}\bigg[-\frac{\dot
B}{B}-\frac{\dot A}{A}+\frac{\dot
G}{G}\bigg]-\frac{1}{6A}\bigg(\frac{\dot{f}_R}{f_R}\bigg),\\\label{3p9}
&\sigma_{II}=-\frac{1}{3A}\bigg(\frac{\dot C}{C}-\frac{\dot
B}{B}\bigg)-\frac{2G^2}{3A(r^2A^2B^2+G^2)}\bigg[\frac{\dot
B}{B}+\frac{\dot A}{A}-\frac{\dot
G}{G}\bigg]-\frac{1}{6A}\bigg(\frac{\dot{f}_R}{f_R}\bigg)
+\frac{-G^2}{2A\sqrt{r^2A^2B^2+G^2}}\bigg(\frac{f'_R}{f_R}\bigg),\\\label{3p10}
&\Theta_P=V_{;\gamma}^{\gamma}=\frac{2}{A}\bigg(\frac{\dot
C}{2C}+\frac{\dot B}{B}\bigg)
-\frac{G^2}{(A^2r^2B^2+G^2)A}\bigg(\frac{\dot B}{B}+\frac{\dot
A}{A}-\frac{\dot G}{G}\bigg)
+\frac{\dot{f}_R}{f_R}\bigg(\frac{2}{A}\bigg).
\end{align}
In contrast to the previous cases, it is significant to mention that
there are two non-zero components of vorticity-tensor in Palatini
$f(R)$ formalism
\begin{align}\nonumber
\Omega_{12}&=-\frac{1}{2}\bigg[2\frac{G}{A}\bigg(\frac{A'}{A}\bigg)
-\frac{G'}{A}+\frac{G}{2A}\bigg(\frac{f'_R}{f_R}\bigg)\bigg];~~
\Omega_{01}=-\frac{A}{4}\bigg(\frac{f'_R}{f_R}\bigg); \text{ along
with scalar}\\\label{3p11}\Omega&=-\frac
{G}{2B\sqrt{A^2r^2B^2+G^2}}\bigg[\bigg(\frac{2A'}{A}-\frac{G'}{G}\bigg)^2
-\frac{A^2r^2B^2+G^2}{4G^2}\bigg(\frac{{f'}^2_R}{{f}^2_R}\bigg)\bigg]^{\frac{1}{2}}.
\end{align}
The last term describes the Palatini effects due to connection
symbols along with the said geometric variables. However, in metric
formalism and with the condition that $f(\hat{R})=R$, the above
findings will reduce to the results presented in the study
\cite{99}. We would like to highlight the key
note for this framework as:
\begin{itemize}
\item
In order to evaluate the relativistic equations in the Palatini $f(R)$
context, the connection symbols have to be used instead of the
christoffel's symbols. Henceforth, the Palatini $f(R)$ equations
could be acquired by utilizing the connections expressed to be
$\Gamma^{\beta}_{\gamma\nu}=\{^{\beta}_{\gamma\nu}\}+ \frac{1}{2f_R}
\left[(\delta^{\beta}_{\gamma} \partial_{\nu}+ \delta^{\beta}_{\nu}
\partial_{\gamma})f_R-g_{\gamma\nu} g^{\beta\sigma} \partial_{\sigma}f_R\right]$,
determining the significant difference between all of the
consequences of the metric and Palatini $f(R)$ version. In contrast
to the consequences of EGT presented in
\cite{99}, our Palatini based study includes all
the consequential effects of the EGT along with the strong
gravitational-field interaction because of higher curvature
corrections at large cosmic scales. These higher curvature
corrections demonstrate the consequences of the dark constituents of
cosmos's energy-composition, thereby, illustrating the current
accelerating expansion scenario. Also the equations of motion that
are deduced through the $f(R)$ gravity in Palatini context, are of
quadratic order unlike to the the metric $f(R)$ counterparts as well
as Einstein's gravity. Therefore, the Palatini version is the current demand to explore the hidden secrets of the current scenario. Moreover, all our study based astrophysical consequences would be recovered in EGT
under acceptable limits.
\end{itemize}
Next, we evaluate the structure scalars by using Eq.
\eqref{28} in Palatini based framework as follows:
\begin{align}\label{3p12}
X_{\alpha\gamma}=-E_{\alpha\gamma}-\frac{\kappa}{f_R}\bigg[\frac{\Pi_{\alpha\gamma}}{2}+
\mu\frac{h_{\alpha\gamma}}{3}
+(f-\hat{R}f_R)\frac{h_{\alpha\gamma}}{6}+\Psi_1\bigg],
\end{align}
here
\begin{align}\nonumber
&\Psi_1=\frac{1}{8f_R}\epsilon_{\alpha}^{\lambda\varrho}
\bigg[\bigg(\nabla^{\pi}\nabla_{\lambda}f_R-\frac{3}{2f_R}
\nabla^{\pi}f_R\nabla_{\lambda}f_R\bigg)\epsilon_{\gamma\pi
\varrho}+\bigg(-\nabla^{\pi} \nabla_{\lambda}f_R
+\frac{3}{2f_R}\nabla^{\pi}f_R
\nabla_{\lambda}f_R\bigg)\epsilon_{\gamma\pi\varrho}
+\bigg(\frac{3}{2f_R}\nabla^{\beta}f_R\nabla_{\lambda}f_R
\\\nonumber&-\nabla^{\beta}\nabla_{\lambda}f_R\bigg)\epsilon_{\beta\gamma\varrho}
+\bigg(\nabla^{\beta}\nabla_{\varrho}f_R
+\frac{3}{2f_R}\nabla^{\beta}f_R
\nabla_{\varrho}f_R\bigg)\epsilon_{\gamma\beta\lambda}\bigg].
\end{align}
From Eq. \eqref{3p12}, the four scalar variables are
\begin{align}\label{3p13}
&X_T=\frac{\kappa}{f_R}\bigg\{\mu+\frac{1}{2}(f-\hat{R}f_R)
+\Psi^{*}_1\bigg\};~
X_{I}=-\varepsilon_I-\frac{\kappa}{2f_R}\Pi_I;~
X_{II}=-\varepsilon_{II}-\frac{\kappa}{2f_R}\Pi_{II};~
X_{KL}=-\varepsilon_{KL}-\frac{\kappa}{2f_R}\Pi_{KL}.
\end{align}
The trace part $X_T=X^{\gamma}_{\gamma}$, describing the
energy-inhomogeneity of the cosmic structure along with the effects
of the dark sources constituents in the context of Palatini
$f(\hat{R})$ formalism. The remaining components are corresponding
to the space-like vectors that elucidate the impact of the system's
anisotropy. In similar way $Y_{\alpha\gamma}$ can be expressed as
\begin{align}\label{3p14}
&Y_{\alpha\gamma}=E_{\alpha\gamma}+\frac{\kappa}{2f_R}
\bigg[(3\mu+P)V_{\alpha}V_{\gamma}+(\mu+P)g_{\alpha\gamma}
+\Pi_{\alpha\gamma}+\frac{2}{3}h_{\alpha\gamma}\bigg\{3P-\mu
+2(f-\hat{R}f_R)\bigg\}+\Psi_2\bigg],
\end{align}
here
\begin{align}\nonumber
&\Psi_2=\bigg[h_{\alpha\gamma}\bigg\{\frac{1}{2}\bigg(\nabla
f_R\bigg)^2 +(\nabla_{\alpha}f_R
\nabla_{\gamma}f_R)\bigg\}+\bigg\{\nabla_{\alpha}\nabla_{\gamma}f_R
-(\nabla_{\alpha}\nabla_{\beta}f_R)V_{\gamma}V^{\beta}
-(\nabla^{\lambda}\nabla_{\alpha}f_R)V_{\gamma}V_{\lambda}
+(\nabla^{\lambda}\nabla_{\beta}f_R)g_{\alpha\gamma}V_{\lambda}V^{\beta}\bigg\}
\\\nonumber&-3\bigg\{(\nabla_{\alpha}f_R
\nabla_{\gamma}f_R)+(\nabla_{\alpha}f_R\nabla_{\beta}f_R)V_{\gamma}V^{\beta}
+(\nabla^{\lambda}f_R\nabla_{\alpha}f_R)V_{\gamma}V_{\lambda}
-(\nabla^{\lambda}f_R\nabla_{\beta}f_R)g_{\alpha\gamma}
V_{\lambda}V^{\beta}\bigg\}\bigg].
\end{align}
From Eq. \eqref{3p14}, the scalar variables are
\begin{align}\label{3p15}
&Y_T=\frac{\kappa}{2f_R}\bigg\{2(f-\hat{R}f_R)+\mu+3P+\Psi^*_2\bigg\}
;~Y_{I}=\varepsilon_I-\frac{\kappa}{2f_R}\Pi_I;~
Y_{II}=\varepsilon_{II}-\frac{\kappa}{2f_R}\Pi_{II};
&Y_{KL}=\varepsilon_{KL}-\frac{\kappa}{2f_R}\Pi_{KL}.
\end{align}
These scalars are significant in the sense that they control the
evolution of the kinematical variables, particularly, expansion and
shear evolution. It should be noted that the trace component
$Y_T=Y_{\gamma}^{\gamma}$ does not explicitly reveal the effect of
energy-inhomogeneity. It also designs the impact of pressure
isotropy as well as the dark sources constituents in high-curvature
regimes. Finally
\begin{align}\label{3p16}
Z_{\alpha\gamma}&=H_{\alpha\gamma}+\frac{\kappa}{2f_R}\bigg[q^{\lambda}
\epsilon_{\alpha\lambda\gamma}+\Psi_3\bigg]; \text{ with }
\Psi_3=\bigg(-1+\frac{3}{2f_R}\bigg)\bigg\{\nabla^{\lambda}
\nabla_{\beta}f_R\bigg\}V^{\beta}\epsilon_{\alpha
\lambda\gamma}.
\end{align}
Along with related scalar variables
\begin{align}\label{3p17}
&Z_{I}=H_1-\frac{\kappa}{2f_R}q_{II};\quad\quad Z_{II}
=H_1+\frac{\kappa}{2f_R}q_{II};\quad\quad
Z_{III}=H_2-\frac{\kappa}{2f_R}q_{I};\quad\quad
Z_{IV}=H_2+\frac{\kappa}{2f_R}q_{I}.
\end{align}
These scalars correspond to heat dissipation as well as the
magnetic effects of the axially symmetric cosmic structures.
Consequently, one can say that these scalars control the anisotropy
of the system in some way. This is the amazing feature of axially
symmetric cosmic configuration to reveal the magnetic aspects of the
Weyl-tensor in terms of structure scalars in contrast to the
spherical cosmic structures \cite{79}.

\subsection{Palatini $f(R)$ Thermo-inertial Effects of the
Dissipative Axial Source}

In order to understand the thermo-inertial effects of the
dissipative axially symmetric self-gravitating cosmic structures,
the non-zero components of Eq. \eqref{32} in Palatini based
setting get the form as provided below.
\begin{align}\label{3p18}
&\frac{\tau}{f_R}\bigg[\dot{q}_{I}-Aq_{II}\bigg\{{\Omega}^2
+\frac{1}{16B^2}\bigg(\frac{{f'}^2_R}{f^2_R}\bigg)\bigg\}^\frac{1}{2}
+\frac{q_{I}}{2}\dot{f}_R\bigg]+{q_{I}}^{\textrm{eff}}
=-Ak\bigg\{\frac{1}{B}(\mathbb{T}Ba_I+\mathbb{T}')+\frac{\mathbb{T}^2}{2}\bigg(\frac{\tau
V^{\gamma}} {k
\mathbb{T}^2}\bigg)_{;\gamma}{q_{I}}^{\textrm{eff}}\bigg\},\\\nonumber
&\frac{\tau}{f_R}\bigg[\dot{q}_{II}+Aq_{I}\bigg\{{\Omega}^2
+\frac{1}{16B^2}\bigg(\frac{{f'}^2_R}{f^2_R}\bigg)\bigg\}^\frac{1}{2}
+\frac{\dot{f}_R}{2}q_{II}\bigg]+{q_{II}}^{\textrm{eff}}
=-Ak\bigg\{\mathbb{T}a_{II}+\frac{G\dot{\mathbb{T}}+A^2\mathbb{T}_{,\theta}}{A\sqrt{A^2r^2B^2+G^2}}+
\frac{\mathbb{T}^2}{2}\bigg(\frac{\tau V^{\gamma}} {k
\mathbb{T}^2}\bigg)_{;\gamma}{q_{II}}^{\textrm{eff}}
\\\label{3p19}&+\frac{G\mathbb{T}}{2\sqrt{A^2r^2B^2+G^2}}\bigg(\frac{\dot{f}_R}{f_R}\bigg)\bigg\}.
\end{align}
In the above-said equations, the effective scalars include the usual
heat fluxes as well as their corresponding Palatini effects combined
with the vorticity. Now, we concentrate on the properties of the
system right after it leaves the thermal equilibrium configuration,
i.e., the shortest time for which the modifications can be seen. As
a result, we assume that time is zero at the start and that the
system is in thermal equilibrium in the $\theta$-direction before to
that, resulting in the value of $q_{II}$ being zero. However, the time
derivative values of the corresponding quantities do not vanish,
those are considered to be of very low order. Using these
assumptions along with Tolman's condition
\cite{131}, the Eq. \eqref{3p19} reads
\begin{align}\label{3p20}
\dot{q_{II}}=-Aq_I\Omega+\frac{1}{f_R\sqrt{A^2r^2B^2+G^2}}
\bigg\{\frac{\mathcal{T}_{02}}{rB}+\frac{\mathcal{T}_{00}}{A^2}\bigg\},
\end{align}
revealing that no heat flow in corresponding $r$-direction, though
the extra curvature effects propagating in all direction. Imposing
the same concept in the opposite direction
\begin{align}\label{3p21}
\dot{q_{I}}=Aq_{II}\Omega+\frac{1}{rAB}\bigg(\frac{\mathcal{T}_{01}}{f_R}\bigg),
\end{align}
It is to be noted that only the lack of vorticity ensures the
time-propagation of the relating thermal equilibrium configuration
(either in $r$-direction or $\theta$-direction). Instead that, it
needed initial thermal-equilibrium for both manners. To study
the thermodynamical aspects of the self-gravitating matter
configuration coupled with vorticity may be rather difficult.
However, certain interesting consequences might be analyzed through
the transport and the modified Euler equation for the proposed
radiating structures.
In last decades, the radiating collapsing stelar structures through
diffusion approximation have been brought in discussion a lot in
various relativistic dynamics
\cite{134,36,129,
135}.
The distinctive consequences of bouncing related to the diminishing
of the effective-inertial-mass-density (e.i.m.d.) have also been
explored within the context of EGT but in modified relativistic
dynamics, it seems rare. For the first time, Herrera and his
co-researchers \cite{136} explored the e.i.m.d. for
the heat conductivity of the relativistic source when it departed
from hydro-static stage. After that, it has been discussed for the
various radiating cosmic systems in the setting of EGT
\cite{137,138,139}
where the inertial effects emerged in the casuality of corresponding
dissipation. The e.i.m.d. value had been analyzed from the
connection of hyperbolic dissipation with the relevant dynamical
equations. In our study, the combined effects of the Palatini $f(R)$
heat equation \eqref{32} with the generalized-Euler equation
(Provided in Appendix C) help to study the e.i.m.d. of the
dissipative cosmic structures which have remarkable potential effects
on the evolution of radiating cosmic objects. Now, combining
Eq. \eqref{32} and Eq. (C7) to study the e.i.m.d. of the
radiating system under consideration, this factor reads
\begin{align}\nonumber
&\frac{1}{f_R}\bigg[a_\gamma\bigg\{1-\frac{kT}{\tau(P+\mu)}
\bigg\}(P+\mu)+h^{\beta}_{\gamma}\Pi^{\alpha}_{\beta;\alpha}+\nabla_{\gamma}P
-q^{\alpha}\bigg(\sigma_{\gamma\alpha}+\Omega_{\gamma\alpha}
\bigg)\bigg] -\frac{k}{\tau}\nabla_{\gamma}T-
\bigg[\frac{1}{2}D_t\bigg\{\ln\bigg(\frac{\tau}{kT^2}\bigg)
+\frac{1}{\tau}\bigg\}\bigg]\\\nonumber&\times{q_\gamma}^{\textrm{eff}}
+\bigg(\frac{5}{6}\Theta_P
+\frac{1}{4B}\frac{f'_R}{f_R}\bigg)q_{\gamma}
=\frac{1}{2f_R}\bigg[\frac{1}{A^2r^2B^2+G^2} \bigg\{GA^2(f
-\hat{R}f_R)+rB\bigg(f-\hat{R}f_R\bigg)_{,\theta}\bigg\}
+\bigg(f-\hat{R}f_R\bigg)'\bigg]\\\label{3p22}&+\chi_2+\chi_3,
\end{align}
where $D_{,t}f=\dot{D}f\equiv f_{,\alpha}V^{\alpha}$ and
$\nabla_{\gamma}P\equiv h^{\beta}_{\gamma}P_{,\beta}$. The function
$f$ governs the time evolution within the problem under
consideration. In the above equation, first term on the left with the
factor $(P+\mu)$ describes the e.i.m.d. by four-acceleration and the
remaining ones elucidate the hydro-dynamic forces beside the terms
involving dissipative constituents in the Palatini framework. However,
the terms on the right of the equation demonstrate the gravitational
effects in high curvature regimes. Consequently, the obtained
findings include the heat conducting factors, diminishing e.i.m.d.
value in comparison to the dissipation-free configurations. More
precisely, such diminishing of e.i.m.d. is found to be applicable
throughout evolutionary phases, whereas the gravitational force
constituents appearing in dynamical laws are diminished by that
factor as well. The extra factors on the right of above equation
contribute for the more gravitational effects in high curvature
domains.

\subsection{Palatini $f(R)$ Based Evolution Equations of Axial
System and Their Physical Aspects}

In this section, we discuss some
basic equations relevant to our proposed problem in Palatini
approach. In particular, we study the time-propagation equations for
the kinematical variables likewise expansion scalar $\Theta_P$,
Shear $\sigma_{\alpha\gamma}$, and the vorticity
$\Omega_{\alpha\gamma}$. We would also like to analyze whether such
time-propagation equations are controlled by the anisotropy of the
fluids, we will also check the impact of structure scalars on their
evolution in non-geodesic problems within the Palatini $f(R)$
background.

\subsubsection{Propagation Equation for the Expansion Scalar}
For the four-velocity vector $V^{\gamma}$, the Ricci identities are
expressed as $V_{\alpha;\gamma;\mu}-V_{\alpha;\mu;\gamma}=V_{\nu}\hat{R}^{\nu}_{\alpha\gamma\mu}$
with
\begin{align}\label{3p23}
&V_{\alpha;\gamma}=\sigma_{\alpha\gamma}+\frac{1}{3}
h_{\alpha\gamma}\Theta_{P}-a_{\alpha}V_{\gamma}
+\Omega_{\alpha\gamma},\\\label{3p24}
&\frac{1}{2}\hat{R}^{\nu}_{\alpha\gamma\mu}V_{\nu}=a_{\alpha;[\mu}V_{\gamma]}
+a_{\alpha}V_{[\gamma;\mu]}+\sigma_{\alpha[\gamma;\mu]}
+\Omega_{\alpha[\gamma;\mu]}+\frac{1}{3}\bigg(\Theta_{P,[\mu}h_{\gamma]\alpha}
+\Theta_Ph_{\alpha[\gamma;\mu]}\bigg).
\end{align}
As the effects of vorticity does not vanish in axially symmetric
configuration, therefore the covariant derivative of velocity vector
also includes the effects of vorticity. Our problem yields
$2\sigma^2=\sigma^{\alpha\gamma}\sigma_{\alpha\gamma}$ and
$2\Omega^2=\Omega^{\alpha\gamma}\Omega_{\alpha\gamma}$. Contracting
Eq. \eqref{3p24} with $V^{\alpha}g^{\mu\nu}$ and then indices $\alpha$
and $\mu$, we have
\begin{align}\label{3p25}
&\Theta_{P;\mu}V^{\mu}+\frac{1}{3}\Theta^2_{P}-2(\Omega^2-\sigma^2)-a^{\mu}_{;\mu}
+\frac{\kappa}{2}(\mu+3P)=\frac{1}{\kappa}\bigg[3\Box
f_R+\frac{3}{f_R}(\nabla f_R)^2
+\frac{3}{2f_R}\nabla^{\mu}f_R\nabla_{\mu}\bigg]-\frac{2}{\kappa}f_R(f-\hat{R}f_R).
\end{align}
All the kinematical quantities are carried out the dark source
constituents, describing the effects of high curvature regimes due
to the generalization of the gravitational theory. In contrast to
the Eq. (A1) studied in \cite{99}, the
time-propagation of the expansion scalar is not equal to zero in our
problem which reveals the results in high curvature domains in
generic case, i.e., non-geodesic within the regime of Palatini
$f(R)$ gravity.

\subsubsection{Propagation Equation for the Shear}

To compute the time-propagation equation for the shear, we contract
Eq. \eqref{3p24} by $V^{\gamma}h_{\omega}^{\alpha}h_{\lambda}^{\mu}$,
the resulting equation gets the form
\begin{align}\label{3p26}
&h^{\lambda}_{\alpha}h^{\nu}_{\gamma}\sigma_{\lambda\nu;\beta}V^{\beta}
+\sigma^{\nu}_{\alpha}\sigma_{\gamma\nu}
+\frac{2}{3}\bigg(\frac{a^{\beta}_{;\beta}}{2}-\frac{\Omega^2}{2}-\sigma^2\bigg)h_{\alpha\gamma}
-\frac{1}{2}\bigg(a_{\alpha}a_{\gamma}-\omega_{\alpha}\omega_{\gamma}\bigg)
+E_{\alpha\gamma}+\frac{2}{3}\sigma_{\alpha\gamma}\Theta_P
=\frac{\kappa}{2}\Pi^{\textrm{eff}}_{\alpha\gamma}.
\end{align}
It is important to emphasize that the time-propagation of shear is
controlled by the electric effects of the Weyl-tensor and the
anisotropy of the fluids through pressure. In above expression,
$\Pi^{\textrm{eff}}_{\alpha\gamma}$ is carried the usual as well as
the extra constituents of the relevant quantity. On further
evaluation, one can observe that the shear evolution may be
controlled by the anisotropy of the fluids using structure
scalars $Y_I, Y_{II}$ and $Y_{KL}$ which reveals the key role of
structure scalars to analyze the physical characteristics of the
self-gravitating source.

\subsubsection{Propagation Equations for the Vorticity}

The role of vorticity is highly significant for the dynamic as well
as QS evolution of the axial source. To assess the evolutionary
aspects, the time propagation equation for the vorticity
$\Omega_{\alpha\gamma}$ in our study reads as
\begin{align}\label{3p27}
h^{\gamma}_{\omega}h^{\alpha}_{\lambda}\Omega_{\gamma\alpha;\beta}V^{\beta}
+2\sigma_{\gamma[\omega}\Omega^{\gamma}_{\lambda]}
-h^{\gamma}_{[\omega}h^{\alpha}_{\lambda]}a_{\gamma;\alpha}=-
\frac{2}{3}\Omega_{\omega\lambda}\Theta_P.
\end{align}
Additionally, the subsequent constraint equations for our relativistic system follows
\begin{align}\label{3p28}
&h^{\alpha}_{\gamma}\bigg(\frac{2}{3}\Theta_{P;\alpha}-\sigma^{\alpha}_{\gamma;\alpha}
+\Omega^{\alpha}_{\gamma;\alpha}\bigg)
+a^{\alpha}(\sigma_{\gamma\alpha+\Omega_{\gamma\alpha}}) =\kappa
q^{\textrm{eff}}_{\gamma},\\\label{3p29}
&2\omega_{(\gamma}a_{\alpha)}+h^{\mu}_{(\gamma}h_{\alpha)\beta}\bigg(\sigma_{\mu\delta}
+\Omega_{\mu\delta}\bigg)_{;\nu} \eta^{\beta\varrho
\nu\delta}V_{\varrho}=H_{\gamma\alpha},
\end{align}
former equation describes the relation between the Palatini
formulated kinematical variables and the generalized dissipative
fluxes, while the latter reveals the role of anisotropy in terms
of magnetic part $H_{\gamma\alpha}$ of the Weyl-tensor From these
basic equations, we formulate the scalar equations in the framework
of Palatini $f(R)$ gravity. These equations are helpful to
understand the evolutionary stages of axially symmetric anisotropic
and radiating axial structures. By contracting Eq. \eqref{3p26} with
\textbf{KK, KL} and \textbf{LL} , the shear propagation reads
\begin{align}\label{3p30}
&\sigma_{I,\mu}V^{\mu}+
\frac{1}{3}(\sigma_I+2\Theta_P)\sigma_I+2\bigg(\frac{a^{\beta}_{;\beta}}{2}
-\frac{\Omega^2}{2}-\sigma^2\bigg)-3\bigg[
\bigg\{\frac{1}{2B}\bigg(\frac{2A'}{A}+\frac{f'_R}{f_R}
\bigg)\bigg\}^2+K^{\lambda}K^{\nu}a_{\lambda;\nu}\bigg]
+\varepsilon_I=\frac{\kappa}{2f_R}\Pi_I,\\\label{3p31}&
-\frac{\Omega}{3}(\sigma_{II}-\sigma_{I})
-\bigg\{\frac{1}{2B}\bigg(\frac{2A'}{A}+\frac{f'_R}{f_R}
\bigg)\bigg\}a_{II}-K^{\lambda}L^{\nu}a_{\lambda;\nu}
+\varepsilon_{KL}=\frac{\kappa}{2f_R}\Pi_{KL}
,\\\nonumber &\sigma_{II,\mu}V^{\mu}+
\frac{1}{3}(\sigma_{II}+2\Theta_P)\sigma_{II}
+2\bigg(\frac{a^{\beta}_{;\beta}}{2}
-\frac{\Omega^2}{2}-\sigma^2\bigg)-3\bigg[
\frac{A^2}{A^2r^2B^2+G^2}\bigg\{\frac{A_{,\theta}}{A}
-\frac{G}{A^2}\bigg(\frac{\dot{A}}{A}
-\frac{\dot{G}}{G}\bigg)\bigg\}^2+L^{\lambda}L^{\nu}a_{\lambda;\nu}\bigg]
\\\label{3p32}&+\varepsilon_{II}=\frac{\kappa}{2f_R}\Pi_{II}.
\end{align}
The above shear evolution equations are controlled by $Y_I, Y_{II}$
and $Y_{KL}$, respectively. Thus, one should say that the shearing
motion of the axially symmetric self-gravitating cosmic structures
is controlled by the combined effects of these structure scalars
and the palatini-formulated vorticity scalar. Similarly, we
can formulate the vorticity evolution by the possible combinations
of the corresponding tetrad vectors. Therefore, contracting
Eq. \eqref{3p28} with the vector \textbf{K} and \textbf{L}, we have the
following results, respectively,
\begin{align}\nonumber
&\frac{2}{3B}\Theta_P'-\Omega_{;\beta}L^{\beta}-\bigg(L^{\beta}_{;\beta}
-L_{\beta;\alpha}K^{\alpha}K^{\beta}\bigg)\Omega
+\frac{\sigma_I}{3}\bigg\{\frac{1}{2B}\bigg(\frac{2A'}{A}+\frac{f'_R}{f_R}
\bigg)\bigg\}-\frac{\sigma_{I;\beta}}{3} K^{\beta}-\Omega a_{II}
-\frac{1}{3}\bigg[K^{\beta}_{;\beta}
-\frac{1}{3}\bigg\{\frac{1}{2B}\\\label{3p33}&\bigg(\frac{2A'}{A}+\frac{f'_R}{f_R}
\bigg)\bigg\}\bigg](\sigma_{II}+2\sigma_I)
+\frac{2}{3}\bigg(\frac{\sigma_{I}}{2}+\sigma_{II}\bigg)
\bigg(L_{\gamma;\alpha} L^{\alpha}K^{\gamma}-\frac{1}{3}a_I\bigg)
=\frac{\kappa}{f_R}\bigg\{q_I-\frac{\mathcal{T}_{01}}{AB}\bigg\},\\\nonumber
&\frac{2A}{3\sqrt{A^2r^2B^2+G^2}}\bigg\{\Theta_{P,\theta}+\frac{G}{A^2}\dot{\Theta}_P\bigg\}
+K^{\beta}\Omega_{;\beta}+\frac{\sigma_{II}}{3}a_{II}
+\Omega\bigg\{L_{\beta;\mu}L^{\mu}K^{\beta}+K^{\nu}_{;\nu}\bigg\}+
\bigg\{\frac{1}{2B}\bigg(\frac{2A'}{A}+\frac{f'_R}{f_R}
\bigg)\bigg\}\Omega\\\nonumber&-\frac{1}{3}\sigma_{II;\beta}L^{\beta}
+\frac{2}{3}\bigg(\sigma_{II} +2\sigma_I\bigg)
\bigg[L_{\beta;\lambda}K^{\lambda}K^{\beta}
+\frac{A}{3\sqrt{A^2r^2B^2+G^2}}\bigg\{\frac{A_{,\theta}}{A}
-\frac{G}{A^2}\bigg(\frac{\dot{A}}{A}-\frac{\dot{G}}{G}\bigg)\bigg\}\bigg]
+\frac{1}{9}(\sigma_{I}+2\sigma_{II})\bigg\{a_{II}
\\\label{3p34}&-3L^{\alpha}_{;\alpha}\bigg\}=\frac{\kappa}{f_R}\bigg\{q_{II}-\frac{1}{\sqrt{A^2r^2B^2+G^2}}
\bigg(\frac{\mathcal{T}_{02}}{A^2}
+\frac{\mathcal{T}_{00}}{A^2}\bigg)\bigg\}.
\end{align}
To analyze the magnetic aspects of the Weyl tensor, we project the
Eq. \eqref{3p29} with the tetrad vector \textbf{KS} and \textbf{LS},
and obtain the following equations for magnetic components $H_1$ and
$H_2$ of the Weyl tensor
\begin{align}\label{3p35}
H_1&=\frac{1}{2}\bigg(-K^{\alpha}S_{\lambda}-K_{\lambda}S^{\alpha}\bigg)\bigg(\sigma_{\alpha\delta}
+\Omega_{\alpha\delta}\bigg)_{;\nu}\epsilon^{\beta\nu\delta}
-\Omega\bigg\{\frac{1}{2B}\bigg(\frac{2A'}{A}+\frac{f'_R}{f_R}
\bigg)\bigg\},\\\label{3p36}
H_2&=\frac{1}{2}\bigg(-L^{\alpha}S_{\lambda}-L_{\lambda}S^{\alpha}\bigg)\bigg(\sigma_{\alpha\delta}
+\Omega_{\alpha\delta}\bigg)_{;\nu}\epsilon^{\beta\nu\delta}-
\frac{A\Omega}{3\sqrt{A^2r^2B^2+G^2}}\bigg\{\frac{A_{,\theta}}{A}
-\frac{G}{A^2}\bigg(\frac{\dot{A}}{A}
-\frac{\dot{G}}{G}\bigg)\bigg\},
\end{align}
demonstrating that stable configuration of shear free motion is
influenced by the combined effects of the $H_{1}, H_{2}$ and the
generalized heat conducting vectors. It is also significant to
mention here that shear free motion of the axially symmetric
configuration is directly linked with the free-vorticity constraint.
Though the extra constituents imposed the high curvature effects on
the evolving cosmic structures in this situation as well.

\section{Conclusions}

We have addressed the concepts of dynamic as well as QS pattern of evolution of axially and reflection symmetric self-gravitating source within the domains of various high curvature regimes i.e., a comparison-based analysis of the impact of various curvature factors on the propagation of the axial source is exhibited. For doing se, we have chosen the most generic
representation of stress energy-tensor as given in Eq.(\ref{2}). We developed concerning field equations to study the motion of relativistic source. To analyze the basic aspects of fluid contents, we discussed the
kinematical quantities including $a_{\lambda}$ (four-acceleration),
$\Theta$ (expansion-scalar), $\Omega$ (vorticity-scalar) and
$\sigma_{\lambda\omega}$ (shear-tensor).
Moreover, we have discussed the magnetic and electric parts
of the Weyl-tensor. Five scalars $H_1, H_2$ and $\varepsilon_I,
\varepsilon_{II}, \varepsilon_{KL}$ illuminating the magnetic and
electric-parts, respectively, are also established.
The set of invariant-velocities are defined for the comprehension of kinematics as well as for the concept of the pattern of QS evolution.
These scalar functions basically hold the relation
$V_{(1)}+V_{(2)}+V_{(3)}=\Theta$ in metric based contexts.
It can be seen from Eqs. \eqref{20}--\eqref{23} that
geometrical and physical demonstration of such specific-velocities is governed by kinematical-variables with the unit space-like vectors
$K_{\gamma}, L_{\gamma}$ and $S_{\gamma}$.
The relevant axial evolutionary pattern is examined at three distinct large-scales, which is summarized in the following aspects:
We have commenced by evaluating the consequential impact of $f(R,T)$ gravity over the evolving system. We established the following in this case:\\
The set of seven field equations for our $(1+3)$ formalism are calculated
by using Eq. \eqref{1p3}, then QS
constraints defined in second section are imposed to evaluate
the concerned approximation.
The QS pattern of evolution of the field and dynamical consequences are presented in
Eqs. \eqref{1p16}--\eqref{1p25}, including
the extra-curvature terms due to the effects of $f(R, T)$ gravity as compared to the results presented in \cite{7}.
The modified heat-fluxes are also executed to examine
the thermodynamic aspects of self-gravitating evolving fluid through
the proposed approximation. The significant role played
by kinematical as well as modified heat-fluxes is
clearly revealed through the Eqs. \eqref{1p31}--\eqref{1p35}.
The above mentioned constituents lead to the pattern of
various structures. It is to be noted that in the scenario of $f(R, T)$ gravity, the modified heat-fluxes and extra curvature terms are emerged as shown in Eqs. \eqref{1p13}--\eqref{1p15}, however, in Einstein's
gravity \cite{7}, these were governed with usual heat-fluxes
together with $(TA)'=0, \text{ and }
(TA)_{,\theta}=0.$
Moreover, we were exclusively focused on the time propagation of the relativistic source, whose evolution was controlled by the dynamics of the structure scalars under the shearing motion. Thereby, evolutionary aspects of the shear, expansion, and vorticity parameters for the system under investigation have been identified.
It is noticed from Eqs. (A9)--(A10) that shear evolution is coupled with $\Omega^2+2\sigma^2$ together with $f(R,T)$ structure scalars, revealing the gravitational interaction at large-scale.
Consequently, it can be said that the shear free deviation is directly connected with the vorticity-free state. However,
to enhance the stability of the dynamic evolution, we have enforced the QS approximation over the proposed system. As a result, the system gains simplicity in its pattern of evolution, deploying slowly evolving regimes.\\
Next, this case extends to $f(R,T^2)$ gravity, illustrating the impact of energy momentum squared-gravity domains and being comparable to Einstein's gravity in the vacuum background. More specifically, effects of this theory are more perceivable in domains with strong curvature. In this particular scenario, the computed $f(R,T^2)$ constituents work together to
form a structure of exceptional properties.
In the case of geodesics, vorticity-evolution is too constrained,
as shown in \cite{99} (full discussion is
available there).
one can notice that the evolution equation (A11)
is also governed by the $f(R,T^2)$ scalar $Y_{KL}$. In the evolutionary phase, this equation exposes the significance of $Y_{KL}=\varepsilon_{KL}-\frac{\kappa}{2f_R}(1-2\mu
f_{T^2})\Pi_{KL},$ revealing the effects of anisotropy as well as the
electric component of the Weyl-tensor. In the case of geodesic, the above evolutionary aspect with $\Omega=0$ i.e., in vorticity-free zone, becomes $Y_{KL}=0$. comparing this result with Eq. (B3) in \cite{99}, one should satisfy that $\varepsilon_{KL}=4\pi\Pi_{KL}$
under these constraints. However,
in the study of energy-momentum squared gravity, one should satisfy that
$\varepsilon_{KL}=\frac{\kappa}{2f_R}(1-2\mu
f_{T^2})\Pi_{KL}$ under consideration, where $\kappa=8\pi$.
This reveals the impact of energy momentum
squared-gravity on the evolution of the axially
symmetric self-gravitating objects. Therefore, we can say that if
$\varepsilon_{KL}\neq\frac{\kappa}{2f_R}(1-2\mu
f_{T^2})\Pi_{KL}$, it means that $Y_{KL}\neq0$, consequently,
there is no geodesic and vorticity free evolution.
As our metric-element is not diagonal,
$L^0$ is not zero. The zero-vorticity requirement, on the other
hand, is not effective in the presence of ordinary dissipative
fluxes, as shown by Eq. (B12) in Appendix B, but the extra elements of modified heat fluxes cause the developing system to settle into a more stable
configuration. Furthermore, Eqs. \eqref{1p45} and \eqref{1p46}
demonstrate that shear-free and vanishing of vorticity are
interconnected for fluid, resulting in $H1 \text{ and } H2$ being disappeared in the vorticity-free situation in an identical manner to $f(R,T)$ gravity case. With no dissipation, Eqs. (B9) can be used to find the inverse of this. We determined that the
absence of $H_1, H_2$ in dissipation does not necessarily imply
zero-vorticity, and we have also noted certain $f(R, T^2)$
modifications.
Its evolution is based on $f(R, T^2)$ scalar variables.
In the geodesic condition, $Y_{I, II, KL}$ are accountable for
shear evolution. Although, in the generic instance, $Z_{I, II, KL}$
as well as the $f(R, T^2)$ modifications, have a healthy impact.\\
Thirdly, importance of the Palatini-based theories in the context of dynamical self-gravitating systems could never be overstated and Palatini $f(R)$ gravity is one of the most straightforward methods of this kind. We have developed this theory to understand the thermodynamical characteristics of the topic under consideration as a third large-scale domain.
It is worthy noting that Palatini corrections affected the kinematical quantities as well, in contrast to their counterparts in metric $f(R)$ and Einstein's gravity. As a result, the effects of the Palatini approach on the dynamical progression of shear, expansion, and vorticity have been found in relation to the two metric-based theories presented in the prior sections. This is due to the fact that this kind of formalism uses connections rather than Christoffel's symbols to distinguish between all of the consequences of the metric and Palatini $f(R)$ versions. In this setting:\\
We have explored the two non-zero components of the $\Omega_{\alpha\gamma}$
instead of one as in metric formalism, as in \eqref{18}. This enhances the
pattern of vorticity evolution in palatini case because $\Omega_{01}$ and $\Omega_{12}$ have their contribution in the evolution of the relativistic configuration. In order to analyze the shearing motion on the proposed relativistic structures in non-geodesic case, i.e., $a_{\gamma}\neq 0$,
we have also developed the non-zero components of the shear, and it is observed that these scalars also contain dark source constituents in their expression. For the analysis of evolutionary phases of our problem, we determined the set of Palatini $f(R)$ structure scalars with the help of the Riemann-tensor, and expressed them in terms of trace and trace-free parts of the system. To study the thermo-inertial effects of the dissipative axially
symmetric cosmic structures, we computed the heat conduction equation
by using the usual dissipation theory \cite{124} in the framework of Palatini $f(R)$ theory. The relaxation time is
the key parameter relating to the dissipative mechanism, and is demonstrated as the time taken by the self-gravitating fluid for getting its steady position after facing some fluctuations.
It is worthy mentioning that investigation of the dissipative
aspects of the self-gravitating matter configuration coupled with vorticity may be rather difficult, however, we have analyzed certain interesting findings through Eqs. \eqref{3p18} and (C7) for the proposed dissipative structures and checked the effects in high curvature domains as shown in Eq. \eqref{3p22}. With coupling of both above mentioned equations, we studied the e.i.m.d. factor of the dissipative source that has key potential effects on the evolution of radiating cosmic systems, and concluded that the thermo-inertial effects tend to dislocate to the center, thereby diminishing any outward oriented heat-flow.
Moreover, we believe that the QS evolution of an axial source could be worth considering within the regime of Palatini $f(R)$ gravity. Last but not least, Palatini-based novel exact solutions of the specific velocities in $f(R)$ gravity would serve as a proving ground for patterning QS evolution.\\
As a definition of QS evolution, a system evolves slowly at all times, and it can be said to be in an equilibrium state in such an evolution. Thereby; the metric, kinematical, specific velocity, and fluid parameters have a decisive role in defining these constraints. So, the parameters assuming order $O(\epsilon^2)$ have been disregarded throughout the QS study. Subsequently, one can get the simplest mode of the dynamic evolution of an axial system in these scenarios, where the level of scales could be estimated through the influence of the gravitational interaction of the relating AGTs scenarios.
Even though QS evolution is the simplest pattern of evolution for axially and reflectionally self-gravitating structures,
continuing along the same lines of reasoning. However, it is unclear if alternative evolutionary patterns could potentially play a part in the basic pattern of evolution.
After all of this, it should be noted that there are several other candidates that may be put forward in place of the one listed above for patterning the QS evolution of a self-gravitating system in different alternatives. These comments suggest some queries and open issues that, in our opinion, should be analyzed further in the future.:
\begin{itemize}
\item
Are there any other concepts of QS evoloution of axially symmetric system, except from the one that was developed in \cite{105}?
\item
What would be the consequential impact of $f(R,T)$ gravity, when a specific cosmic model will be employed in this scenario?
\item
Could we describe any other kind of evolution pattern in addition to the QS domains that would be considered the simplest for an axial self-gravitating system?
\item
Within the domain of energy momentum squared-gravity, are there any other
concepts of patterning QS evolution of the proposed problem different from the one that was adopted in \cite{103}?
\item
Is there any connection between the pattern of Qs evolution and the e.i.m.d. factor within the proposed matter-field coupling bounds, and is there any particular dissipative regime that could be deemed this factor to be the simplest one?
\item
How should we demonstrate the pattern of QS evolution in light of the other modified gravity theories that have not yet been taken into account?
\item
Although the QS study for a changed axial fluid has not yet been proposed, does it exist or not in the evolution of such a complex configuration?
\item
What QS evolution share the physical aspects of the system if Palatini $f(R)$ technique is to be used?
\item
Is there another method to propose the pattern of QS evolution in an axial self-gravitating system besides using the particular velocities technique?
\item
How much would Palatini-based $f(R)$ exact solutions of the invariant velocities be significant for the QS study of the axial dissipative source?
\item
What would be the consequential impact of the QS evolution of shear, vorticity, and Raychaudhuri-equation in geodesic and non-geodesic scenarios under the context of the Palatini $f(R)$ approach? Is there any physically relevant system revealing such consequences? If yes, then what would be the physical meanings of the corresponding solutions?
\item
How does QS pattern evolve? Do any physically significant axial system
favor this kind of evolution over dynamic case?
\end{itemize}
In a nutshell, a comparison based summary table
of axially symmetric system's QS evolution in distinct scenario is provided below, which may be helpful for the readers to grasp the consequential impact of this review.
\begin{table}[h!]
\caption{ A summary table, describing the numerous relativistic consequences within the distinct gravitational context.
This could assist readers in understanding the significant implications of this study.}
 \label{Table:} {\small\centering
\begin{tabular}{|c| c| c| c| c| c|}
 \hline\hline
\textbf{Sr. No.}  &\textbf{Gravitational Context} & \textbf{Remarks}\\ [0.5ex]
\hline\
01.&$f(R,T)$ gravity&The QS-pattern of field, dynamical, and propagation equations has been brought out\\ [0.5ex]
&&together with the aspects of higher-order curvature factors. The significant role of the\\ [0.5ex]
&&kinematical
as well as $f(R,T)$ heat-fluxes has been revealed to describing the strong\\ [0.5ex]
&&gravitational interaction. The dynamics of the propagating system is studied under \\ [0.5ex]
&&shearing motion, coupling of vorticity and shear scalar is observed in shear evolution\\ [0.5ex]
&& along with the generic structure scalars. The QS-approximation of Raychaudhuri\\ [0.5ex]
&&equation is determined also the stability of the system is enhanced by
enforcing the \\ [0.5ex]
&&QS-approximation of evolution. Moreover, certain astrophysical results along with the\\ [0.5ex]
&& physical meanings to better grasp the current cosmic time scenario, has been studied.\\ [0.5ex]
\hline
02.&$f(R,T^2)$ gravity&In contrast to previous case, astrophysical consequences have been evaluated for minimal \\ [0.5ex]
&& geometry matter coupling scenario of quadratic curvature corrections. This case study\\ [0.5ex]
&& describes more powerful results and stability of the dynamic evolution is boosted, which \\ [0.5ex]
&&disclosing more meaningful aspects of QS-pattern of evolution under the  bounds of \\ [0.5ex]
&&higher-curvature regimes. Anisotropy, inhomogeneity, and thermodynamical aspects\\ [0.5ex]
&&of the system has increased with the inclusion of $f(R, T^2)$ scalars. The QS-Pattern of\\ [0.5ex]
&& of approximation in this case demonstrates more stable aspects of the system. \\ [0.5ex]
\hline
03.& Palatini $f(R)$ gravity& Dynamical aspects of the axial anisotropic and dissipative system has analyzed via Palatini\\ [0.5ex]
&& based approach. In contrast to above cases, the consequential impact of Palatini approach \\ [0.5ex]
&&over kinematical variables is observed as well due to the inclusion of connections. Time\\ [0.5ex]
&& evolution of dissipative source is controlled by the function, describing thermo-inertial\\ [0.5ex]
&& effects by powerful gravitational interactions via Palatini-based corrections. Propagation\\ [0.5ex]
&& of shear, expansion, and vorticity scalar deployed the some intriguing outcomes under this\\ [0.5ex]
&& scenario. \\ [0.5ex]
\hline
04. & Proposed Problems & These comments suggest some queries and open issues that, in our opinion, should be
\\ [0.5ex]
&& analyzed further in the future, which would contributes some more meaningful and\\ [0.5ex]
&& alternative consequences in the field of astrophysics and cosmology.\\ [0.5ex]
\hline\hline
\end{tabular}}
\end{table}

\section*{Acknowledgement}
The work of KB was supported in part by the
JSPS KAKENHI Grant Number JP21K03547.

\section*{Data Availability Statement}
This manuscript has no associated data
or the data will not be deposited. [Authors comment: This is a theoretical
study and no experimental data is used.]

\section*{Appendix A}
\subsection*{$f(R,T)$ Field and Dynamical Equations for Axial Self-gravitating
System} For the spacetime \eqref{1} and matter content
\eqref{2}, equation \eqref{1p3} reads:
\begin{align}\tag{A1}
&G_{00}=\frac{\kappa}{f_R}\bigg[A^2(\mu
-\frac{1}{2}(f-Rf_R))+\chi_{0}\bigg];~~~~
G_{01}=\frac{\kappa}{f_R}\bigg[-AB(1+f_T)q_{I}+\chi_1\bigg],
\\\tag{A2}
&G_{02}=\frac{\kappa}{f_R}\bigg[\mu
G-\sqrt{r^2A^2B^2+G^2}(1+f_T)q_{II}
+\frac{G}{2}(f-Rf_R)+\chi_2\bigg],\\\tag{A3}
&G_{12}=\frac{\kappa}{f_R}\bigg[\frac{B}{A}{\sqrt{r^2A^2B^2+G^2}
(1+f_T)\Pi_{KL}+Gq_{I}}+\chi_3\bigg],\\\tag{A4}
&G_{11}=\frac{\kappa}{f_R}\bigg[B^2\bigg({(1+f_T)(P+\frac{\Pi_{I}}{3})
+\mu f_T+\frac{1}{2}(f-Rf_R)}\bigg)+\chi_4\bigg],
\\\nonumber
&G_{22}=\frac{\kappa}{f_R}\bigg[(1+f_T)\bigg\{\frac{\mu
G^2}{A^2} +\frac{r^2A^2B^2+G^2}{A^2}(P+\frac{\Pi_{II}}{3})
+\frac{2G\sqrt{r^2A^2B^2+G^2}}{A^2}q_{II}\bigg\}\bigg]
+\frac{\kappa}{f_R}\bigg\{r^2B^2\bigg(\mu f_T+\frac{1}{2}(f\\\tag{A5}&-Rf_R)\bigg)+\chi_5\bigg\};~~~~
G_{33}=\frac{\kappa}{f_R}\bigg[C^2(1+f_T)\bigg(P-\frac{1}{3}(\Pi_{I}
+\Pi_{II})\bigg)
+C^2\bigg(\mu f_T+\frac{1}{2}(f-Rf_R)\bigg)+\chi_6\bigg].
\end{align}
In $f(R,T)$ gravity, energy conservation laws
$(T^{\mu}_{\nu;\mu})^{\textrm{eff}}\neq0$ leads to the continuity equation:
\begin{align}\nonumber
&\frac{1}{f_R}\bigg[V^{\beta}\mu_{;\beta}+(1+f_T)\bigg(\Theta(\mu
+P)+\frac{\Pi_I}{9}(2\sigma_I
+\sigma_{II})+\frac{\Pi_{II}}{9}(\sigma_I+2\sigma_{II})+
q^{\beta}_{;\beta}+q^{\beta}a_{\beta}\bigg)\bigg]
=\frac{\chi_7}{f_R}
+\frac{1}{f_R}(f-Rf_R)\\\tag{A6}&\bigg[\frac{G^2}{r^2A^2B^2
+G^2}\bigg(\frac{\dot{A}}{A}
+\frac{\dot{B}}{B}+\frac{\dot{G}}{G}
\bigg)\bigg],
\end{align}
while the generalized Euler-equation in our case is
\begin{align}\nonumber
&\frac{1}{f_R}\bigg[(1+f_T)\bigg\{a_{\mu}(\mu+P)
+h^{\gamma}_{\mu}\bigg(P_{;\gamma}
+\Pi^{\nu}_{\gamma;\nu}+q_{\gamma;\nu}V^{\nu}\bigg)+
q^{\gamma}\bigg(\sigma_{\mu\gamma}+\Omega_{\mu\gamma}+\frac{4}{3}
h_{\mu\gamma}\Theta\bigg)\bigg\}\bigg]
=\bigg(\frac{-1}{2f_R}(f-Rf_R)\bigg)'
\\\tag{A7}&-\frac{1}{2f_R}(f-Rf_R)-\frac{1}{f_R}\bigg(\chi_8+\chi_9\bigg).
\end{align}
The extra terms appear in these equations are termed as $\chi_i$'s,
which are given below.
\begin{align}\nonumber
&\chi_0=\ddot{f_R}-\frac{r^2A^2B^2}{r^2A^2B^2+G^2}\ddot{f_R}
+\frac{A^2}{B^2}{f_R}''
+\frac{A^4}{r^2A^2B^2+G^2}{f_{R\theta\theta}}
+\frac{2A^2G}{r^2A^2B^2+G^2}{\dot{f_{R\theta}}}
+\frac{r^2A^2B^2}{r^2A^2B^2+G^2}
\dot{f_R}\bigg[
\frac{G^2}{r^2A^2B^2+G^2}\\\nonumber&\times\bigg(\frac{\dot{A}}{A}\bigg)
-\frac{2r^2A^2B^2+3G^2}{r^2A^2B^2+G^2}\bigg(\frac{\dot{B}}{B}\bigg)
-\frac{\dot{C}}{C}
+\frac{G^2}{r^2A^2B^2+G^2}\bigg(\frac{\dot{G}}{G}\bigg)
-\frac{A^2G}{r^2A^2B^2+G^2}\bigg(\frac{A_{,\theta}}{A}
+\frac{B_{,\theta}}{B}-\frac{G_{,\theta}}{G}\bigg)
-\frac{G}{r^2B^2}\bigg(\frac{A_{,\theta}}{A}\\\nonumber&-\frac{B_{,\theta}}{B}
+\frac{C_{,\theta}}{C}\bigg)
-\frac{G^2}{r^2A^2B^2}\bigg]\bigg(\frac{\dot{G}}{G}\bigg)
+\frac{r^2A^2B^2}{r^2A^2B^2+G^2}{f_R}'\bigg[\frac{A^2}{B^2}\bigg(\frac{A'}{A}
+\frac{B'}{B}+\frac{1}{r}\bigg)-\frac{r^2A^2B^2
+G^2}{r^2B^4}\bigg(\frac{A'}{A}+\frac{B'}{B}
-\frac{C'}{C}\bigg)\bigg]
\\\nonumber&+\frac{G^2}{r^2B^4}\bigg(\frac{G'}{G}\bigg)
+\frac{r^2A^2B^2}{r^2A^2B^2+G^2}{f_{R\theta}}\bigg[
\frac{G}{r^2B^2}\bigg(\frac{\dot{A}}{A}+\frac{\dot{B}}{B}+\frac{\dot{C}}{C}
-\frac{\dot{G}}{G}\bigg)
-\frac{GA^2}{r^2A^2B^2+G^2}\bigg(\frac{\dot{A}}{A}
-\frac{\dot{B}}{B}-\frac{\dot{G}}{G}\bigg)
-\frac{A^2}{r^2B^2}\bigg(\frac{A_{,\theta}}{A}-
\\\nonumber&\frac{B_{,\theta}}{B}-\frac{C_{,\theta}}{C}\bigg)
-\frac{2A^2G}{r^2A^2B^2+G^2}\bigg\{\frac{\dot{B}}{B}
-\frac{1}{rB^2}\bigg(\frac{A_{,\theta}}{A}\bigg)
+\frac{A^2}{2G}\bigg(\frac{B_{,\theta}}{B}\bigg)+\frac{G}{2r^2B^2}
\bigg(\frac{G_{,\theta}}{G}\bigg)\bigg\}\bigg],\\\nonumber
&\chi_1=\dot{f_R}'-\frac{r^2A^2B^2}{r^2A^2B^2+G^2}\bigg[\bigg(\frac{A'}{A}
+\frac{G^2}{2r^2A^2B^2}\frac{G'}{G}\bigg)\dot{f_R}+\bigg(\frac{r^2A^2B^2
+G^2}{r^2A^2B^2}\bigg)\frac{\dot{B}}{B}f_R'
+\frac{G}{r^2B^2}\bigg(\frac{A'}{A}-\frac{G'}{2G}\bigg){f_{R\theta}}\bigg],\\\nonumber
&\chi_2={\dot{f_{R\theta}}}-G\bigg[\frac{-1}{r^2A^2B^2+G^2}\bigg(r^2B^2\ddot{f_R}
-2G{\dot{f_{R\theta}}}-A^2{f_{R\theta\theta}}\bigg)+\frac{1}{B^2}{f_R}''
+\frac{r^2A^2B^2}{r^2A^2B^2+G^2}\dot{f_R}\bigg\{\bigg(\frac{\dot{A}}{A}
-\frac{\dot{B}}{B}
+\frac{G^2}{r^2A^2B^2}\frac{\dot{G}}{G}\bigg)
\\\nonumber&\times\frac{r^2B^2}{r^2A^2B^2+G^2}
-\frac{\dot{C}}{A^2C}
+\frac{G}{r^2A^2B^2}\bigg(\frac{B_{,\theta}}{B}+\frac{C_{,\theta}}{C}\bigg)
-\frac{G}{r^2A^2B^2+G^2}\bigg(\frac{B_{,\theta}}{B}
-\frac{G_{,\theta}}{G}\bigg)\bigg\}
+\frac{r^2A^2B^2}{r^2A^2B^2+G^2}{f'_R}
\bigg\{\frac{1}{B^2}\bigg(\frac{A'}{A}\\\nonumber&+\frac{(Br)'}{Br}\bigg)
+\frac{r^2A^2B^2+G^2}{r^2A^2B^4}\bigg(\frac{C'}{C}\bigg)
+\frac{G^2}{r^2A^2B^4}\bigg(\frac{G'}{G}\bigg)\bigg\}
+\frac{r^2A^2B^2}{r^2A^2B^2+G^2}
{f_{R\theta}}\{\frac{-G}{r^2A^2B^2+G^2}\bigg(\frac{\dot{A}}{A}
-\frac{\dot{G}}{G}+\frac{\dot{B}}{B}\bigg)
+\frac{G}{r^2A^2B^2}\\\nonumber&\times\bigg(\frac{\dot{B}}{B}
+\frac{\dot{C}}{C}\bigg)
+\frac{1}{r^2B^2}\bigg(\frac{B_{,\theta}}{B}+\frac{C_{,\theta}}{C}\bigg)
-\bigg(\frac{2A_{,\theta}}{A}-\frac{G_{,\theta}}{G}\bigg)
\frac{G^2}{r^2B^2(r^2A^2B^2+G^2)}\bigg\}\bigg],\\\nonumber
&\chi_3={f'_{R\theta}}-\frac{r^3B^2G}{r^2A^2B^2+G^2}
\dot{f_R}\bigg\{\frac{(Br)'}{Br}-\frac{G'}{2G}\bigg\}
-\frac{B_{,\theta}}{B}f'_R-\frac{r^2A^2B^2}{r^2A^2B^2+G^2}{f_{R\theta}}
\bigg\{\frac{(Br)'}{Br}+\frac{G^2}{r^2A^2B^2}
\bigg(\frac{G'}{2G}\bigg)\bigg\},\\\nonumber
&\chi_4=\frac{r^2A^2B^2}{r^2A^2B^2+G^2}\bigg\{\frac{B^2}{A^2}\ddot{f_R}
-\frac{1}{r^2}f_{R\theta\theta}
-\frac{2G}{A^2r^2}\dot{f_{R\theta}}\bigg\}
+\frac{r^2A^2B^2}{r^2A^2B^2+G^2}\dot{f_R}\bigg[
\frac{r^2B^4}{r^2A^2B^2+G^2}\bigg\{-\frac{\dot{A}}{A}
-\bigg(\frac{G^2}{r^2A^2B^2}\bigg)\frac{\dot{G}}{G}\\\nonumber&+\frac{\dot{B}}{B}
\bigg(\frac{r^2A^2B^2+2G^2}{r^2A^2B^2}\bigg)\bigg\}
+\frac{B^2}{A^2}\bigg(\frac{\dot{C}}{C}\bigg)
+\frac{GB^2}{r^2A^2B^2+G^2}\bigg(\frac{B_{,\theta}}{B}
-\frac{G_{,\theta}}{G}\bigg)
-\frac{G}{A^2r^2}\bigg(\frac{C_{,\theta}}{C}\bigg)\bigg]
-\frac{r^2A^2B^2}{r^2A^2B^2+G^2}{f_R}'\bigg[\frac{A'}{A}
\\\nonumber&+\frac{(Br)'}{Br}
+\frac{r^2A^2B^2+G^2}{r^2A^2B^2}\bigg(\frac{C'}{C}\bigg)
+\frac{G^2}{r^2A^2B^2}\bigg(\frac{G'}{G}\bigg)\bigg]
+\frac{r^2A^2B^2}{r^2A^2B^2+G^2}{f_{R\theta}}
\bigg[\frac{GB^2}{r^2A^2B^2+G^2}\bigg\{\frac{\dot{A}}{A}
+\frac{\dot{B}}{B}-\frac{\dot{G}}{G}\bigg\}
-\frac{G}{A^2r^2}\bigg(\frac{\dot{C}}{C}\bigg)
\\\nonumber&-\frac{A^2B^2}{r^2A^2B^2+G^2}\bigg\{\frac{r^2A^2B^2
+2G^2}{r^2A^2B^2}\bigg(\frac{A_{,\theta}}{A}\bigg)
-\frac{B_{,\theta}}{B}-\frac{G^2}{r^2A^2B^2}
\bigg(\frac{G_{,\theta}}{G}\bigg)\bigg\}-\frac{1}{r^2}
\bigg(\frac{C_{,\theta}}{C}\bigg)\bigg],\\\nonumber
\chi_5&=\frac{r^2A^2B^2}{r^2A^2B^2+G^2}\bigg\{\frac{B^2r^2}{A^2}
\ddot{f_R}-f_{R\theta\theta}
-\frac{2G}{A^2}\dot{f_{R\theta}}\bigg\}-r^2f''_R+f_{R\theta\theta}
-\frac{r^2A^2B^2}{r^2A^2B^2+G^2}\dot{f_R}
\bigg[\frac{r^4B^4}{r^2A^2B^2+G^2}\bigg\{\frac{\dot{A}}{A}
+\frac{G^2}{A^2r^2B^2}\bigg(\frac{\dot{G}}{G}\bigg)
\\\nonumber&-\frac{r^2A^2B^2+2G^2}{r^2A^2B^2}\bigg(\frac{\dot{B}}{B}\bigg)
\bigg\}
-\frac{B^2r^2}{A^2}\bigg(\frac{\dot{C}}{C}\bigg)
+\frac{r^2B^2G}{r^2A^2B^2+G^2}\bigg(\frac{B_{,\theta}}{B}
-\frac{G_{,\theta}}{G}\bigg)
-\frac{G}{A^2}\bigg(\frac{B_{,\theta}}{2B}+\frac{C_{,\theta}}{C}
-\frac{G_{,\theta}}{G}\bigg)\bigg]
-\frac{r^2A^2B^2}{r^2A^2B^2+G^2}{f_R}'
\\\nonumber&\times\bigg[\frac{r^2B^2}{B^2}
\bigg(\frac{A'}{A}
+\frac{(Br)'}{Br} +\frac{G^2}{A^2r^2B^2}\bigg(\frac{G'}{G}\bigg)\bigg)
-\frac{r^2A^2B^2+2G^2}{A^2B^2}\bigg\{\frac{(Br)'}{Br}+\frac{B'}{B}
-\frac{C'}{C}\bigg\}\bigg]
+\frac{r^2A^2B^2}{r^2A^2B^2+G^2}{f_{R\theta}}
\bigg[\frac{Gr^2B^2}{r^2A^2B^2+G^2}
\\\nonumber&\times\bigg\{\frac{\dot{A}}{A}
+\frac{\dot{B}}{B}-\frac{\dot{G}}{G}\bigg\}
-\frac{G}{A^2}\frac{\dot{C}}{C}
-\frac{r^2A^2B^2}{r^2A^2B^2+2G^2}\bigg\{\frac{B_{,\theta}}{B}
+\bigg(\frac{A_{,\theta}}{A}\bigg)\frac{r^2A^2B^2+2G^2}{r^2A^2B^2}
-\frac{G^2}{r^2A^2B^2}\bigg(\frac{G_{,\theta}}{G}\bigg)\bigg\}
-\frac{C_{,\theta}}{C}
\\\nonumber&-\frac{G^2}{r^2A^2B^2}\bigg(\frac{G_{,\theta}}{G}\bigg)\bigg],\\\nonumber
\chi_6&=\frac{r^2C^2B^2}{r^2A^2B^2+G^2}\bigg\{\ddot{f_R}
-\frac{A^2}{B^2r^2}f_{R\theta\theta}
-\frac{2G}{B^2r^2}\dot{f_{\theta}}\bigg\}
+\frac{r^2A^2B^2}{r^2A^2B^2+G^2}\dot{f_R}\bigg[
\frac{r^2C^2B^2}{r^2A^2B^2+G^2}
\bigg\{-\frac{G^2}{A^2r^2B^2}\bigg(\frac{\dot{G}}{G}\bigg)
-\frac{\dot{A}}{A}\\\nonumber&+\frac{r^2A^2B^2
+2G^2}{r^2A^2B^2}\bigg(\frac{\dot{B}}{B}\bigg)\bigg\}
+\frac{C^2}{A^2}\frac{\dot{B}}{B}
+\frac{GC^2}{r^2A^2B^2+G^2}\bigg\{\frac{A_{,\theta}}{A}
+\frac{B_{,\theta}}{B}-\frac{G_{,\theta}}{G}\bigg\}
-\frac{GC^2}{A^2r^2B^2}\bigg(\frac{B_{,\theta}}{B}\bigg)\bigg]
+\frac{r^2A^2B^2}{r^2A^2B^2+G^2}{f_R}'
\\\nonumber&\times\bigg[\frac{r^2A^2B^2+G^2} {r^2A^2B^4}\bigg\{\frac{B'}{B}
-\frac{G^2}{r^2A^2B^2+G^2}\bigg(\frac{G'}{G}\bigg)\bigg\}C^2
-\frac{C^2}{B^2}\bigg(\frac{A'}{A}
+\frac{(Br)'}{(Br)}\bigg)\bigg]
+\frac{r^2A^2B^2}{r^2A^2B^2+G^2}{f_{R\theta}}
\bigg[\frac{GC^2}{r^2A^2B^2+G^2}
\bigg\{\frac{\dot{A}}{A}\\\nonumber&+\frac{\dot{B}}{B}-\frac{\dot{G}}{G}\bigg\}
-\frac{A^2C^2}{r^2A^2B^2+G^2}
\bigg\{\frac{r^2A^2B^2+2G^2}{r^2A^2B^2}\bigg(\frac{A_{,\theta}}{A}\bigg)
-\frac{G^2}{r^2A^2B^2}
\bigg(\frac{B_{,\theta}}{B}-\frac{G_{,\theta}}{G}\bigg)\bigg\}\bigg].
\end{align}
The quantities emerging in Eqs. (A6)--(A7) are computed as follows:
\begin{align}\nonumber
&\chi_7=\frac{\chi_0}{f_R(r^2A^2B^2+G^2)}\bigg[(-2r^2B^2)\frac{\dot{A}}{A}
+\frac{r^2A^2B^2+G^2}{A^2}
\bigg(\frac{\dot{3B}}{B}-\bigg(\frac{\dot{C}}{C}\bigg)
-\frac{\dot{f_R}}{f_R}\bigg)
-2\frac{G^2}{A^2}\bigg(\frac{\dot{G}}{G}\bigg)
-G\frac{A_{,\theta}}{A}+\frac{r^2A^2B^2+G^2}{r^2A^2B^2}
\\\nonumber&\times\bigg(\frac{G_{,\theta}}{G}
-\frac{C_{,\theta}}{C}\bigg)G\bigg]
-\frac{1}{f_R}\bigg[\frac{\dot{\chi_0}}{A^2}
+\frac{G}{r^2A^2B^2}\chi_{0_{\theta}}\bigg]
+\frac{1}{B^2f_R}\bigg[\chi'_1
+\chi_1\bigg\{\frac{f_R'}{R}+\frac{A'}{A}+\frac{(Br)'}{Br}
+\frac{C'}{C}\bigg\}-\frac{\dot{B}}{B}\chi_4\bigg]
+\frac{1}{rBf_R}\chi_2\bigg[2\bigg(\frac{A_{,\theta}}{A}
\\\nonumber&+\frac{B_{,\theta}}{B}\bigg)
+\frac{C_{,\theta}}{C}
-\frac{Gr^2B^2}{r^2A^2B^2+G^2}\bigg\{\frac{\dot{A}}{A}
+\frac{\dot{B}}{B}-\frac{\dot{G}}{G}
+\frac{G^2}{r^2B^2}\bigg(\frac{G_{,\theta}}{G}\bigg)\bigg\}\bigg]
+\frac{A}{r^2Bf_R}\bigg[\chi_{2_{\theta}}
+\frac{G}{B^2}\frac{A'}{A}\chi_3
+\frac{G^2}{r^2A^2B^2+G^2}
\bigg\{\frac{\dot{A}}{A}-\frac{\dot{G}}{G}
\\\nonumber&-\frac{A^2}{G}\frac{A_{,\theta}}{A}\bigg\}
\chi_5\bigg]+\mu\frac{\dot{f_R}}{f_R^2}
-\frac{1}{f_R}\bigg(\dot{R}f_R-R\dot{f_R}\bigg)
-\frac{A}{B}q_If'_T
-\frac{A^2}{\sqrt{r^2A^2B^2+G^2}}q_{II}f_{T\theta}
+\frac{G(1+f_T)}{f_R\sqrt{r^2A^2B^2+G^2}}\Pi_{KL}
\bigg[\frac{A}{B}\bigg(\frac{A'}{A}-\frac{G'}{2G}\bigg)\bigg],\\\nonumber
&\chi_8=\frac{B}{A}q_I\bigg\{\dot{f_T}-(1+f_T)\frac{\dot{f_R}}{f_R}\bigg\}
+\dot{f_T}\bigg(P+\frac{\Pi_I}{3}\bigg)
+\frac{\sqrt{r^2A^2B^2+G^2}}{r^2B^2}f_{T\theta}\Pi_{KL}
+\frac{1}{B}\bigg(\chi'_4
-\frac{f'_R}{f_R}\chi_4\bigg)
+\frac{r^2A^2B^2}{r^2A^2B^2+G^2}\bigg[\chi_0\bigg(\frac{A'}{A}
\\\nonumber&+\frac{G^2}{2r^2B^2}\frac{G'}{G}\bigg)
+\frac{\chi_5}{C^2}\bigg(\frac{A'}{A}+\frac{(Br)'}{Br}\bigg)\bigg]
+\frac{\chi_2}{r}\frac{r^2A^2B^2}{r^2A^2B^2+G^2}\bigg[\frac{(Br)'}{Br}
+\frac{G^2}{2r^2B^2}\frac{G'}{G}\bigg],\\\nonumber
&\chi_9=\mu
f'_T-\frac{f'_R}{f_R}(1+f_T)\bigg(P+\frac{\Pi_I}{3}\bigg)
+\frac{\mu}{rB}
f_T+\frac{1}{B^2}
\frac{f_{R\theta}}{f_R}\bigg\{\frac{G}{r^2}q_If_T-\frac{(1
+f_T)}{\sqrt{r^2A^2B^2+G^2}}\Pi_{KL}\bigg\}
+\frac{r^2A^2B^2}{r^2A^2B^2+G^2}\bigg[\chi_1\bigg
(\frac{\dot{A}}{A}
+\frac{\dot{2B}}{B}
\\\nonumber&+\frac{G^2}{2r^2B^2}\frac{\dot{G}}{G}
+\frac{1}{rB}\bigg(\frac{A_{,\theta}}{A}-\frac{B_{,\theta}}{2B}\bigg)
+\frac{G^2}{r^2B^2C}\frac{G_{,\theta}}{G}\bigg)\bigg]
+\frac{1}{r^2B}\chi_3\bigg[\frac{B_{,\theta}}{B}
+\frac{r^2A^2B^2}{r^2A^2B^2+G^2}\bigg\{
\frac{A_{,\theta}}{A}+\frac{G^2}{r^2B^2}\frac{G_{,\theta}}{G}
+\frac{C_{,\theta}}{C}\bigg\}\bigg].
\end{align}
The extra terms $\psi_i's$ appearing in Eqs. \eqref{1p4}, \eqref{1p7},
and \eqref{1p12} are
\begin{align}\nonumber
&\psi_1=\frac{\kappa}{8f_R}\epsilon_{\mu}^{\epsilon \varrho}
\bigg[(\nabla^{\pi}\nabla_{\epsilon}f_R) \epsilon_{\nu\pi
\varrho}-(\nabla^{\pi}\nabla_{\varrho}f_R)\epsilon_{\nu\pi\epsilon}
-(\nabla^{\theta}\nabla_{\epsilon}f_R)\epsilon_{\theta\nu\varrho}
+(\nabla^{\theta}\nabla_{\varrho}f_R)\epsilon_{\theta\nu\epsilon}\bigg],\\\nonumber
&\psi_2=\nabla_{\mu}\nabla_{\nu}f_R-\frac{3}{2}V_{\mu}V_{\nu}(f-Rf_R)
-\nabla_{\mu}\nabla_{\varrho}f_RV^{\varrho}V_{\nu}
+2V_{\mu}V_{\nu}\Box
f_R-(\nabla^{\theta}\nabla_{\nu}f_R)V_{\mu}V^{\theta}
+g_{\mu\nu}-(\nabla^{\theta}\nabla_{\varrho}f_R)V_{\theta}V^{\varrho},\\\nonumber
&\psi_3=\nabla^{\mu}\nabla_{\mu}f_R+\frac{3}{2}(f-Rf_R)
-(\nabla^{\mu}\nabla_{\varrho}f_R)V^{\varrho}V_{\mu}
-2\Box f_R-(\nabla^{\theta}\nabla_{\mu}f_R)V^{\mu}V_{\theta}
+4(\nabla^{\theta}\nabla_{\varrho}f_R)V_{\theta}V^{\varrho},\\\nonumber
&\psi_4=(\nabla^{\varrho}\nabla_{\gamma}f_R)V^{\gamma}\epsilon_{\mu
\varrho\nu}.
\end{align}
The effective terms of matter contents appearing in corresponding
differential equations, modified scalar equations are found to be
\begin{align}\nonumber
&\mu^{\textrm{eff}}=\frac{1}{f_R}\bigg[\mu-\frac{1}{2}(f-Rf_R)
+\chi_0\bigg];\quad\quad\quad
q^{\textrm{eff}}_I=\frac{1}{f_R}\bigg[q_I(1
+f_T)-\frac{1}{AB}\chi_1\bigg],\\\nonumber
&q^{\textrm{eff}}_{II}=\frac{1}{f_R}q_{II}(1+f_T)
-\frac{1}{f_R\sqrt{r^2A^2B^2+G^2}}\bigg[
G\chi_0+\chi_2\bigg],\\\nonumber
&\Pi^{\textrm{eff}}_{KL}=\frac{1}{f_R}(1+f_T)\Pi_{KL}
+\frac{1}{f_R\sqrt{r^2A^2B^2+G^2}}\bigg[
G\bigg(q_If_T-\frac{1}{AB}\chi_1\bigg)+\frac{A}{B}\chi_3\bigg],\\\nonumber
&(P+\frac{\Pi_I}{3})^{\textrm{eff}}=\frac{1}{f_R}(1
+f_T)\bigg(P+\frac{\Pi_I}{3}\bigg)+\frac{1}{f_R}\bigg[\mu
f_T+\frac{1}{2}(f-Rf_R)+\frac{1}{B^2}\chi_4\bigg],\\\nonumber
&(P+\frac{\Pi_{II}}{3})^{\textrm{eff}}=\frac{1}{f_R}(1
+f_T)(P+\frac{\Pi_{II}}{3})
+\frac{1}{f_R\sqrt{r^2A^2B^2+G^2}}\bigg[\bigg\{\frac{1}{2}(f-Rf_R)
(G-rB^2)-G\chi_0\bigg\}
\\\nonumber&+2\bigg\{\sqrt{r^2A^2B^2+G^2}q_{II}f_T+G(\chi_0
+\frac{\chi_2}{r})\bigg\}+rB^2\bigg(\mu
f_T+\frac{\chi_5}{C^2}\bigg)\bigg].
\end{align}

\subsection*{Scalar Equations and Some Ricci-Identities in $f(R,T)$ Scenario}

The evolution of expansion-scalar (Raychaudhuri-equation) indicates the
motion of matter elements. In the non-geodesic case of our formalism, it is
calculated as
\begin{align}\tag{A8}
\Theta_{;\mu}V^{\mu}+\frac{\Theta^2}{3}-2(\Omega^2-\sigma^2)
-a^{\mu}_{;\mu}+Y^{*}_T=0.
\end{align}
The shear equations inform of its two free scalars are
\begin{align}\tag{A9}
&\sigma_{I,\alpha}V^{\alpha}+\frac{\sigma_I}{3}(\sigma_I+2\Theta)
+(a^{\alpha}_{;\alpha}
-2\sigma^2-\Omega^2)-3(a^2_I+K^{\delta}K^{\gamma}a_{\delta;\gamma})+Y^{*}_I=0,
\\\tag{A10}&\sigma_{{II},\alpha}V^{\alpha}+\frac{\sigma_{II}}{3}(\sigma_I
+2\Theta)+(a^{\alpha}_{;\alpha}
-2\sigma^2-\Omega^2)-3(a^2_{II}+L^{\delta}L^{\gamma}a_{\delta;\gamma})
+Y^{*}_{II}=0.
\end{align}
The significance of $Y^{*}_{KL}$ in the evolutionary phase is depicted by
\begin{align}\tag{A11}
\frac{\Omega}{3}(\sigma_I-\sigma_{II})-a_{II}a_I
-K^{(\delta}L^{\gamma)}a_{\delta;\gamma}+Y^{*}_{KL}=0,
\end{align}
and vorticity-evolution
\begin{align}\tag{A12}
\Omega_{,\alpha}V^{\alpha}+\frac{\Omega}{3}(\sigma_I
+\sigma_{II}+2\Theta)+K^{[\delta}L^{\gamma]}a_{\delta;\gamma}=0,
\end{align}
In the generic case, the succeeding
scalar equations are obtained by projecting on every viable
combination of tetrad vectors, which are \textbf{V}, \textbf{K},
\textbf{L}, and \textbf{S}.
Using Eq. \eqref{7} and the fact from the field equations
of $f(R,T)$ gravity follows ${{T}_{03}}^{\textrm{eff}}=0$, we also
obtain
\begin{align}\tag{A13}
h^{\nu}_{\mu}\bigg(\frac{2}{3}\Theta_{;\nu}-\sigma^{\gamma}_{\nu;\gamma}
+\Omega^{\gamma}_{\nu;\gamma}\bigg)+a^{\nu}(\sigma_{\mu\nu+\Omega_{\mu\nu}})
=\kappa q^{\textrm{eff}}_{\mu},\\\tag{A14}
2\omega_{(\mu}a_{\nu)}+h^{\gamma}_{(\mu}h_{\nu)\beta}(\sigma_{\gamma\alpha}
+\Omega_{\gamma\alpha})_{;\delta} \eta^{\beta k
\delta\alpha}V_{k}=H_{\mu\nu}.
\end{align}
On contraction of Eq. (A13) by
the tetrad vector \textbf{K} and \textbf{L}, the obtained equations
show that the magnetic part of the Weyl-tensor and effective heat
flux constituents that play a dominant role in the stable configuration
of shear-free evolution. These are given as
\begin{align}\nonumber
&\frac{2}{3B}\Theta'-\Omega\bigg(L^{\alpha}_{;\alpha}
-L_{\gamma;\alpha}K^{\alpha}K^{\gamma}\bigg)
-\Omega_{;\alpha}L^{\alpha}+\frac{a_I}{3}\sigma_I-\frac{1}{3}
\sigma_{I;\alpha}K^{\alpha}-a_{II}\Omega
-\frac{1}{3}(K^{\alpha}_{;\alpha}-\frac{1}{3}a_I)(\sigma_{II}+2\sigma_I)
-\frac{1}{3}(\sigma_{I}+2\sigma_{II})
\\\tag{A15}\times&\bigg(L_{\gamma;\alpha}
L^{\alpha}K^{\gamma}-\frac{1}{3}a_I\bigg)=\kappa
{q_I}^{\textrm{eff}},\\\nonumber
&\frac{2}{3\sqrt{r^2A^2B^2+G^2}}\bigg(A\Theta_{,\theta}
+\frac{G}{A}\dot{\Theta}\bigg)
+\Omega_{;\alpha}K^{\alpha}+\frac{a_{II}}{3}\sigma_{II}
+\Omega\bigg(L_{\gamma;\alpha}L^{\alpha}K^{\gamma}
+K^{\alpha}_{;\alpha}\bigg)
+a_{I}\Omega-\frac{1}{3}\sigma_{II;\alpha}L^{\alpha}
+\frac{1}{3}a_I(\sigma_{II}+2\sigma_I)
\\\tag{A16}\times&\bigg(L_{\gamma;\alpha}K^{\alpha}K^{\gamma}
+\frac{1}{3}a_{II}\bigg)
-\frac{1}{3}(\sigma_{I}+2\sigma_{II})\bigg(L^{\alpha}_{;\alpha}
-\frac{1}{3}a_{II}\bigg)=\kappa {q_{II}}^{\textrm{eff}}.
\end{align}
The differential equations are provided as follows:
\begin{align}\nonumber
&h^{\rho}_{(\mu}h^{\beta}_{\nu)}E_{\rho\beta;\alpha}V^{\alpha}
+E_{\mu\nu}\Theta
+E_{\rho\beta}h_{\mu\nu}\sigma^{\rho\beta}
+h^{\rho}_{(\mu}\eta^{\alpha\gamma\delta}_{\nu)}V_{\alpha}H_{\gamma\rho;\delta}
-3E_{\rho(\mu}\sigma^{\rho}_{\nu)}-E_{\alpha(\mu}\Omega^{\alpha}_{\nu)}
-2H^{\rho}_{(\mu}\eta_{\nu)\alpha\delta\rho}V^{\alpha}a^{\delta}
=\frac{\kappa}{2}{(\tilde{\mu}+P)}^{\textrm{eff}}\sigma_{\mu\nu}
\\\nonumber&-\frac{\kappa}{6f_R}(1+f_T)\Theta\Pi_{\mu\nu}
-\frac{\kappa}{2f_R}h^{\rho}_{(\mu}h^{\beta}_{\nu)}(1+f_T)
\Pi_{\rho\beta;\alpha}V^{\alpha}-\frac{\kappa}{2}
\Omega^{\rho}_{(\mu}{\Pi_{\nu)\rho}}^{\textrm{eff}}
-\kappa a_{(\mu}{q_{\nu)}}^{\textrm{eff}}
+\frac{\kappa}{6f_R}(1+f_T)
\bigg(\Pi_{\rho\beta}\sigma^{\rho\beta}+a_{\rho}{q^{\rho}}^{\textrm{eff}}+
\\\tag{A17}&\frac{1}{f_R}(1+f_T)q^{\rho}_{;\rho}\bigg)h_{\mu\nu}
-\frac{\kappa}{2f_R}h^{\rho}_{(\mu}
h^{\beta}_{\nu)}(1+f_T)q_{\beta;\rho},\\\nonumber
&h^{\rho}_{\mu}h^{\beta\nu}E_{\rho\beta;\nu}
-{\eta_{\mu}}^{\alpha\beta\delta}V_{\alpha}
\sigma^{\gamma}_{\beta}H_{\delta\gamma}+3H_{\mu\nu}\omega^{\nu}=
\frac{\kappa}{3f_R}(1+f_T)\mu_{;\nu}h^{\nu}_{\mu}
-\frac{\kappa}{2f_R}\bigg(1+f_T\bigg)h^{\nu}_{\mu}
h^{\rho\beta}\Pi_{\nu\beta;\rho}
-\frac{\kappa}{2}\bigg(\frac{2}{3}
h^{\nu}_{\mu}\Theta-\sigma^{\nu}_{\mu}
\\\tag{A18}&+3\Omega^{\nu}_{\mu}\bigg){q_{\nu}}^{\textrm{eff}},\\\tag{A19}
&(\sigma_{\mu\alpha}E^{\alpha}_{\nu}
+3\Omega_{\mu\alpha}E^{\alpha}_{\nu})
{\epsilon_{\delta}}^{\mu\nu}+a^{\beta}H_{\beta\delta}
-H^{\beta\alpha}_{~~;\alpha}h_{\beta\delta}=
\frac{\kappa}{2}{(\tilde{\mu}+P)}^{\textrm{eff}}
\Omega_{\mu\nu}{\epsilon_{\delta}}^{\mu\nu}+
\frac{\kappa}{2f_R}(1+f_T)\bigg[q_{\mu;\nu}
+\Pi_{\beta\mu}(\sigma^{\beta}_{\nu}
+\Omega^{\beta}_{\nu})\bigg]{\epsilon_{\delta}}^{\mu\nu} ,\\\nonumber
&2a_{\nu}E_{\rho\delta}{\epsilon_{\gamma}}^{\rho\nu}-E_{\beta\nu;\alpha}
h^{\beta}_{\delta}{\epsilon_{\gamma}}^{\alpha\nu}
+E^{\alpha}_{\nu;\alpha}{\epsilon_{\gamma\delta}}^{\nu}
+\frac{2}{3}H_{\delta\gamma}\Theta
+H^{\mu}_{\beta;\alpha}V^{\alpha}h^{\beta}_{\delta}h_{\mu\gamma}
-H^{\alpha}_{\gamma}(\sigma_{\delta\alpha}
+\Omega_{\delta\alpha})+(\sigma_{\nu\alpha}
+\Omega_{\nu\alpha})
H^{\mu}_{\rho}{\epsilon_{\delta\mu}}^{\alpha}{\epsilon_{\gamma}}^{\rho\nu}
\\\nonumber&+\frac{\Theta}{3}H^{\mu}_{\rho}
{\epsilon_{\delta\mu}}^{\alpha}{\epsilon_{\gamma\alpha}}^{\rho}
=\frac{\kappa}{6f_R}(1+f_T)\tilde{\mu}_{,\nu}{\epsilon_{\gamma\delta}}^{\nu}
+\frac{\kappa}{2f_R}(1+f_T)
\Pi_{\rho\beta;\nu}h^{\beta}_{\delta}{\epsilon_{\gamma}}^{\rho\nu}+
\frac{\kappa}{2}\bigg[\Omega_{\rho\nu}{q_{\delta}}^{\textrm{eff}}
+{q_{\rho}}^{\textrm{eff}}
(\sigma_{\delta\nu}+\Omega_{\delta\nu}+
\\\tag{A20}&\frac{\Theta}{3}h_{\delta\nu})\bigg]{\epsilon_{\gamma}}^{\rho\nu}.
\end{align}
Contracting Eq. (A17) by tetrad vectors
\textbf{KK}, \textbf{KL}, \textbf{LL} and \textbf{SS}, we attain
the fruitful results to study the aspects of $f(R,T)$ structure scalars.
\begin{align}\nonumber
&-\frac{1}{3}\bigg(X^{*}_I-\frac{\kappa}{2f_R}(1
+f_T)\mu\bigg)_{,\alpha}V^{\alpha}
-\frac{\varepsilon_I}{9}(\sigma_I-\sigma_{II}-3\Theta)
+\frac{\varepsilon_{II}}{9}(\sigma_I+\sigma_{II})
-K_{\beta}[H_{1,\delta}S_{\gamma}+H_1
S_{\gamma;\delta}
+H_2(L_{\mu;\delta}S_{\gamma}K^{\mu}
\\\nonumber&+S_{\mu;\delta}L_{\gamma}K^{\mu})]+\Omega X^{*}_{KL}=2a_{II}H_1
-\frac{\kappa}{6}(P+\mu
+\frac{\Pi_I}{3})^{\textrm{eff}}(\Theta+\sigma_I)
-\kappa
a_I{q_I}^{\textrm{eff}}
-\frac{\kappa}{2B}({q_I}^{\textrm{eff}})'
-\frac{\kappa A{q_{II}}^{\textrm{eff}}}{2\sqrt{r^2A^2B^2+G^2}}
\bigg(\frac{G\dot{B}}{A^2B}\\\tag{A21}&
+\frac{B_{,\theta}}{B}\bigg),\\\nonumber
&-X^{*}_{KL,\alpha}V^{\alpha}
-\frac{\Omega}{6}(X^{*}_I-X^{*}_{II})+\frac{X^{*}_{KL}}{2}
(\sigma_I+\sigma_{II}-2\Theta)-(a_{II}H_2-a_IH_1)
-\frac{1}{2}\bigg\{\bigg(H_{1,\delta}S_{\gamma}
+H_1(S_{\gamma;\delta}
+S_{\mu;\delta}K_{\gamma}K^{\mu})
\\\nonumber&+H_2S_{\gamma}L_{\mu;\delta}K^{\mu}\bigg)
\epsilon^{\nu\gamma\delta}\bigg\}
+\frac{1}{2}\bigg[\bigg(H_1K^{\mu}S_{\gamma}
+H_2S^{\mu}L_{\gamma}\bigg)L_{\mu;\delta}\epsilon^{\nu\gamma\delta}K_{\nu}\bigg]
-\frac{1}{2}(H_2S_{\gamma;\delta}
+H_{2;\delta}S_{\gamma})\epsilon^{\nu\gamma\delta}K_{\nu}
=-\frac{\kappa}{3}{\Pi_{KL}}^{\textrm{eff}}\bigg(\sigma_I
\\\tag{A22}&+\sigma_{II}-\Theta\bigg)
-\frac{\kappa}{2}(a_{II}{q_I}^{\textrm{eff}}+a_I{q_{II}}^{\textrm{eff}})
-\frac{\kappa}{4f_R}(K^{\beta}L^{\mu}
+L^{\beta}K^{\mu})q_{\beta;\mu},\\\nonumber
&\frac{1}{3}\bigg(-X^{*}_{II}+\frac{\kappa}{2f_R}(1
+f_T)\mu\bigg)_{;\alpha}V^{\alpha}
-\frac{\varepsilon_{II}}{9}(\sigma_{II}-\sigma_{I}-3\Theta)
+\frac{\varepsilon_{I}}{9}(2\sigma_I+\sigma_{II})-\Omega X^{*}_{KL}
-\bigg[H_{2;\delta}S_{\gamma}
-H_1(S_{\gamma}L_{\mu;\delta}K^{\mu}
+L_{\mu;\delta}\\\tag{A23}&S^{\mu}K_{\gamma})
+H_2S_{\gamma;\delta}\bigg]\epsilon^{\nu\gamma\delta}L_{\nu}
+2H_2a_I=-\frac{\kappa}{6}(\Theta+\sigma_{II})(P+\mu
+\frac{\Pi_{II}}{3})^{\textrm{eff}}
-\frac{\kappa}{2f_R}(1+f_T)L^{\mu}L_{\beta}q^{\beta}_{;\mu}-\kappa
a_{II}{q_{II}}^{\textrm{eff}},\\\nonumber
&\frac{1}{3}\bigg(X^{*}_I+X^{*}_{II}+\frac{\kappa}{2f_R}(1
+f_T)\mu\bigg)_{;\alpha}V^{\alpha}
+\frac{1}{3}(X^{*}_I+X^{*}_{II})(\sigma_{II}+\sigma_{I}+\Theta)
+\frac{1}{9}\{\varepsilon_I(2\sigma_I+\sigma_{II})
+\varepsilon_{II}(\sigma_I+2\sigma_{II})\}
-(H_{1,\delta}K_{\gamma}\\\nonumber&+H_{2,\delta}L_{\gamma}
+H_1K_{\gamma;\delta}+H_2L_{\gamma;\delta})\epsilon^{\nu\gamma\delta}S_{\nu}
+2(a_{II}H_I-a_IH_2)
=-\frac{\kappa}{6}{(\mu+P)}^{\textrm{eff}}(\sigma_I
+\sigma_{II}-\Theta)
-\frac{\kappa}{9}(2\sigma_I+2\sigma_{II}+\Theta)\bigg(\Pi_I
\\\tag{A24}&+\Pi_{II}\bigg)^{\textrm{eff}}
-\frac{\kappa
A{q_{II}}^{\textrm{eff}}}{2\sqrt{r^2A^2B^2+G^2}}
(\frac{G\dot{C}}{A^2C}+\frac{C_{,\theta}}{C})-\frac{\kappa C'}{2BC}{q_{I}}^{\textrm{eff}}.
\end{align}
Contraction of Eq. (A18) with \textbf{K} and \textbf{L},
reveals the ramification of inhomogeneity of the energy-density in the absence of the concerned scalar variables, hence
\begin{align}\nonumber
&\frac{1}{3}X^{*}_{I,\nu}K^{\nu}-X^{*}_{KL,\nu}L^{\nu}-\frac{1}{3}(2X^{*}_I
+X^{*}_{II})(K^{\nu}_{;\nu}-a_{\beta}K^{\beta})
-\frac{1}{3}(2X^{*}_I+X^{*}_{II})(L_{\mu;\nu}L^{\nu}K^{\mu})
-X^{*}_{KL}(L_{\mu;\nu}K^{\nu}K^{\mu}
+L^{\nu}_{;\nu}-a_{\nu}L^{\nu})
\\\tag{A25}&-\frac{H_2}{3}(\sigma_I+2\sigma_{II})-3\Omega H_1
=\frac{\kappa}{3f_R}(1+f_T)\tilde{\mu}_{;\nu}K^{\nu}
+\frac{3\kappa}{2}\Omega{q_{II}}^{\textrm{eff}}
+\frac{\kappa}{6f_R}(\sigma_{I}-2\Theta){q_{I}}^{\textrm{eff}},\\\nonumber
&\frac{1}{3}X^{*}_{II,\nu}L^{\nu}-X^{*}_{KL,\nu}K^{\nu}-\frac{1}{3}(X^{*}_I
+2X^{*}_{II})(L^{\nu}_{;\nu}-a_{\beta}L^{\beta})
-\frac{1}{3}(2X^{*}_I+X^{*}_{II})(K_{\mu;\nu}K^{\nu}L^{\mu})
-X^{*}_{KL}(K_{\mu;\nu}L^{\nu}L^{\mu}+K^{\nu}_{;\nu}-a_{\nu}K^{\nu})
\\\tag{A26}&-\frac{H_1}{3}(2\sigma_I+\sigma_{II})-3\Omega H_2
=\frac{\kappa}{3f_R}(1+f_T)\tilde{\mu}_{;\nu}L^{\nu}
-\frac{3\kappa}{2}\Omega{q_{I}}^{\textrm{eff}}
+\frac{\kappa}{6f_R}(\sigma_{II}-2\Theta){q_{II}}^{\textrm{eff}}.
\end{align}
Also by contracting Eq. (A19) with \textbf{S}, leads
\begin{align}\nonumber
&\frac{1}{3}X^{*}_{KL}(\sigma_I-\sigma_{II})+(a_IH_1+a_{II}H_2)
-H_{1,\alpha}K^{\alpha}-H_{2,\alpha}L^{\alpha}
-H_1(K^{\alpha}_{;\alpha}+K^{\beta}_{;\alpha}S^{\alpha}S_{\beta})
-H_2(L^{\alpha}_{;\alpha}+L^{\beta}_{;\alpha}S^{\alpha}S_{\beta})
\\\nonumber&=\bigg[\kappa\bigg\{(\mu+P)^{\textrm{eff}}
-\frac{1}{3f_R}(1+f_T)(\Pi_I+\Pi_{II})\bigg\}
-(Y^{*}_I+Y^{*}_{II})\bigg]\Omega-(1+f_T)
\frac{\kappa
A}{2Bf_R\sqrt{r^2A^2B^2+G^2}}\bigg[(q_IB)_{,\theta}-
\\\tag{A27}&(\frac
{q_{II}\sqrt{r^2A^2B^2+G^2}}{A})'\bigg].
\end{align}
However, contracting Eq. (A20) by \textbf{SK} and \textbf{SL},
the leading equations are disclosing the impact of scalar variables
together with the effective components of physical variables.
\begin{align}\nonumber
&-\frac{2a_{II}}{3}\varepsilon_I+2a_I\varepsilon_{KL}-E^{\alpha}_{2;\alpha}L^2
-\frac{AY^{*}_{I,\theta}}{3\sqrt{r^2A^2B^2+G^2}}+\frac{(Y^{*}_{KL})'}{B}
-\bigg[\frac{1}{3}(2Y^{*}_I+Y^{*}_{II})K_{\nu;\alpha}
+\frac{1}{3}(Y^{*}_I
+2Y^{*}_{II})L_{\nu;\alpha}K^{\beta}L_{\nu}
\\\nonumber&+Y^{*}_{KL}(L_{\beta;\alpha}K^{\beta}K_{\nu}
+L_{\nu;\alpha})\bigg]\epsilon^{\gamma\alpha\nu}S_{\gamma}
+H_{1,\alpha}V^{\alpha}-\frac{1}{3}H_1(\sigma_I-\sigma_{II}-\Theta)
+\Omega
H_2=-\frac{\kappa}{6f_R}(1+f_T)\mu_{\theta}L^{2}
+\frac{3\kappa}{2}\Omega{q_{I}}^{\textrm{eff}}
\\\tag{A28}&+\frac{\kappa}{6f_R}(\sigma_{I}
+\Theta){q_{II}}^{\textrm{eff}},\\\nonumber
&\frac{2a_{I}}{3}\varepsilon_{II}-2a_I\varepsilon_{KL}
-E^{\alpha}_{\nu;\alpha}K^{\nu}
+\frac{(Y^{*}_{II})'}{3B}-\frac{AY^{*}_{KL,\theta}}{\sqrt{r^2A^2B^2+G^2}}
-\bigg[\frac{1}{3}(Y^{*}_I+2Y^{*}_{II})L_{\nu;\alpha}
-\frac{1}{3}(2Y^{*}_I+Y^{*}_{II})L_{\beta;\alpha}
K^{\beta}K_{\nu}-Y^{*}_{KL}
\\\nonumber&\times\bigg(L_{\beta;\alpha}K^{\beta}K_{\nu}-K_{\nu;\alpha}
\bigg)\bigg]\epsilon^{\gamma\alpha\nu}S_{\gamma}
+H_{2,\alpha}V^{\alpha}-\frac{1}{3}H_2(\sigma_{II}-\sigma_{I}-3\Theta)
-\Omega H_2=\frac{\kappa}{6f_R}(1+f_T)\mu_{\nu}L^{\nu}
+\frac{3\kappa}{2}\Omega{q_{II}}^{\textrm{eff}}
\\\tag{A29}&-\frac{\kappa}{6f_R}(\sigma_{II}+\Theta){q_{I}}^{\textrm{eff}}.
\end{align}
Now, by contracting Eq. (A14) with \textbf{KS} and \textbf{LS},
we get, accordingly
\begin{align}\tag{A30}
H_1=-\frac{1}{2}(K^{\alpha}S_{\beta}+K_{\beta}S^{\alpha})(\sigma_{\alpha\gamma}
+\Omega_{\alpha\gamma})_{;\delta}\epsilon^{\beta\delta\gamma}
-a_{I}\Omega,\\\tag{A31}
H_2=-\frac{1}{2}(L^{\alpha}S_{\beta}+L_{\beta}S^{\alpha})(\sigma_{\alpha\gamma}
+\Omega_{\alpha\gamma})_{;\delta}\epsilon^{\beta\delta\gamma}-a_{II}\Omega.
\end{align}

\section*{Appendix B}
\subsection*{$f(R,T^2)$ Field and Dynamical Equations for Axial Self-gravitating System}
The $f(R,T^2)$ field equations for our relativistic object are
calculated by employing Eqs. \eqref{1}, \eqref{2} and \eqref{2p3}
\begin{align}\tag{B1}
&G_{00}=\frac{\kappa
A^2}{f_R}\bigg[\mu-\frac{1}{2}(f-Rf_R)-\bigg\{2\mu^2-7(-\mu+3P)^2\bigg\}f_{T^2}
+\frac{\chi_{0}}{A^2}\bigg];~~~
G_{01}=-AB\frac{\kappa}{f_R}\bigg[q_{I}(1+2\mu
f_{T^2})-\frac{\chi_1}{AB}\bigg],\\\nonumber
&G_{02}=\frac{\kappa}{f_R}\bigg[-\mu G-\sqrt{A^2r^2B^2+G^2}q_{II}
+\frac{G}{2}(f-Rf_R)+f_{T^2}\bigg\{2\mu^2 G +q_{II}\mu
\sqrt{A^2r^2B^2+G^2}
+7G(-\mu+3P)^2\bigg\}\\\tag{B2}&+\chi_2\bigg];~~~~~
G_{12}=\frac{\kappa B}{Af_R}\bigg[\bigg\{\Pi_{KL}\sqrt{A^2r^2B^2+G^2}
+Gq_{I}\bigg\}(1-2\mu f_{T^2})+\chi_3\bigg],\\\tag{B3}
&G_{11}=\frac{\kappa B^2}{f_R}\bigg[(1-2\mu
f_{T^2})\bigg(P+\frac{\Pi_{I}}{3}\bigg)-7f_{T^2}(-\mu+3P)^2
+\frac{1}{2}(f-Rf_R)+\frac{\chi_4}{B^2}\bigg],\\\nonumber
&G_{22}=\frac{\kappa}{A^2f_R}\bigg[\bigg\{\mu
G^2+(A^2r^2B^2+G^2)\bigg(P+\frac{\Pi_{II}}{3}\bigg)
+2Gq_{II}\sqrt{A^2r^2B^2+G^2}\bigg\}(1-2\mu
f_{T^2})
-7\frac{r^2B^2}{A^2}f_{T^2}(-\mu+3P)^2\\\tag{B4}&+\chi_5\bigg];~~~
G_{33}=\frac{\kappa C^2}{f_R}\bigg[(1-2\mu
f_{T^2})\bigg(P-\frac{1}{3}(\Pi_{I}+\Pi_{II})\bigg)+\frac{1}{2}(f-Rf_R)
-7f_{T^2}(-\mu+3P)^2
+\chi_6\bigg].
\end{align}
The values of $\chi_1-\chi_6$ are computed same as to their values
presented in Appendix A. In this case,
the energy conservation law
$(T^{\omega}_{\lambda;\omega})^{\textrm{eff}}\neq0$, leads to the
continuity equation in the perspective of $f(R,T^2)$ gravity as
\begin{align}\nonumber
&\frac{1}{f_R}\bigg[V^{\lambda}\mu_{;\lambda}
+(1+2\mu f_{T^2})\bigg\{\Theta(\mu+P)+\frac{\Pi_I}{9}(2\sigma_I
+\sigma_{II})+\frac{\Pi_{II}}{9}(\sigma_I+2\sigma_{II})+
q^{\lambda}_{;\lambda}+q^{\lambda}a_{\lambda}\bigg\}\bigg]
=\frac{1}{f_R}(f-Rf_R)
\\\tag{B5}&\bigg[\frac{G^2}{r^2A^2B^2+G^2}\bigg(\frac{\dot{A}}{A}
+\frac{\dot{B}}{B}+\frac{\dot{G}}{G}
\bigg)\bigg]+\frac{7G}{f_R(r^2A^2B^2+G^2)}(-\mu+3P)^2f_{T^2}
+\frac{\chi_7}{f_R},
\end{align}
and the generalized Euler's equation for our relativistic system is
\begin{align}\nonumber
&\frac{1}{f_R}\bigg[(1+2\mu f_{T^2})\bigg\{a_{\omega}(\mu+P)
+h^{\gamma}_{\omega}(P_{;\gamma}
+\Pi^{\lambda}_{\gamma;\lambda}+q_{\gamma;\lambda}V^{\lambda})+
q^{\gamma}(\sigma_{\omega\gamma}+\Omega_{\omega\gamma}
+\frac{4}{3}h_{\omega\gamma}\Theta)\bigg\}\bigg]
=\bigg(\frac{-1}{2f_R}(f-Rf_R)\bigg)'
\\\tag{B6}&-\frac{1}{2f_R}(f-Rf_R)-\frac{A}{B}\bigg(q_I \mu f'_{T^2}
+\frac{\mu}{rB}q_{II}f_{T^2_{,\theta}}\bigg)
+\frac{7A}{\sqrt{r^2A^2B^2+G^2}}
(-\mu+3P)^2(f_{T^2_{,\theta}}-f_{T^2})
-\frac{1}{f_R}\bigg(\chi_8+\chi_9\bigg),
\end{align}
along with
\begin{align}\nonumber
&\chi_7=\frac{\chi_0}{f_R(r^2A^2B^2+G^2)}\bigg[(-2r^2B^2)\frac{\dot{A}}{A}
+\frac{r^2A^2B^2+G^2}{A^2}
\bigg(\frac{\dot{3B}}{B}-\frac{\dot{C}}{C}-\frac{\dot{f_R}}{f_R}\bigg)
-2\frac{G^2}{A^2}\frac{\dot{G}}{G}
-G\frac{A_{,\theta}}{A}
+\frac{r^2A^2B^2+G^2}{r^2A^2B^2}\bigg(\frac{G_{,\theta}}{G}
\\\nonumber&-\frac{C_{,\theta}}{C}\bigg)G\bigg]
-\frac{1}{f_R}\bigg[\frac{\dot{\chi_0}}{A^2}+\frac{G}{r^2A^2B^2}
\chi_{0,\theta}\bigg]
+\frac{1}{B^2f_R}\bigg[\chi'_1
+\chi_1\{\frac{f_R'}{R}+\frac{A'}{A}+\frac{(Br)'}{Br}
+\frac{C'}{C}\}-\frac{\dot{B}}{B}\chi_4\bigg]
+\frac{1}{rBf_R}\chi_2\bigg[2\bigg(\frac{A_{,\theta}}{A}
+\frac{B_{,\theta}}{B}\bigg)
\\\nonumber&+\frac{C_{,\theta}}{C}
-\frac{Gr^2B^2}{r^2A^2B^2+G^2}\bigg\{\frac{\dot{A}}{A}
+\frac{\dot{B}}{B}-\frac{\dot{G}}{G}
+\frac{G^2}{r^2B^2}(\frac{G_{,\theta}}{G})\bigg\}\bigg]
+\frac{A}{r^2Bf_R}\bigg[\chi_{2,\theta}
+\frac{G}{B^2}\frac{A'}{A}\chi_3
+\frac{G^2}{r^2A^2B^2+G^2}
\{\frac{\dot{A}}{A}-\frac{\dot{G}}{G}-\frac{A^2}{G}
+\\\nonumber&\frac{A_{,\theta}}{A}\}
\chi_5\bigg]+\mu\frac{\dot{f_R}}{f_R^2}-\frac{1}{f_R}(\dot{R}f_R-R\dot{f_R})
-\frac{A}{B}q_I\mu f'_{T^2}
-\frac{A^2}{\sqrt{r^2A^2B^2+G^2}}\bigg(q_{II}\mu+7(-\mu+3P)^2\bigg)
f_{T^2}+\frac{G(1-2\mu f_{T^2})}{f_R\sqrt{r^2A^2B^2+G^2}}\Pi_{KL}
\\\nonumber&\bigg[\frac{A}{B}\bigg(\frac{A'}{A}
-\frac{G'}{2G}\bigg)\bigg],\\\nonumber
&\chi_8=\frac{B}{A}q_I\mu\{\dot{f_{T^2}}-(1-2\mu
f_{T^2})\frac{\dot{f_R}}{f_R}\}+\mu\dot{f_{T^2}}\bigg(P+\frac{\Pi_I}{3}\bigg)
+\frac{\sqrt{r^2A^2B^2+G^2}}{r^2B^2}f_{T,\theta}\Pi_{KL}+\frac{1}{B}\bigg(\chi'_4
-\frac{f'_R}{f_R}\chi_4\bigg)
+\frac{r^2A^2B^2}{r^2A^2B^2+G^2}\\\nonumber&\bigg[\chi_0\bigg(\frac{A'}{A}
+\frac{G^2}{2r^2B^2}\frac{G'}{G}\bigg)
+\frac{\chi_5}{C^2}\bigg(\frac{A'}{A}+\frac{(Br)'}{Br}\bigg)\bigg]
+\frac{\chi_2}{r}\frac{r^2A^2B^2}{r^2A^2B^2+G^2}\bigg[\frac{(Br)'}{Br}
+\frac{G^2}{2r^2B^2}\frac{G'}{G}\bigg]+\frac{1}{r^2B^2}\mu
f_{{T^2},\theta}\Pi_{KL}\sqrt{r^2A^2B^2+G^2}\\\nonumber&-\frac{A}{B}q_I\mu
f'_{T^2}-7\frac{A}{\sqrt{r^2A^2B^2+G^2}}\mu f_{T^2},\\\nonumber
&\chi_9=\frac{\mu}{r^2B^2}f'_{T^2}-\frac{f'_R}{f_R}(1-2\mu
f_{T^2})\bigg(P+\frac{\Pi_I}{3}\bigg)
+\frac{\mu}{A^3r^3rB^2}f_{T^2}+\frac{1}{r^2B^2}
\frac{f_{R,\theta}}{f_R}\bigg\{\frac{G}{r^2}q_I\mu f_{T^2}-\frac{(1-2\mu
f_{T^2})}{\sqrt{r^2A^2B^2+G^2}}\Pi_{KL}\bigg\}
\\\nonumber&+\frac{r^2A^2B^2}{r^2A^2B^2+G^2}\bigg[\chi_1\bigg(\frac{\dot{A}}{A}
+\frac{\dot{2B}}{B}
+\frac{G^2}{2r^2B^2}\frac{\dot{G}}{G}
+\frac{1}{rB}\bigg(\frac{A_{,\theta}}{A}-\frac{B_{,\theta}}{2B}\bigg)
+\frac{G^2}{r^2B^2C}\frac{G_{,\theta}}{G}\bigg)\bigg]
+\frac{1}{r^2B}\chi_3\bigg[\frac{B_{,\theta}}{B}
+\frac{r^2A^2B^2}{r^2A^2B^2+G^2}\bigg\{
\frac{A_{,\theta}}{A}\\\nonumber&+\frac{G^2}{r^2B^2}\frac{G_{,\theta}}{G}
+\frac{C_{,\theta}}{C}\bigg\}\bigg]+7\frac{A}{\sqrt{r^2A^2B^2+G^2}}(-\mu+3P)^2
f_{T^2}-\frac{1}{rB^2}q_{II,\theta}\mu f_{T^2}.
\end{align}
The extra terms $\zeta_i's$ appearing in Eqs. \eqref{2p5}, \eqref{2p8},
\eqref{2p9} and \eqref{2p11} are
\begin{align}\nonumber
&\psi_1=\frac{\kappa}{8f_R}\epsilon_{\omega}^{\epsilon \varrho}
\bigg[(\nabla^{\pi}\nabla_{\epsilon}f_R) \epsilon_{\lambda\pi
\varrho}-(\nabla^{\pi}\nabla_{\varrho}f_R)\epsilon_{\lambda\pi\epsilon}
-(\nabla^{\theta}\nabla_{\epsilon}f_R)\epsilon_{\theta\lambda\varrho}
+(\nabla^{\theta}\nabla_{\varrho}f_R)
\epsilon_{\theta\lambda\epsilon}\bigg],\\\nonumber
&\psi_2=\nabla_{\omega}\nabla_{\lambda}f_R-\frac{3}{2}V_{\omega}V_{\lambda}
(f-Rf_R)-\nabla_{\omega}\nabla_{\varrho}f_RV^{\varrho}V_{\lambda}
+2V_{\omega}V_{\lambda}\Box
f_R-(\nabla^{\theta}\nabla_{\lambda}f_R)V_{\omega}V^{\theta}
+g_{\omega\lambda}
\\\nonumber&-(\nabla^{\theta}\nabla_{\varrho}f_R)V_{\theta}V^{\varrho},\\\nonumber
&\psi_3=\nabla^{\omega}\nabla_{\omega}f_R+\frac{3}{2}(f-Rf_R)-(\nabla^{\omega}\nabla_{\varrho}f_R)V^{\varrho}V_{\omega}
-2\Box f_R-(\nabla^{\theta}\nabla_{\omega}f_R)V^{\omega}V_{\theta}
+4(\nabla^{\theta}\nabla_{\varrho}f_R)V_{\theta}V^{\varrho},\\\nonumber
&\psi_4=(\nabla^{\varrho}\nabla_{\gamma}f_R)V^{\gamma}\epsilon_{\omega
\varrho\lambda}.
\end{align}
The effective terms of matter contents emerging in modified field as
well as scalar equations are
\begin{align}\nonumber
&\mu^{\textrm{eff}}=\frac{1}{f_R}\bigg[\mu-\frac{1}{2}(f-Rf_R)
-f_{T^2}\bigg(2\mu^2-7(-\mu+3P)^2\bigg)+\frac{\chi_0}{A^2}\bigg],\\\nonumber
&\Pi^{\textrm{eff}}_{KL}=\frac{1}{f_R}(1-2\mu
f_{T^2})\Pi_{KL}+\frac{1}{ABf_R\sqrt{r^2A^2B^2+G^2}}\bigg[
A^2\chi_3-G\chi_1\bigg],\\\nonumber
&(P+\frac{\Pi_I}{3})^{\textrm{eff}}=\frac{1}{f_R}\bigg[(1-2\mu
f_{T^2})\bigg(P+\frac{\Pi_I}{3}\bigg)-7f_{T^2}(-\mu+3P)^2\bigg],\\\nonumber
&(P+\frac{\Pi_{II}}{3})^{\textrm{eff}}
=\frac{1}{f_R}\bigg[\frac{G^2}{r^2A^2B^2+G^2}\bigg\{
-\frac{1}{2}(f-Rf_R)+7f_{T^2}(-\mu+3P)^2\bigg\}+(1-2\mu
f_{T^2})(P+\frac{\Pi_{II}}{3})\bigg]-\frac{1}{f_R(r^2A^2B^2+G^2)}
\\\nonumber
&\bigg[-3\frac{G^2}{A^2}\chi_0-2G\chi_2
+A^2\chi_5\bigg].
\end{align}

\subsection*{Scalar Equations and Some Ricci-Identities in $f(R,T^2)$ Scenario}
In this case, the differential equations for the proposed system read the forms given below.
\begin{align}\nonumber
&h^{\rho}_{(\omega}h^{\beta}_{\lambda)}E_{\rho\beta;\alpha}V^{\alpha}
+E_{\omega\lambda}\Theta
+E_{\rho\beta}h_{\omega\lambda}\sigma^{\rho\beta}
+h^{\rho}_{(\omega}\eta^{\alpha\gamma\delta}_{\lambda)}V_{\alpha}H_{\gamma\rho;\delta}
-3E_{\rho(\omega}\sigma^{\rho}_{\lambda)}-E_{\alpha(\omega}\Omega^{\alpha}_{\lambda)}
-2H^{\rho}_{(\omega}\eta_{\lambda)\alpha\delta\rho}V^{\alpha}a^{\delta}
=\frac{\kappa}{2}{(\mu+P)}^{\textrm{eff}}\\\nonumber&\sigma_{\omega\lambda}
-\frac{\kappa}{6f_R}(1-2\mu f_{T^2})\Theta\Pi_{\omega\lambda}
-\frac{\kappa}{2f_R}h^{\rho}_{(\omega}h^{\beta}_{\lambda)}(1-2\mu f_{T^2})
\Pi_{\rho\beta;\alpha}V^{\alpha}-\frac{\kappa}{2}
\Omega^{\rho}_{(\omega}{\Pi_{\lambda)\rho}}^{\textrm{eff}}
-\kappa a_{(\omega}{q_{\lambda)}}^{\textrm{eff}}
+\frac{\kappa}{6f_R}(1-2\mu f_{T^2})
\bigg(\Pi_{\rho\beta}\sigma^{\rho\beta}\\\tag{B7}&+a_{\rho}{q^{\rho}}^{\textrm{eff}}+
\frac{1}{f_R}(1-2\mu f_{T^2})q^{\rho}_{;\rho}\bigg)h_{\omega\lambda}
-\frac{\kappa}{2f_R}h^{\rho}_{(\omega}
h^{\beta}_{\lambda)}(1-2\mu f_{T^2}) q_{\beta;\rho},\\\nonumber
&h^{\rho}_{\omega}h^{\beta\lambda}E_{\rho\beta;\lambda}-{\eta_{\omega}}^{\alpha\beta\delta}V_{\alpha}
\sigma^{\gamma}_{\beta}H_{\delta\gamma}+3H_{\omega\lambda}\hat{\omega}^{\lambda}=
\frac{\kappa}{3f_R}(1-2\mu f_{T^2})\mu_{;\nu}h^{\nu}_{\omega}
-\frac{\kappa}{2f_R}(1-2\mu f_{T^2})h^{\nu}_{\omega}h^{\rho\beta}\Pi_{\nu\beta;\rho}
-\frac{\kappa}{2}
\bigg(\frac{2}{3}h^{\nu}_{\omega}\Theta-\sigma^{\nu}_{\omega}
\\\tag{B8}&+3\Omega^{\nu}_{\omega}\bigg){q_{\nu}}^{\textrm{eff}},\\\nonumber
&(\sigma_{\omega\alpha}E^{\alpha}_{\lambda}
+3\Omega_{\omega\alpha}E^{\alpha}_{\lambda})
{\epsilon_{\delta}}^{\omega\lambda}+a^{\beta}H_{\beta\delta}
-H^{\beta\alpha}_{~~;\alpha}h_{\beta\delta}=
\frac{\kappa}{2}{(\mu+P)}^{\textrm{eff}}\Omega_{\omega\lambda}{\epsilon_{\delta}}^{\omega\lambda}+
\frac{\kappa}{2f_R}(1-2\mu f_{T^2})\bigg[q_{\omega;\lambda}
+\Pi_{\beta\omega}(\sigma^{\beta}_{\lambda}
\\\tag{B9}&+\Omega^{\beta}_{\lambda})\bigg]{\epsilon_{\delta}}^{\omega\lambda},\\\nonumber
&2a_{\nu}E_{\rho\delta}{\epsilon_{\gamma}}^{\rho\nu}-E_{\beta\nu;\alpha}
h^{\beta}_{\delta}{\epsilon_{\gamma}}^{\alpha\nu}
+E^{\alpha}_{\nu;\alpha}{\epsilon_{\gamma\delta}}^{\nu}+\frac{2}{3}H_{\delta\gamma}\Theta
+H^{\mu}_{\beta;\alpha}V^{\alpha}h^{\beta}_{\delta}h_{\mu\gamma}
-H^{\alpha}_{\gamma}(\sigma_{\delta\alpha}+\Omega_{\delta\alpha})+
(\sigma_{\nu\alpha}+\Omega_{\nu\alpha})
H^{\mu}_{\rho}{\epsilon_{\delta\mu}}^{\alpha}{\epsilon_{\gamma}}^{\rho\nu}
\\\nonumber&+\frac{\Theta}{3}H^{\mu}_{\rho}{\epsilon_{\delta\mu}}^{\alpha}{\epsilon_{\gamma\alpha}}^{\rho}
=\frac{\kappa}{6f_R}(1-2\mu f_{T^2})\mu_{,\nu}{\epsilon_{\gamma\delta}}^{\nu}
+\frac{\kappa}{2f_R}(1-2\mu f_{T^2})
\Pi_{\rho\beta;\nu}h^{\beta}_{\delta}{\epsilon_{\gamma}}^{\rho\nu}
+\frac{\kappa}{2}\bigg[\Omega_{\rho\nu}{q_{\delta}}^{\textrm{eff}}
+{q_{\rho}}^{\textrm{eff}}
(\sigma_{\delta\nu}+\Omega_{\delta\nu}
\\\tag{B10}&+\frac{\Theta}{3}h_{\delta\nu})\bigg]{\epsilon_{\gamma}}^{\rho\nu}
\end{align}
Some Ricci-Identities in this case are
\begin{align}\nonumber
&\frac{2}{3B}\Theta'-\Omega\bigg(L^{\lambda}_{;\lambda}
-L_{\gamma;\lambda}K^{\lambda}K^{\gamma}\bigg)
-\Omega_{;\lambda}L^{\lambda}+\frac{a_I}{3}\sigma_I-\frac{1}{3}
\sigma_{I;\lambda}K^{\lambda}-a_{II}\Omega
-\frac{1}{3}(K^{\lambda}_{;\lambda}-\frac{1}{3}a_I)(\sigma_{II}+2\sigma_I)
-\frac{1}{3}(\sigma_{I}+2\sigma_{II})
\\\tag{B11}&\bigg(L_{\gamma;\lambda}
L^{\lambda}K^{\gamma}-\frac{1}{3}a_I\bigg)=\kappa
{q_I}^{\textrm{eff}},\\\nonumber
&\frac{2}{3\sqrt{r^2A^2B^2+G^2}}\bigg(A\Theta_{,\theta}
+\frac{G}{A}\dot{\Theta}\bigg)
+\Omega_{;\lambda}K^{\lambda}+\frac{a_{II}}{3}\sigma_{II}
+\Omega\bigg(L_{\gamma;\lambda}L^{\lambda}K^{\gamma}
+K^{\lambda}_{;\lambda}\bigg)
+a_{I}\Omega-\frac{1}{3}\sigma_{II;\lambda}L^{\lambda}
+\frac{1}{3}a_I(\sigma_{II}+2\sigma_I)
\\\tag{B12}&\bigg(L_{\gamma;\lambda}K^{\lambda}K^{\gamma}
+\frac{1}{3}a_{II}\bigg)
-\frac{1}{3}(\sigma_{I}+2\sigma_{II})\bigg(L^{\lambda}_{;\lambda}
-\frac{1}{3}a_{II}\bigg)=\kappa {q_{II}}^{\textrm{eff}}.
\end{align}
Furthermore, on the contraction of Eq. (B7) by tetrad vectors
\textbf{KK}, \textbf{KL}, \textbf{LL} and \textbf{SS}, we obtain
nice outcomes to study the evolution of the modified scalar
variables of the tensors, i.e., $X_{\omega\lambda}$ and
$Z_{\omega\lambda}$. Therefore,
\begin{align}\nonumber
&-\frac{1}{3}\bigg(X_I-\frac{\kappa}{2f_R}(1+f_T^2)\mu\bigg)_{,\lambda}V^{\lambda}
-\frac{\varepsilon_I}{9}(\sigma_I-\sigma_{II}-3\Theta)
+\frac{\varepsilon_{II}}{9}(\sigma_I+\sigma_{II})
-K_{\beta}[H_{1,\delta}S_{\alpha}
+H_1S_{\alpha;\delta}
+H_2\bigg(L_{\lambda;\delta}S_{\gamma}K^{\lambda}
\\\nonumber&+S_{\lambda;\delta}L_{\gamma}K^{\lambda}\bigg)]+\Omega X_{KL}=2a_{II}H_1
-\frac{\kappa}{6}(P+\mu+\frac{\Pi_I}{3})^{\textrm{eff}}(\Theta+\sigma_I)
-\kappa
a_I{q_I}^{\textrm{eff}}
-\frac{\kappa}{2B}({q_I}^{\textrm{eff}})'
-\frac{\kappa A{q_{II}}^{\textrm{eff}}}{2\sqrt{r^2A^2B^2+G^2}}
\bigg(\frac{G\dot{B}}{A^2B}\\\tag{B13}&+\frac{B_{,\theta}}{B}\bigg),\\\nonumber
&-X_{KL,\gamma}V^{\gamma}
-\frac{\Omega}{6}(X_I-X_{II})+\frac{X_{KL}}{2}
(\sigma_I+\sigma_{II}-2\Theta)-(a_{II}H_2-a_IH_1)
-\frac{1}{2}\bigg\{\bigg(H_{1,\delta}S_{\alpha}+H_1(S_{\alpha;\delta}
+S_{\lambda;\delta}K_{\gamma}K^{\lambda})
\\\nonumber&+H_2S_{\gamma}L_{\lambda;\delta}K^{\lambda}\bigg)
\epsilon^{\nu\gamma\delta}\bigg\}
+\frac{1}{2}\bigg[\bigg(H_1K^{\mu}S_{\gamma}
+H_2S^{\mu}L_{\gamma}\bigg)L_{\mu;\delta}\epsilon^{\nu\gamma\delta}K_{\nu}\bigg]
-\frac{1}{2}(H_2S_{\gamma;\delta}
+H_{2;\delta}S_{\gamma})
\epsilon^{\nu\gamma\delta}K_{\nu}
=-\frac{\kappa}{3}{\Pi_{KL}}^{\textrm{eff}}\bigg(\sigma_I
\\\tag{B14}&+\sigma_{II}-\Theta\bigg)
-\frac{\kappa}{2}(a_{II}{q_I}^{\textrm{eff}}+a_I{q_{II}}^{\textrm{eff}})
-\frac{\kappa}{4f_R}(K^{\beta}L^{\mu}+L^{\beta}K^{\mu})q_{\beta;\mu},\\\nonumber
&\frac{1}{3}\bigg(-X_{II}+\frac{\kappa}{2f_R}(1-2\mu f_{T^2})\mu\bigg)_{;\alpha}V^{\alpha}
-\frac{\varepsilon_{II}}{9}(\sigma_{II}-\sigma_{I}-3\Theta)
+\frac{\varepsilon_{I}}{9}(2\sigma_I+\sigma_{II})-\Omega X_{KL}
-\bigg[H_{2;\delta}S_{\gamma}
-H_1(S_{\gamma}L_{\mu;\delta}K^{\mu}
\\\tag{B15}&+L_{\lambda;\delta}S^{\lambda}K_{\gamma})
+H_2S_{\gamma;\delta}\bigg]\epsilon^{\nu\gamma\delta}L_{\nu}
+2H_2a_I=-\frac{\kappa}{6}(\Theta+\sigma_{II})(P+\mu+\frac{\Pi_{II}}{3})^{\textrm{eff}}
-\frac{\kappa}{2f_R}(1-2\mu f_{T^2})L^{\lambda}L_{\beta}q^{\beta}_{;\lambda}-\kappa
a_{II}{q_{II}}^{\textrm{eff}},\\\nonumber
&\frac{1}{3}\bigg(X_I+X_{II}+\frac{\kappa}{2f_R}(1-2\mu f_{T^2})\mu\bigg)_{;\gamma}V^{\gamma}
+\frac{1}{3}(X_I+X_{II})(\sigma_{II}+\sigma_{I}+\Theta)
+\frac{1}{9}\bigg\{\varepsilon_I(2\sigma_I+\sigma_{II})
+\varepsilon_{II}(\sigma_I+2\sigma_{II})\bigg\}
\\\nonumber&-\bigg(H_{1,\delta}K_{\alpha}+H_{2,\delta}L_{\alpha}
+H_1K_{\alpha;\delta}+H_2L_{\alpha;\delta}\bigg)\epsilon^{\nu\alpha\delta}S_{\nu}
+2(a_{II}H_I-a_IH_2)
=-\frac{\kappa}{6}{(\mu+P)}^{\textrm{eff}}(\sigma_I
+\sigma_{II}-\Theta)
-\frac{\kappa}{9}\bigg(2\sigma_I
\\\tag{B16}&+2\sigma_{II}+\Theta\bigg)(\Pi_I+\Pi_{II})^{\textrm{eff}}
-\frac{\kappa
A{q_{II}}^{\textrm{eff}}}{2\sqrt{r^2A^2B^2+G^2}}
\bigg(\frac{G\dot{C}}{A^2C}
+\frac{C_{,\theta}}{C}\bigg)
-\frac{\kappa C'}{2BC}{q_{I}}^{\textrm{eff}}.
\end{align}
Now, contracting Eq. (B8) with \textbf{K} and \textbf{L}, we
can see the effects of inhomogeneity energy density in the exclusion
of the relevant scalar variables.
\begin{align}\nonumber
&\frac{1}{3}X_{I,\omega}K^{\omega}-X_{KL,\omega}L^{\omega}-\frac{1}{3}(2X_I
+X_{II})(K^{\omega}_{;\omega}-a_{\beta}K^{\beta})
-\frac{1}{3}(2X_I+X_{II})(L_{\lambda;\omega}L^{\omega}K^{\lambda})
-X_{KL}(L_{\lambda;\omega}K^{\omega}K^{\lambda}
+L^{\omega}_{;\lambda}-a_{\lambda}L^{\lambda})
\\\tag{B17}&-\frac{H_2}{3}(\sigma_I+2\sigma_{II})-3\Omega H_1
=\frac{\kappa}{3f_R}\bigg(1-2\mu f_{T^2}\bigg)\mu_{;\lambda}K^{\lambda}
+\frac{3\kappa}{2}\Omega{q_{II}}^{\textrm{eff}}
+\frac{\kappa}{6f_R}(\sigma_{I}-2\Theta){q_{I}}^{\textrm{eff}},\\\nonumber
&\frac{1}{3}X_{II,\lambda}L^{\lambda}-X_{KL,\lambda}K^{\lambda}
-\frac{1}{3}(X_I+2X_{II})(L^{\omega}_{;\omega}-a_{\beta}L^{\beta})
-\frac{1}{3}(2X_I
+X_{II})(K_{\omega;\lambda}K^{\omega}L^{\lambda})
-X_{KL}(K_{\omega;\lambda}L^{\lambda}L^{\omega}
+K^{\lambda}_{;\lambda}-a_{\lambda}K^{\lambda})
\\\tag{B18}&-\frac{H_1}{3}(2\sigma_I+\sigma_{II})-3\Omega H_2
=\frac{\kappa}{3f_R}(1-2\mu f_{T^2})\mu_{;\lambda}L^{\lambda}
-\frac{3\kappa}{2}\Omega{q_{I}}^{\textrm{eff}}
+\frac{\kappa}{6f_R}(\sigma_{II}-2\Theta){q_{II}}^{\textrm{eff}}.
\end{align}
Moreover by contracting Eq. (B9) with \textbf{S}, we obtain
\begin{align}\nonumber
&\frac{1}{3}X_{KL}(\sigma_I-\sigma_{II})+(a_IH_1+a_{II}H_2)
-H_{1,\gamma}K^{\gamma}-H_{2,\gamma}L^{\gamma}
-H_1(K^{\gamma}_{;\gamma}+K^{\beta}_{;\gamma}S^{\gamma}S_{\beta})
-H_2(L^{\gamma}_{;\gamma}
+L^{\beta}_{;\gamma}S^{\gamma}S_{\beta})
=\bigg[\kappa\bigg\{(\mu+P)^{\textrm{eff}}
\\\tag{B19}&-\frac{1}{3f_R}(1-2\mu f_{T^2})(\Pi_I+\Pi_{II})\bigg\}
-(Y_I+Y_{II})\bigg]\Omega(1-2\mu f_{T^2})
\frac{\kappa
A}{2Bf_R\sqrt{A^2r^2B^2+G^2}}\bigg[(q_IB)_{,\theta}-\bigg(\frac
{q_{II}\sqrt{A^2r^2B^2+G^2}}{A}\bigg)'\bigg].
\end{align}
When Eq. (B10) is contracted by \textbf{SK} and \textbf{SL}, the
resulting equations indicate the influence of scalar variables as
well as the effective components of matter variables.
\begin{align}\nonumber
&-\frac{2a_{II}}{3}\varepsilon_I+2a_I\varepsilon_{KL}-E^{\gamma}_{2;\gamma}L^2
-\frac{AY_{I,\theta}}{3\sqrt{r^2A^2B^2+G^2}}+\frac{(Y_{KL})'}{B}
-\bigg[\frac{1}{3}(2Y_I
+Y_{II})K_{\lambda;\alpha}
+\frac{1}{3}(Y_I
+2Y_{II})L_{\lambda;\alpha}K^{\beta}L_{\lambda}
\\\nonumber&+Y_{KL}(L_{\beta;\alpha}K^{\beta}K_{\lambda}
+L_{\lambda;\alpha})\bigg]\epsilon^{\gamma\alpha\lambda}S_{\gamma}
+H_{1,\alpha}V^{\alpha}-\frac{1}{3}H_1(\sigma_I-\sigma_{II}-\Theta)
+\Omega H_2
=-\frac{\kappa}{6f_R}(1-2\mu f_{T^2})\mu_{,\theta}L^{2}
+\frac{3\kappa}{2}\Omega{q_{I}}^{\textrm{eff}}
\\\tag{B20}&+\frac{\kappa}{6f_R}(\sigma_{I}
+\Theta){q_{II}}^{\textrm{eff}},\\\nonumber
&\frac{2a_{I}}{3}\varepsilon_{II}-2a_I\varepsilon_{KL}
-E^{\alpha}_{\lambda;\alpha}K^{\lambda}
+\frac{(Y_{II})'}{3B}-\frac{AY_{KL,\theta}}{\sqrt{r^2A^2B^2+G^2}}
-\bigg[\frac{1}{3}(Y_I+2Y_{II})L_{\lambda;\alpha}
-\frac{1}{3}(2Y_I
+Y_{II})L_{\beta;\alpha}K^{\beta}K_{\lambda}-Y_{KL}
\\\nonumber&\bigg(L_{\beta;\alpha}K^{\beta}K_{\lambda}
-K_{\lambda;\alpha}\bigg)\bigg]\epsilon^{\gamma\alpha\lambda}S_{\gamma}
+H_{2,\alpha}V^{\alpha}-\frac{1}{3}H_2(\sigma_{II}-\sigma_{I}-3\Theta)
-\Omega H_2=\frac{\kappa}{6f_R}(1-2\mu f_{T^2})\mu_{\lambda}L^{\lambda}
+\frac{3\kappa}{2}\Omega{q_{II}}^{\textrm{eff}}
\\\tag{B21}&-\frac{\kappa}{6f_R}(\sigma_{II}+\Theta){q_{I}}^{\textrm{eff}}.
\end{align}

\section*{Appendix C}

\subsection*{Palatini $f(R)$ Field and Dynamical Equations for Axial Self-gravitating System}
For the system \eqref{1},
the non-zero palatini $f(R)$ equations-of-motion \eqref{3p6}
are
\begin{align}\tag{C1}
&\mu^{\textrm{eff}}=\frac{\kappa}{f_R}\bigg\{\mu
+\frac{\mathcal{T}_{00}}{A^2}-
\frac{1}{2\kappa^2}(f-\hat{R}f_R)f_R\bigg\};\quad\quad
q_I^{\textrm{eff}}=\frac{\kappa}{f_R}\bigg\{q_I
-\frac{\mathcal{T}_{01}}{AB}\bigg\},\\\tag{C2}
&q_{II}^{\textrm{eff}}=\frac{\kappa}{f_R}\bigg\{q_{II}
-\frac{1}{\sqrt{A^2r^2B^2+G^2}}
\bigg(\frac{\mathcal{T}_{02}+\mathcal{T}_{00}}{A^2}\bigg)\bigg\},\\\tag{C3}
&(P+\frac{\Pi_I}{3})^{\textrm{eff}}=\frac{\kappa}{f_R}\bigg\{\bigg(P
+\frac{\Pi_I}{3}\bigg)
+\frac{\mathcal{T}_{11}}{B^2}+\frac{1}{2\kappa^2}(f
-\hat{R}f_R)f_R\bigg\},\\\tag{C4}
&\Pi^{\textrm{eff}}_{KL}=\frac{\kappa}{f_R}\bigg\{\Pi_{KL}
+\frac{1}{AB\sqrt{A^2r^2B^2+G^2}}
\bigg(A^2\mathcal{T}_{12}-G\mathcal{T}_{01}\bigg)\bigg\}
,\\\tag{C5}
&(P+\frac{\Pi_{II}}{3})^{\textrm{eff}}=\frac{\kappa}{f_R}\bigg\{\bigg(P
+\frac{\Pi_{II}}{3}\bigg)
-\frac{1}{A^2r^2B^2+G^2}\bigg(A^2\mathcal{T}_{22}-G^2\mathcal{T}_{00}
-\frac{(A^2G^2-A^2R^2B^2)}{2\kappa^2}
(f-\hat{R}f_R)f_R\bigg)\bigg\}.
\end{align}
In the Palatini based background, the energy conservation law leads to the continuity equation
\begin{align}\nonumber
&\frac{1}{f_R}\bigg[V^{\gamma}\mu_{;\gamma}
+\Theta_P(\mu+P)+\frac{1}{9}\bigg\{\Pi_I(2\sigma_I
+\sigma_{II})+\Pi_{II}(\sigma_I+2\sigma_{II})
\bigg\}+q^{\gamma}a_{\gamma}+
q^{\gamma}_{;\gamma}\bigg]
=
\frac{1}{f_R}\bigg[\frac{1}{A^2r^2B^2+G^2}\bigg\{-r^2B^2\mathcal{\dot{T}}_{00}
\\\tag{C6}&+G\mathcal{\dot{T}}_{02}+\mathcal{T}_{12}
+G^2(\mathcal{T}_{00}+\mathcal{T}_{00,\theta})
+\frac{G}{r^2B}\mathcal{T}_{22}\bigg\}
+\frac{B_{,\theta}}{B^2}\mathcal{T'}_{01}\bigg]
-\frac{r^2B^2}{2(A^2r^2B^2+G^2)}(f-\hat{R}f_R)_{,\theta}
+\frac{1}{rBf_R}(f-\hat{R}f_R).
\end{align}
In this case, the generalized Euler equation becomes
\begin{align}\nonumber
&\frac{1}{f_R}\bigg[a_{\gamma}(\mu+P)
+h^{\beta}_{\gamma}\bigg\{P_{;\beta}
+\Pi^{\alpha}_{\beta;\alpha}+q_{\beta;\alpha}V^{\alpha}\bigg\}+
q^{\alpha}\bigg(\sigma_{\gamma\alpha}+\Omega_{\gamma\alpha}
+\frac{4}{3}h_{\gamma\alpha}\Theta_P\bigg)\bigg]
=\frac{1}{2f_R}\bigg[(f-\hat{R}f_R)'+\frac{1}{A^2r^2B^2+G^2}
\\\tag{C7}&\bigg\{GA^2(f-\hat{R}f_R)+rB(f-\hat{R}f_R)_{,\theta}\bigg\}\bigg]+\chi_2
+\chi_3,
\end{align}
where
\begin{align}\nonumber
\chi_2&=\frac{1}{f_R}\bigg[\frac{1}{B^2}\bigg(-2\dot{B}\mathcal{T}_{01}+
\mathcal{\dot{T}}_{01}-2\frac{B'}{B}\mathcal{T}_{11}+\mathcal{T'}_{11}
\bigg)+\frac{1}{A^2r^2B^2+G^2}\bigg(G_{,\theta}\mathcal{T}_{01}
+A_{,\theta}\mathcal{T}_{12}\bigg)\bigg],\\\nonumber
\chi_3&=\frac{1}{\sqrt{A^2r^2B^2+G^2}}\bigg(\dot{A}\mathcal{T}_{02}+
AB\mathcal{\dot{T}}_{02}\bigg)+\frac{1}{A^2r^2B^2+G^2}
\bigg[2\dot{A}\mathcal{T}_{22}+B\mathcal{\dot{T}}_{22}+A\mathcal{T'}_{12}
+\frac{G_{,\theta}}{rB}\mathcal{T}_{02}\bigg].
\end{align}
The values of $\mathcal{T}_{ij}$'s appearing in Eqs. (C1)--(C7) are computed as:
\begin{align}\nonumber
&\mathcal{T}_{00}=\frac{1}{\kappa}\bigg[\ddot{f_R}
-\frac{\dot{f_R}}{{A^2r^2B^2+G^2}}\bigg({A^2r^2B^2\dot{A}}-G\dot{G}-GAA_{,\theta}
+A^2r^2B^2\frac{\dot{f_R}}{2f_R}\bigg)-\frac{\dot{f}^2_R}{f_R}
+\frac{A^2G}{A^2r^2B^2+G^2}\frac{f_{R,\theta}}{2f_R}-A^2\chi_1
\\\nonumber&-\frac{3A^2}{2f_R}\bigg\{
(\nabla f_R)^2-\dot{f}^2_R\bigg\}\bigg];~~~~
\mathcal{T}_{11}=\frac{1}{\kappa}\bigg[f''_R-\frac{\dot{f_R}}{{A^2r^2B^2+G^2}}\bigg\{
B^3R^2\dot{B}-GBB_{,\theta}-B^4r^2\frac{\dot{f}_R}{2f_R}\bigg\}
-f'_R\bigg(\frac{B'}{B}+\frac{f'_R}{2f_R}
\bigg)\\\nonumber&-B^2\chi_1
+\frac{3}{4f_R}\bigg\{B^2\bigg(\frac{r^2B^2}{A^2r^2B^2+G^2}\dot{f}^2_R
+2\frac{f'^2_R}{B^2}-\frac{f'^2_R}{f_R}\bigg)\bigg\}\bigg];~~~~
\mathcal{T}_{22}=\frac{1}{\kappa}\bigg[\frac{B^2r^2}{A^2r^2B^2+G^2}\dot{f_R}
\bigg\{G_{,\theta}-B^2r^2\dot{B}+\frac{B_{,\theta}}{B}-B^2r^2\frac{\dot{f_R}}{f_R}\bigg\}
\\\nonumber&-r^2f'_R\bigg\{\frac{(Br)'}{Br}+\frac{f'_R}{2f_R}\bigg\}
-\frac{r^2B^2}{A^2r^2B^2+G^2}f_{R,\theta}
\bigg\{B\dot{B}r^2-\frac{B_{,\theta}}{B}-\frac{G_{,\theta}}{r}\bigg\}
-r^2B^2\chi_1+\frac{3}{4f_R}\bigg\{\frac{1}{A^2r^2B^2+G^2}\bigg(-B^4r^4\dot{f^2_R}
\\\nonumber&+A^2r^2B^2f^2_{R,\theta}+r^2f'^2_R\bigg)\bigg\}\bigg];~~~~
\mathcal{T}_{12}=\frac{1}{\kappa}\bigg[\frac{-r^2B^2}{A^2r^2B^2+G^2}\dot{f}_R
\bigg\{\frac{G'}{2}+\frac{G}{r}+G\frac{B'}{B}\bigg\}-\frac{B_{,\theta}}{B}\bigg(
f'_R+f'_{R,\theta}\bigg)\bigg],\\\nonumber
&\mathcal{T}_{02}=\frac{1}{\kappa}\bigg[\dot{f}_{R,\theta}
+\frac{A^2r^2B^2}{A^2r^2B^2+G^2}\dot{f}_R
\bigg\{\frac{A_{,\theta}}{A}+\frac{G\dot{B}}{A^2}
+\frac{G}{2A^2f_R}\bigg\}+\frac{A^2r^2B^2}{A^2r^2B^2+G^2}
\frac{f_{R,\theta}}{2f_R}+f'_R\bigg\{\frac{G'}{2B^2}-
\frac{G}{2B^2}\frac{f'_R}{f_R}\bigg\}
\\\nonumber&-\frac{A^2}{A^2r^2B^2+G^2}\bigg\{B\dot{B}r^2
-G\frac{A_{,\theta}}{A}-\frac{G^2\dot{f}_R}{2A^2f_R}-
Gf_{R,\theta}\bigg\}-f_{R,\theta}\frac{\dot{f}_R}{2f_R}-G\chi_1
+\frac{3}{2f_R}\bigg\{
\frac{G}{2(A^2r^2B^2+G^2)}\bigg(-B^2r^2\dot{f}^2_R+f^2_{R,\theta}
\\\nonumber&+\frac{A^2r^2B^2+G^2}{B^2}f'^2_{R}\bigg)
\bigg\}-\dot{f}_{R,\theta}\bigg];~~~~
\mathcal{T}_{01}=\frac{1}{\kappa}\bigg[\dot{f'}_R
-\frac{\dot{f}_R}{{A^2r^2B^2+G^2}}
\bigg(A^2r^2B^2\frac{A'}{A}+G\frac{G'}{2A}\bigg)
-\frac{\dot{B}}{B}f'_R+3\frac{\dot{f'}_R}{2f_R}
\bigg],\\\nonumber
&\mathcal{T}_{33}=\frac{1}{\kappa}\bigg[\frac{-C^2}{A^2r^2B^2+G^2}\dot{f}_R\bigg(
r^2B^2\frac{\dot{C}}{C}-G\frac{C_{,\theta}}{C}+r^2B^2\frac{\dot{f}_R}{2f_R}
-G_{,\theta}\frac{f_{R,\theta}}{2f_R}\bigg)+C^2\frac{f'_R}{B^2}\bigg(\frac{C'}{C}+
\frac{f'_R}{2f_R}-\bigg)
-C^2\frac{f_{R,\theta}}{A^2r^2B^2+G^2}\\\nonumber&\bigg(G\frac{\dot{C}}{C}
-A^2\frac{C_{,\theta}}{C}-G\frac{\dot{f}_R}{2f_R}-A^2f_{R,\theta}\bigg)
-C^2\chi_1-\frac{3}{2f_R}\bigg\{\frac{C^2}{2(A^2r^2B^2+G^2)}
\bigg(r^2B^2\dot{f^2}_R-\frac{A^2r^2B^2+G^2}{B^2}f'^2_R-
A^2f^2_{R,\theta}\bigg)\bigg\}\bigg],
\end{align}
where
\begin{align}\nonumber
&\chi_1=\frac{1}{A^2r^2B^2+G^2}\bigg\{r^2B^2\ddot{f_R}
+\frac{A^2r^2B^2+G^2}{B^2}f''_R+
A^2f_{R,\theta\theta}\bigg\}-\frac{1}{r^2A^2B^2+G^2}
\bigg\{\frac{r^2B^2G^2}{A^2r^2B^2+G^2}
\bigg(\frac{\dot{f}^2_R}{f_R}\bigg)
+\frac{A^2r^2B^2G^2}{2B^2}\\\nonumber&\bigg(\frac{\dot{f'}^2_R}{2f_R}\bigg)
+2B^2A^2f^2_{R,\theta}
\bigg\}+
\frac{A^2r^2B^2}{A^2r^2B^2+G^2}\dot{f}_R\bigg\{\frac{\dot{B}}{A^2B}+
\frac{Br^2\dot{B}}{A^2r^2B^2+G^2}+
\frac{G^2\dot{B}}{A^2(A^2r^2B^2+G^2)}
+\frac{GB_{,\theta}}{A^2r^2B^3}
+\frac{GB_{,\theta}}{B(A^2r^2B^2+G^2)}
\\\nonumber&+\frac{r^2B^2\dot{A}}{A(A^2r^2B^2+G^2)}
+\frac{G\dot{G}}{A^2(A^2r^2B^2+G^2)}
+\frac{G_{,\theta}}{A^2r^2B^2+G^2}+\frac{\dot{C}}{A^2C}+
\frac{GC_{,\theta}}{A^2r^2B^2C}+\frac{f_{R,\theta}}{2f_R}\bigg(
\frac{2A^2G+G}{A^2(A^2r^2B^2+G^2)}+\frac{G}{A^2r^2}
\\\nonumber&+\frac{1}{2A^2r^2B^2}
\frac{G^3}{A^2r^2B^2+G^2}\bigg)\bigg\}-\frac{A^2r^2}{A^2r^2B^2+G^2}
f'_R\bigg\{\frac{B'}{B}+\frac{(Ar)'}{Ar}+\frac{GG'}{2A^2r^2B^2}\bigg\}
+\frac{A^2r^2B^2}{A^2r^2B^2+G^2}
f_{R,\theta}\bigg\{\frac{1}{A^2r^2B^2+G^2}\bigg(\frac{\dot{B}G}{B}
\\\nonumber&-\frac{GG_{,\theta}}{r^2B^2}-\frac{\dot{A}G}{A}+\frac{\dot{A}G}{A^2}+A^2\frac{A_{,\theta}}{A}
\bigg)\bigg\}.
\end{align}
\section{Appendix D}
The QS evolutionary aspects of Eqs. (B7)--(B10) are reported in the
subsequent mode, which describe the slowly evolutionary aspects of
the axially symmetric source within the bounds of energy-momentum squared gravity. These equations show the more stable scenario of the system due to the inclusion of the higher-order curvature constituents.
The mathematical expressions for these equations are found to be:
\begin{align}\nonumber
&\frac{1}{3A}\bigg(\varepsilon_I+\frac{4\pi}{\tilde{f_R}}(1
+2\mu\tilde{f}_{T^2})(\Pi_I+\mu)\bigg]^{\dot{}}
+\frac{1}{3}(\Theta\varepsilon_I+\tilde{\sigma}\varepsilon_{II})
-\Omega\bigg(\varepsilon_{KL}+\frac{4\pi}{\tilde{f_R}}(1
+2\mu\tilde{f}_{T^2})\Pi_{KL}\bigg)
-\frac{1}{rB}
\bigg(H_1\frac{C_{,\theta}}{C}+H_{1,\theta}\bigg)
\\\tag{D1}&-\frac{H_2}{B}\bigg(\frac{C'}{C}-\frac{(Br)'}{Br}\bigg)
=2a_{II}H_1-\frac{4}{3}\pi(\Theta
+\tilde{\sigma})\bigg(\tilde{\mu}^{\textrm{eff}}
+(\tilde{P}+\frac{\tilde{\Pi}_I}{3})^{\textrm{eff}}\bigg)
-8\pi a_{I}\tilde{q_{I}}^{\textrm{eff}}-\frac{4\pi}{B}
\bigg[\tilde{q_{I}}^{\textrm{eff}}
+\tilde{q_{II}}^{\textrm{eff}}\bigg(\frac{B_{,\theta}}{B}\bigg)\bigg],\\\nonumber
&\frac{1}{A}\bigg[\varepsilon_{KL}+\frac{4\pi}{\tilde{f_R}}(1
+2\mu\tilde{f}_{T^2})\Pi_{KL}\bigg]^{\dot{}}
+\frac{\Omega}{6}\bigg[\varepsilon_{I}-\varepsilon_{II}
+\frac{4\pi}{\tilde{f_R}}(1+2\mu\tilde{f}_{T^2})(\Pi_{I}-\Pi_{II})
\bigg]-(a_{II}H_2
-a_{I}H_1)-\bigg\{\varepsilon_{KL}+\frac{4\pi}{\tilde{f_R}}
\\\nonumber&\bigg(1+2\mu\tilde{f}_{T^2}\bigg)\Pi_{KL}\bigg\}
(\tilde{\sigma}-\Theta)-\frac{1}{2}\bigg[\frac{H_1}{B}\bigg(\frac{(Br)'}{Br}
-\frac{2C'}{C}\bigg)-\frac{H'_1}{B}\bigg]
-\frac{1}{2r}\bigg[\frac{H_{2,\theta}}{B}
-\frac{H_2}{B}\bigg(\frac{B_{,\theta}}{B}
-\frac{2C_{,\theta}}{C}\bigg)\bigg]
=-\frac{8\pi}{3}\bigg(2\tilde{\sigma}-\Theta\bigg)
\\\tag{D2}&\tilde{\Pi}_{KL}^{\textrm{eff}}
-4\pi
\bigg(a_{II}\tilde{q_{I}}^{\textrm{eff}}
+a_{I}\tilde{q_{II}}^{\textrm{eff}}\bigg)
-\frac{2\pi}{B\tilde{f_R}}(1
+2\mu\tilde{f}_{T^2})\bigg(q'_{II}-q_{II}\frac{(Br)'}{Br}\bigg)
-\frac{2\pi}{rB\tilde{f_R}}(1+2\mu\tilde{f}_{T^2})\bigg(q_{I,\theta}
-q_{I,\theta}\frac{B_{,\theta}}{B}\bigg),\\\nonumber
&\frac{1}{3A}\bigg[\varepsilon_{II}+\frac{4\pi}{\tilde{f_R}}(1
+2\mu\tilde{f}_{T^2})(\Pi_{II}+\mu)\bigg]^{\dot{}}
+\frac{1}{3}(\Theta\varepsilon_{II}+\tilde{\sigma}\varepsilon_{I})
+\Omega\bigg(\varepsilon_{KL}+\frac{4\pi}{\tilde{f_R}}(1
+2\mu\tilde{f}_{T^2})\Pi_{KL}\bigg)
+2H_2a_{I}+\frac{1}{B}\bigg(H_2\frac{C'}{C}
\\\nonumber&+H'_2\bigg)+\frac{H_1}{rB}
\bigg(\frac{C_{,\theta}}{C}-\frac{B_{,\theta}}{B}\bigg)=
-\frac{4\pi}{3}(\Theta+\tilde{\sigma})\bigg(\tilde{\mu}^{\textrm{eff}}
+(\tilde{P}+\frac{\tilde{\Pi}_{II}}{3})^{\textrm{eff}}\bigg)
-8\pi a_{I}\tilde{q_{I}}^{\textrm{eff}}
-\frac{4\pi}{\tilde{f_R}}(1+2\mu\tilde{f}_{T^2})\bigg[\frac{q_{II}}{Br}
+\frac{q_{I}}{B}
\\\tag{D3}&\times\frac{(Br)'}{Br}\bigg],\\\nonumber
&\frac{1}{3A}\bigg[-\frac{4\pi}{\tilde{f_R}}(1+2\mu\tilde{f}_{T^2})(-\mu
+\Pi_{I}
+\Pi_{II})-(\varepsilon_{II}+\varepsilon_{I})\bigg]^{\dot{}}
+\frac{4\pi}{9\tilde{f_R}}(1+2\mu\tilde{f}_{T^2})(\Pi_{I}
+\Pi_{II})(2\tilde{\sigma}-\Theta)
-2(a_IH_2-a_{II}H_1)\\\nonumber&-\frac{1}{3}(\Theta
+\tilde{\sigma})(\varepsilon_{I}+\varepsilon_{II})
+\frac{1}{B}\bigg(H_2\frac{(Br)'}{Br}+H'_2\bigg)+\frac{1}{rB}
\bigg(H_{1,\theta}+\frac{B_{,\theta}}{B}H_1\bigg)
=-\frac{4\pi}{3}(\tilde{\mu}^{\textrm{eff}}
+\tilde{P}^{\textrm{eff}})(\Theta-2\tilde{\sigma})
-\frac{4\pi\tilde{q_{I}}^{\textrm{eff}}}{B}\bigg(\frac{C'}{C}\bigg)
\\\tag{D4}&-\frac{4\pi\tilde{q_{II}}^{\textrm{eff}}}{Br}
\bigg(\frac{C_{,\theta}}{C}\bigg),\\\nonumber
&\frac{1}{3B}\bigg[\varepsilon_I+\frac{4\pi}{\tilde{f_R}}(1
+2\mu\tilde{f}_{T^2})\Pi_I\bigg]'
+\frac{1}{Br}\bigg[\frac{4\pi}{\tilde{f_R}}(1
+2\mu\tilde{f}_{T^2})\Pi_{KL}+\varepsilon_{KL}\bigg]_{,\theta}
+\frac{1}{3B}\bigg(\varepsilon_I\frac{4\pi}{\tilde{f_R}}(1+
+2\mu\tilde{f}_{T^2})\Pi_I\bigg)
\bigg(\frac{2C'}{C}+\frac{(Br)'}{Br}\bigg)
\\\nonumber&+\frac{1}{3B}\bigg(\varepsilon_{II}+\frac{4\pi}{\tilde{f_R}}(1
+2\mu\tilde{f}_{T^2})\Pi_{II}\bigg)
\bigg(\frac{C'}{C}-\frac{(Br)'}{Br}\bigg) +\frac{1}{Br}
\bigg(\varepsilon_{KL}+\frac{4\pi}{\tilde{f_R}}\bigg(1
+2\mu\tilde{f}_{T^2}\bigg)\Pi_{KL}\bigg)
\bigg(\frac{C_{,\theta}}{C}-\frac{B_{,\theta}}{B}\bigg)
=\frac{8\pi}{3B\tilde{f_R}}\bigg(1\\\tag{D5}&
+2\mu\tilde{f}_{T^2}\bigg)\mu',\\\nonumber
&\frac{1}{3Br}\bigg[\varepsilon_{II}+\frac{4\pi}{\tilde{f_R}}(1
+2\mu\tilde{f}_{T^2})\Pi_{II}\bigg]_{,\theta}
+\frac{1}{B}\bigg[\frac{4\pi}{\tilde{f_R}}(1+2\mu\tilde{f}_{T^2})\Pi_{KL}
+\varepsilon_{KL}\bigg]'
+\frac{1}{3Br}\bigg(\varepsilon_I+\frac{4\pi}{\tilde{f_R}}
(1+2\mu\tilde{f}_{T^2})\Pi_I\bigg)\bigg(\frac{C_{,\theta}}
{C}-\frac{B_{,\theta}}{B}\bigg)
\\\nonumber&+\frac{1}{3Br}\bigg(\varepsilon_{II}+\frac{4\pi}{\tilde{f_R}}(1
+2\mu\tilde{f}_{T^2})\Pi_{II}\bigg)
\bigg(\frac{2C_{,\theta}}{C}+\frac{B_{,\theta}}{B}\bigg)
+\frac{1}{B}
\bigg(\frac{4\pi}{\tilde{f_R}}(1
+2\mu\tilde{f}_{T^2})\Pi_{KL}+\varepsilon_{KL}\bigg)
\bigg(\frac{2C'}{C}+\frac{(Br)'}{Br}\bigg)
=\frac{8\pi}{3B\tilde{f_R}}\bigg(1
\\\tag{D6}&+2\mu\tilde{f}_{T^2}\bigg)\mu_{,\theta},\\\nonumber
&-\frac{1}{B}\bigg[H_1\bigg(\frac{(Br)'}{Br}+\frac{2C'}{C}\bigg)+H'_1\bigg]
-\frac{1}{rB}\bigg[H_{2,\theta}+H_2\bigg(\frac{B_{,\theta}}{B}
+\frac{2C_{,\theta}}{C}\bigg)\bigg]
+\frac{4\pi}{B\tilde{f_R}}(1+2\mu\tilde{f}_{T^2})
\bigg(q_{II}\frac{(Br)'}{Br}
+q'_{II}\bigg)+\Omega\bigg[8\pi\\\label{2p30}&\bigg(\tilde{\mu}^{\textrm{eff}}
+\tilde{P}^{\textrm{eff}}\bigg)-(\varepsilon_{I}+\varepsilon_{II})
+\frac{4\pi}{3\tilde{f_R}}
(1+\tilde{f_T})\bigg(\Pi_{I}+\Pi_{II}\bigg)\bigg]-\frac{4\pi}{rB\tilde{f_R}}
(1+2\mu\tilde{f}_{T^2})\bigg\{q_{I,\theta}
+q_{I}\frac{B_{,\theta}}{B}\bigg\},\\\nonumber
&-\frac{1}{B}\bigg[\frac{4\pi}{3\tilde{f_R}}(1
+2\mu\tilde{f}_{T^2})\Pi_{KL}\bigg]'
+\frac{1}{rB}\bigg[\frac{4\pi}{3\tilde{f_R}}(1+2\mu\tilde{f}_{T^2})
\Pi_{KL}\bigg]_{,\theta}
-\frac{\varepsilon_I}{3rB}\bigg(\frac{C_{,\theta}}{C}
+\frac{2A_{,\theta}}{A}\bigg)
-\frac{\varepsilon_{II}}{3rB}
\bigg(\frac{2C_{,\theta}}{C}+\frac{A_{,\theta}}{A}\bigg)
-\frac{8\pi}{B\tilde{f_R}}
\\\nonumber&\bigg(1+2\mu\tilde{f}_{T^2}\bigg)\Pi_{KL}\frac{(Br)'}{Br}
-\frac{\varepsilon_{KL}}{B}
\bigg(\frac{C'}{C}-\frac{A'}{A}\bigg)+\frac{4\pi}{3rB\tilde{f_R}}(1
+2\mu\tilde{f}_{T^2})(\Pi_{I}
-\Pi_{II})\frac{B_{,\theta}}{B}+\frac{1}{A}\dot{H_1}
=-\frac{8\pi}{3B\tilde{f_R}}\bigg(1+
\\\tag{D7}&2\mu\tilde{f}_{T^2}\bigg)\mu_{,\theta},\\\nonumber
&-\frac{1}{B}\bigg[\frac{4\pi}{3\tilde{f_R}}(1
+2\mu\tilde{f}_{T^2})\Pi_{II}
-(\varepsilon_{I}-\varepsilon_{II})\bigg]'+\frac{4\pi}{rB\tilde{f_R}}(1
+2\mu\tilde{f}_{T^2})
\bigg(\Pi_{KL}\frac{2B_{,\theta}}{B}+\Pi_{KL,\theta}\bigg)
+\frac{1}{A}\dot{H_2}
+\frac{\varepsilon_{I}}{3B}\bigg(\frac{2C'}{C}
+\frac{A'}{A}\bigg)
\\\tag{D8}&+\frac{\varepsilon_{II}}{3B}\bigg(\frac{C'}{C}+\frac{2A'}{A}\bigg)
+\frac{4\pi}{3B\tilde{f_R}}(1+2\mu\tilde{f}_{T^2})(\Pi_{I}
-\Pi_{II})\frac{(Br)'}{Br}+\frac{\varepsilon_{KL}}{rB}\bigg(\frac{C_{,\theta}}{C}
-\frac{A_{,\theta}}{A}\bigg)
=\frac{4\pi}{3B\tilde{f_R}}(1+2\mu\tilde{f}_{T^2})\mu',
\end{align}
where $\tilde{q_{II}}^{\textrm{eff}}$ indicates the quasi
static-evolution of generalized component of heat-flux i.e.,
$q_{II}^{\textrm{eff}}$. So, one can be easily calculated the quasi
static-evolution of effective components of relativistic-fluid by
utilizing the conditions of the QS approximation as described
earlier.
\vspace{0.5cm}


\end{document}